\pdfoutput=1

\documentclass[12pt,a4paper]{article}

\usepackage{ifthen} 
\newboolean{pdflatex}
\setboolean{pdflatex}{true} 

\newboolean{articletitles}
\setboolean{articletitles}{true} 

\newboolean{uprightparticles}
\setboolean{uprightparticles}{false} 

\def\paperauthors{LHCb collaboration} 
\def\paperasciititle{} 
\def\papertitle{Study of light-meson resonances decaying to $K^0_{\rm S} K \pi$ in the $B \to (K^0_{\rm S} K \pi) K$ channels} 
\def\paperkeywords{{High Energy Physics}, {LHCb}} 
\def\papercopyright{\the\year\ CERN for the benefit of the LHCb collaboration} 
\def\paperlicence{CC BY 4.0 licence}
\def\paperlicenceurl{https://creativecommons.org/licenses/by/4.0/}

\usepackage[top=1in, bottom=1.25in, left=1in, right=1in]{geometry}

\usepackage{microtype}
\usepackage{lineno}  
\usepackage{xspace} 
\usepackage{caption} 

\usepackage{graphicx}  
\usepackage{color}
\usepackage{colortbl}
\graphicspath{{./figs/}} 

\usepackage{amsmath} 
\usepackage{amssymb}
\usepackage{amsfonts}
\usepackage{upgreek} 

\newcommand*\patchAmsMathEnvironmentForLineno[1]{%
\expandafter\let\csname old#1\expandafter\endcsname\csname #1\endcsname
\expandafter\let\csname oldend#1\expandafter\endcsname\csname
end#1\endcsname
 \renewenvironment{#1}%
   {\linenomath\csname old#1\endcsname}%
   {\csname oldend#1\endcsname\endlinenomath}%
}
\newcommand*\patchBothAmsMathEnvironmentsForLineno[1]{%
  \patchAmsMathEnvironmentForLineno{#1}%
  \patchAmsMathEnvironmentForLineno{#1*}%
}
\AtBeginDocument{%
\patchBothAmsMathEnvironmentsForLineno{equation}%
\patchBothAmsMathEnvironmentsForLineno{align}%
\patchBothAmsMathEnvironmentsForLineno{flalign}%
\patchBothAmsMathEnvironmentsForLineno{alignat}%
\patchBothAmsMathEnvironmentsForLineno{gather}%
\patchBothAmsMathEnvironmentsForLineno{multline}%
\patchBothAmsMathEnvironmentsForLineno{eqnarray}%
}

\usepackage{hyperxmp}

\usepackage[pdftex,
            pdfauthor={\paperauthors},
            pdftitle={\paperasciititle},
            pdfkeywords={\paperkeywords},
            pdfcopyright={Copyright (C) \papercopyright},
            pdflicenseurl={\paperlicenceurl}]{hyperref}

\usepackage[colorinlistoftodos,textsize=scriptsize]{todonotes}

\usepackage[bottom,flushmargin,hang,multiple]{footmisc}

\usepackage[all]{hypcap} 

\usepackage{xspace} 
\usepackage{upgreek}

\def\lhcb   {\mbox{LHCb}\xspace}

\def\babar  {\mbox{BaBar}\xspace}

\def\besiii {\mbox{BESIII}\xspace}

\def\MagUp {\mbox{\em Mag\kern -0.05em Up}\xspace}

\ifthenelse{\boolean{uprightparticles}}%
{

 \def\Peta        {\ensuremath{\upeta}\xspace}

 \def\Ppi         {\ensuremath{\uppi}\xspace}

 \def\Ppsi        {\ensuremath{\uppsi}\xspace}

 \def\PDelta      {\ensuremath{\Delta}\xspace}                 
 \def\PXi         {\ensuremath{\Xi}\xspace}                 
 \def\PLambda     {\ensuremath{\Lambda}\xspace}                 
 \def\PSigma      {\ensuremath{\Sigma}\xspace}                 
 \def\POmega      {\ensuremath{\Omega}\xspace}                 
 \def\PUpsilon    {\ensuremath{\Upsilon}\xspace}
 \let\oldPi\Pi
 \def\PPi         {\ensuremath{\oldPi}\xspace}

 \def\PB      {\ensuremath{\mathrm{B}}\xspace}                 
                  
 \def\PD      {\ensuremath{\mathrm{D}}\xspace}

 \def\PJ      {\ensuremath{\mathrm{J}}\xspace}                 
 \def\PK      {\ensuremath{\mathrm{K}}\xspace}

 \def\Pb      {\ensuremath{\mathrm{b}}\xspace}                 
 \def\Pc      {\ensuremath{\mathrm{c}}\xspace}

 \def\Pi      {\ensuremath{\mathrm{i}}\xspace}

 \def\Ps      {\ensuremath{\mathrm{s}}\xspace}

 \def\thebaroffset{0.0em}
}
{

 \def\Peta        {\ensuremath{\eta}\xspace}

 \def\Ppi         {\ensuremath{\pi}\xspace}

 \def\Ppsi        {\ensuremath{\psi}\xspace}                 
                  
 \mathchardef\PDelta="7101
 \mathchardef\PXi="7104
 \mathchardef\PLambda="7103
 \mathchardef\PSigma="7106
 \mathchardef\POmega="710A
 \mathchardef\PUpsilon="7107
 \mathchardef\PPi="7105
                  
 \def\PB      {\ensuremath{B}\xspace}                 
                  
 \def\PD      {\ensuremath{D}\xspace}

 \def\PJ      {\ensuremath{J}\xspace}                 
 \def\PK      {\ensuremath{K}\xspace}

 \def\Pb      {\ensuremath{b}\xspace}                 
 \def\Pc      {\ensuremath{c}\xspace}

 \def\Pi      {\ensuremath{i}\xspace}

 \def\Ps      {\ensuremath{s}\xspace}

 \def\thebaroffset{0.18em}
}
\newcommand{\offsetoverline}[2][\thebaroffset]{\kern #1\overline{\kern -#1 #2}}%

\makeatletter
\ifcase \@ptsize \relax
  \newcommand{\miniscule}{\@setfontsize\miniscule{4}{5}}
\or
  \newcommand{\miniscule}{\@setfontsize\miniscule{5}{6}}
\or
  \newcommand{\miniscule}{\@setfontsize\miniscule{5}{6}}
\fi
\makeatother

\DeclareRobustCommand{\optbar}[1]{\shortstack{{\miniscule (\rule[.5ex]{1.25em}{.18mm})}
  \\ [-.7ex] $#1$}}

\def\squark    {{\ensuremath{\Ps}}\xspace}

\def\cquark    {{\ensuremath{\Pc}}\xspace}

\def\bquark    {{\ensuremath{\Pb}}\xspace}

\def\pion   {{\ensuremath{\Ppi}}\xspace}
\def\piz    {{\ensuremath{\pion^0}}\xspace}
\def\pip    {{\ensuremath{\pion^+}}\xspace}
\def\pim    {{\ensuremath{\pion^-}}\xspace}

\def\pimp   {{\ensuremath{\pion^\mp}}\xspace}

\def\kaon    {{\ensuremath{\PK}}\xspace}
\def\Kbar    {{\ensuremath{\offsetoverline{\PK}}}\xspace}
\def\Kb      {{\ensuremath{\Kbar}}\xspace}
\def\KorKbar {\kern \thebaroffset\optbar{\kern -\thebaroffset \PK}{}\xspace}
\def\Kz      {{\ensuremath{\kaon^0}}\xspace}
\def\Kzb     {{\ensuremath{\Kbar{}^0}}\xspace}
\def\Kp      {{\ensuremath{\kaon^+}}\xspace}
\def\Km      {{\ensuremath{\kaon^-}}\xspace}
\def\Kpm     {{\ensuremath{\kaon^\pm}}\xspace}

\def\KS      {{\ensuremath{\kaon^0_{\mathrm{S}}}}\xspace}

\def\KL      {{\ensuremath{\kaon^0_{\mathrm{L}}}}\xspace}

\def\Kstar   {{\ensuremath{\kaon^*}}\xspace}

\def\Dbar    {{\ensuremath{\offsetoverline{\PD}}}\xspace}
\def\D       {{\ensuremath{\PD}}\xspace}

\def\DorDbar {\kern \thebaroffset\optbar{\kern -\thebaroffset \PD}\xspace}
\def\Dz      {{\ensuremath{\D^0}}\xspace}
\def\Dzb     {{\ensuremath{\Dbar{}^0}}\xspace}
\def\Dp      {{\ensuremath{\D^+}}\xspace}
\def\Dm      {{\ensuremath{\D^-}}\xspace}

\def\DpDm    {\ensuremath{\Dp {\kern -0.16em \Dm}}\xspace}

\def\B       {{\ensuremath{\PB}}\xspace}

\def\BorBbar {\kern \thebaroffset\optbar{\kern -\thebaroffset \PB}\xspace}

\def\Bd      {{\ensuremath{\B^0}}\xspace}

\def\BdorBdbar {\kern \thebaroffset\optbar{\kern -\thebaroffset \Bd}\xspace}
\def\Bu      {{\ensuremath{\B^+}}\xspace}
\def\Bub     {{\ensuremath{\B^-}}\xspace}
\def\Bp      {{\ensuremath{\Bu}}\xspace}

\def\Bs      {{\ensuremath{\B^0_\squark}}\xspace}

\def\BsorBsbar {\kern \thebaroffset\optbar{\kern -\thebaroffset \Bs}\xspace}

\def\jpsi     {{\ensuremath{{\PJ\mskip -3mu/\mskip -2mu\Ppsi}}}\xspace}

\def\etac     {{\ensuremath{\Peta_\cquark}}\xspace}

\def\Y#1S{\ensuremath{\PUpsilon{(#1S)}}\xspace}

\def\LorLbar     {\kern \thebaroffset\optbar{\kern -\thebaroffset \PLambda}\xspace}

\def\to                 {\ensuremath{\rightarrow}\xspace}

\def\AT#1     {\ensuremath{A_{\mathrm{T}}^{#1}}\xspace}           

\def\C#1      {\ensuremath{\mathcal{C}_{#1}}\xspace}                       
\def\Cp#1     {\ensuremath{\mathcal{C}_{#1}^{'}}\xspace}                    
\def\Ceff#1   {\ensuremath{\mathcal{C}_{#1}^{\mathrm{(eff)}}}\xspace}        
\def\Cpeff#1  {\ensuremath{\mathcal{C}_{#1}^{'\mathrm{(eff)}}}\xspace}       
\def\Ope#1    {\ensuremath{\mathcal{O}_{#1}}\xspace}                       
\def\Opep#1   {\ensuremath{\mathcal{O}_{#1}^{'}}\xspace}                    

\newcommand{\aunit}[1]{\ensuremath{\text{\,#1}}}       

\newcommand{\tev}{\aunit{Te\kern -0.1em V}\xspace}
\newcommand{\gev}{\aunit{Ge\kern -0.1em V}\xspace}
\newcommand{\mev}{\aunit{Me\kern -0.1em V}\xspace}
\newcommand{\kev}{\aunit{ke\kern -0.1em V}\xspace}
\newcommand{\ev}{\aunit{e\kern -0.1em V}\xspace}
 
\newcommand{\mevc}{\ensuremath{\aunit{Me\kern -0.1em V\!/}c}\xspace}
\newcommand{\gevc}{\ensuremath{\aunit{Ge\kern -0.1em V\!/}c}\xspace}
\newcommand{\mevcc}{\ensuremath{\aunit{Me\kern -0.1em V\!/}c^2}\xspace}
\newcommand{\gevcc}{\ensuremath{\aunit{Ge\kern -0.1em V\!/}c^2}\xspace}

\def\fb   {\ensuremath{\aunit{fb}}\xspace}
\def\invfb   {\ensuremath{\fb^{-1}}\xspace}

\newcommand{\chisqndf}{\ensuremath{\chi^2/\mathrm{ndf}}\xspace}

\def\gsim{{~\raise.15em\hbox{$>$}\kern-.85em
          \lower.35em\hbox{$\sim$}~}\xspace}
\def\lsim{{~\raise.15em\hbox{$<$}\kern-.85em
          \lower.35em\hbox{$\sim$}~}\xspace}

\newcommand{\Real}{\ensuremath{\mathcal{R}e}\xspace}

\def\pt         {\ensuremath{p_{\mathrm{T}}}\xspace}

\def\evtgen     {\mbox{\textsc{EvtGen}}\xspace}

\def\geant      {\mbox{\textsc{Geant4}}\xspace}

\def\photos     {\mbox{\textsc{Photos}}\xspace}

\def\pythia     {\mbox{\textsc{Pythia}}\xspace}

\def\tell1  {TELL1\xspace}
\def\ukl1   {UKL1\xspace}

\newcommand{\etc}{\mbox{\itshape etc.}\xspace}

\newcommand{\lhcborcid}[1]{\href{https://orcid.org/#1}{\hspace*{0.1em}\raisebox{-0.45ex}{\includegraphics[width=1em]{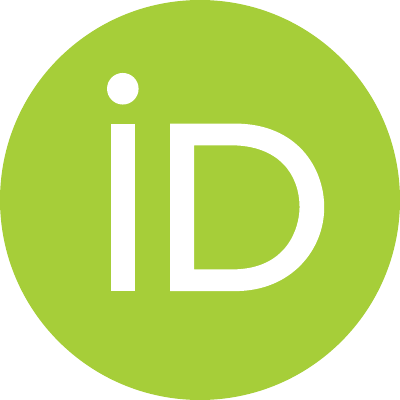}}}}

\usepackage{cite} 
\usepackage{mciteplus}

\usepackage{lscape}
\usepackage{longtable} 
\def\calL         {{\ensuremath{\cal L}\xspace}}
\def\KSLL      {{\ensuremath{\kaon^0_{\mathrm{SLL}}}}\xspace}
\def\KSDD      {{\ensuremath{\kaon^0_{\mathrm{SDD}}}}\xspace}
\def\calB         {{\ensuremath{\cal B}\xspace}}
\def\calR         {{\ensuremath{\cal R}\xspace}}
\def\kskkpi {{\ensuremath{\KS K \pi K}}\xspace}
\def\kskkpip {{\ensuremath{\KS\Km\pip\Kp}}\xspace}
\def\kskkpim {{\ensuremath{\KS\Kp\pim\Kp}}\xspace}
\def\bkskkpip {{\ensuremath{\Bu\to\KS\Km\pip\Kp}}\xspace}
\def\bkskkpim {{\ensuremath{\Bu\to\KS\Kp\pim\Kp}}\xspace}

\def\bkzkkpip {{\ensuremath{\Bu\to\Kz\Km\pip\Kp}}\xspace}
\def\bkzkkpim {{\ensuremath{\Bu\to\Kzb\Kp\pim\Kp}}\xspace}
\def\bkskkpi {{\ensuremath{B\to\KS K \pi K}}\xspace}
\def\kskpi {{\ensuremath{\KS K\pi}}\xspace}

\def\kskpim {{\ensuremath{\KS\Kp\pim}}\xspace}
\def\kskpip {{\ensuremath{\KS\Km\pip}}\xspace}

\newcommand{\etaetapr}{\ensuremath{\Peta^{(\prime)}}\xspace}
\newcommand{\alp}{\ensuremath{\kern 1.0em }}
\newcommand{\al}{\ensuremath{\kern 0.5em }}
\newcommand{\all}{\ensuremath{\kern 0.25em }}
\newcommand{\allm}{\ensuremath{\kern 0.15em }}
\newcommand{\allmm}{\ensuremath{\kern -1.25em }}
\newcommand{\almm}{\ensuremath{\kern -0.75em }}
\newcommand{\aln}{\ensuremath{\kern -0.25em }}
\newcommand{\alm}{\ensuremath{\kern -0.50em }}
\newcommand{\alpp}{\ensuremath{\kern 0.75em }}
\mathchardef\myhyphen="2D

\begin{document}

\renewcommand{\thefootnote}{\fnsymbol{footnote}}
\setcounter{footnote}{1}


\begin{titlepage}
\pagenumbering{roman}

\vspace*{-1.5cm}
\centerline{\large EUROPEAN ORGANIZATION FOR NUCLEAR RESEARCH (CERN)}
\vspace*{1.5cm}
\noindent
\begin{tabular*}{\linewidth}{lc@{\extracolsep{\fill}}r@{\extracolsep{0pt}}}
\ifthenelse{\boolean{pdflatex}}
{\vspace*{-1.5cm}\mbox{\!\!\!\includegraphics[width=.14\textwidth]{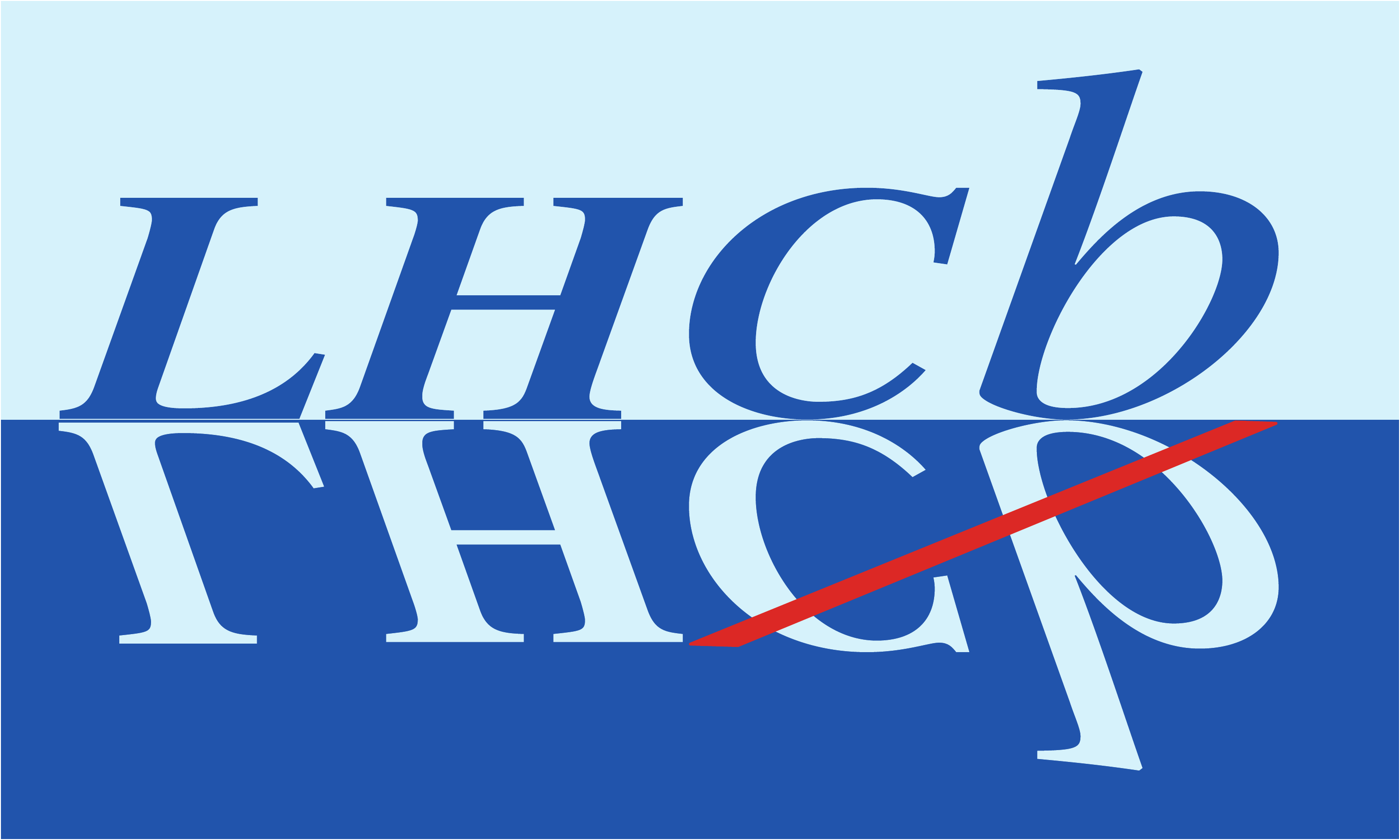}} & &}%
{\vspace*{-1.2cm}\mbox{\!\!\!\includegraphics[width=.12\textwidth]{figs/lhcb-logo.eps}} & &}%
\\
 & & CERN-EP-2024-329 \\  
& & LHCb-PAPER-2024-045\\
 & & May 15, 2025 \\ 
 & & \\
\end{tabular*}

\vspace*{4.0cm}

{\normalfont\bfseries\boldmath\huge
\begin{center}
  \papertitle 
\end{center}
}

\vspace*{2.0cm}

\begin{center}
\paperauthors\footnote{Authors are listed at the end of this paper.}
\end{center}

\vspace{\fill}

\begin{abstract}
  \noindent

A study is presented of \bkskkpip and \bkskkpim decays 
based on the analysis of proton-proton collision data collected with the \lhcb detector at centre-of-mass energies of 7, 8 and 13 \tev, corresponding to an integrated luminosity
of $9 \invfb$.
The \kskpi invariant-mass distributions of both \Bu decay modes show, in the $m(\kskpi)<1.85$ \gev mass region, large activity which is resolved using an amplitude analysis. A simple model, were $J^{PC}$ amplitudes are described by multiple Breit--Wigner functions with appropriate angular distributions provides a good description of the experimental data. In this approach a complex mixture of $J^{PC}=0^{-+}$,$ 1^{++}$ and $1^{+-}$ amplitudes is observed that is dominated by $\eta(1405)$, $\eta(1470)$, $\eta(1760)$, $f_1(1285)$, $f_1(1420)$ and $h_1(1405)$ resonances. The \kskpi Dalitz 
plots are dominated by asymmetric crossing $K^* \Kbar$ bands which are different for the two \Bu decay modes.
This is due to a different interference pattern
between the $1^{++}$ and $1^{+-}$ amplitudes in the two channels.  Branching fractions are measured for each
resonant contribution.   
\end{abstract}


\begin{center}
  Published in Physical Review D 111, (2025) 092009
\end{center}

\vspace{\fill}

{\footnotesize 
\centerline{\copyright~\papercopyright. \href{\paperlicenceurl}{\paperlicence}.}}
\vspace*{2mm}

\end{titlepage}


\newpage
\setcounter{page}{2}
\mbox{~}
%
%
%
%


\renewcommand{\thefootnote}{\arabic{footnote}}
\setcounter{footnote}{0}

\cleardoublepage


\pagestyle{plain} 
\setcounter{page}{1}
\pagenumbering{arabic}


\section{Introduction}
\label{sec:Introduction}
Quantum chromodynamics (QCD) allows, in addition to mesons and baryons, the existence of exotic states such as glueballs, hybrids, and multiquark states. In particular, gluonium states have been extensively searched for over the past few decades in several processes
such as radiative decays of charmonium, central production, $\bar p p$ annihilations, \etc~\cite{Klempt:2007cp}.
The experimental confirmation of
states having valence gluon content 
would provide fundamental information about QCD in the confinement regime and would be a direct test of QCD theory~\cite{Chen:2005mg}. Significant progress on the experimental side has been made,
but many issues remain unresolved~\cite{Klempt:2007cp,Ochs:2013gi}.
In the sector of the pseudoscalar glueball~\cite{Chen:2022asf}, phenomenological models~\cite{Jaffe:1975fd,Carlson:1982er,Chanowitz:1982qj} calculate a mass around $1.4\gev$,\footnote{Natural units with $\hbar = c = 1$ are used throughout this paper.} while 
lattice QCD calculations~\cite{Chen:2005mg} predict a mass around $2.5\gev$. 
The difference in predicted masses is caused by the assumed effective gluon mass inside the hadrons~\cite{Chen:2022asf}.
 
One of the most interesting and disputed questions is the nature of the pseudoscalar structure with a mass around $1.4\gev$, the
so-called ``$\iota(1440)$'' state. This was first observed in the early 1980s by the Mark II and Crystal Ball collaborations~\cite{Scharre:1980zh,Edwards:1982nc} in the $K \Kb \pi$ final state using \jpsi radiative decays. A summary of the experimental and phenomenological status of the subject can be found in Ref.~\cite{Kopke:1988cs}.
The structure has been subsequently confirmed by different experiments and is often interpreted as the combination of three states: two pseudoscalars, the $\eta(1405)$, $\eta(1475)$ and one axial state, the $f_1(1420)$ meson~\cite{Bai:1990hs}.
Pseudoscalar structures in the mass region below $2\gev$ are also observed in the $\etaetapr \pi \pi$ and  $VV$ (where $V$ refers to $\rho, \omega$ vector mesons) final states~\cite{Kopke:1988cs}. Whether these observed pseudoscalar structures originate from the same source remains unclear. However, an amplitude analysis  of $\jpsi \to \gamma \omega \omega$ decays by the BES experiment provides strong evidence of a  pseudoscalar state, labeled as the $\eta(1760)$~\cite{BES:2006nqh}.
Recently, the \besiii experiment  has performed a high-yield partial-wave analysis of the $\KS \KS \piz$ system produced in radiative \jpsi decays~\cite{BESIII:2022chl}, confirming the previously observed resonance composition of the structure around $1.4\gev$.

The $\KS K \pi$ mass region below $1.6\gev$ is further complicated by the presence of two $J^{PC}=1^{++}$ states, $f_1(1420)$ and $f_1(1510)$, both potential candidates for the $s \bar s$ member of the $J^{PC}=1^{++}$ nonet. The $f_1(1420)$ state is observed in $\pim p$ interactions and central production~\cite{WA76:1986sfi}, while the $f_1(1510)$ resonance is mostly seen in $K^- p$ interactions~\cite{Aston:1987ak,Gavillet:1982tv}.
In the latter, strong interference effects are observed between the $J^{PC}=1^{++}$ $f_1(1510)$ and the $J^{PC}=1^{+-}$ $h_1(1415)$ states~\cite{Aston:1987ak}. The presence of two $J^{PC}=1^{++}$ states close in mass suggests the possibility that the $f_1(1420)$ could be a $K^* \Kb$ molecule~\cite{Longacre:1990uc}.

\begin{figure}[htb]
\centering
\small
\includegraphics[width=0.45\textwidth]{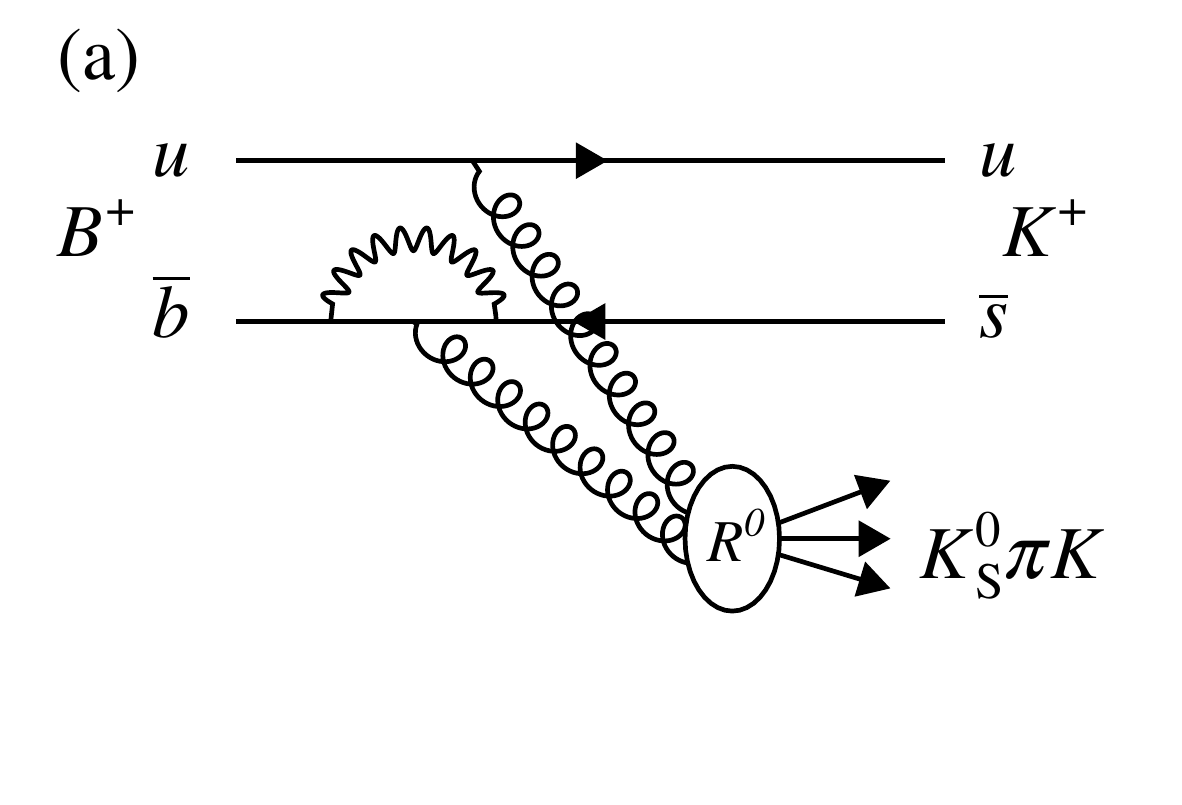}
\includegraphics[width=0.45\textwidth]{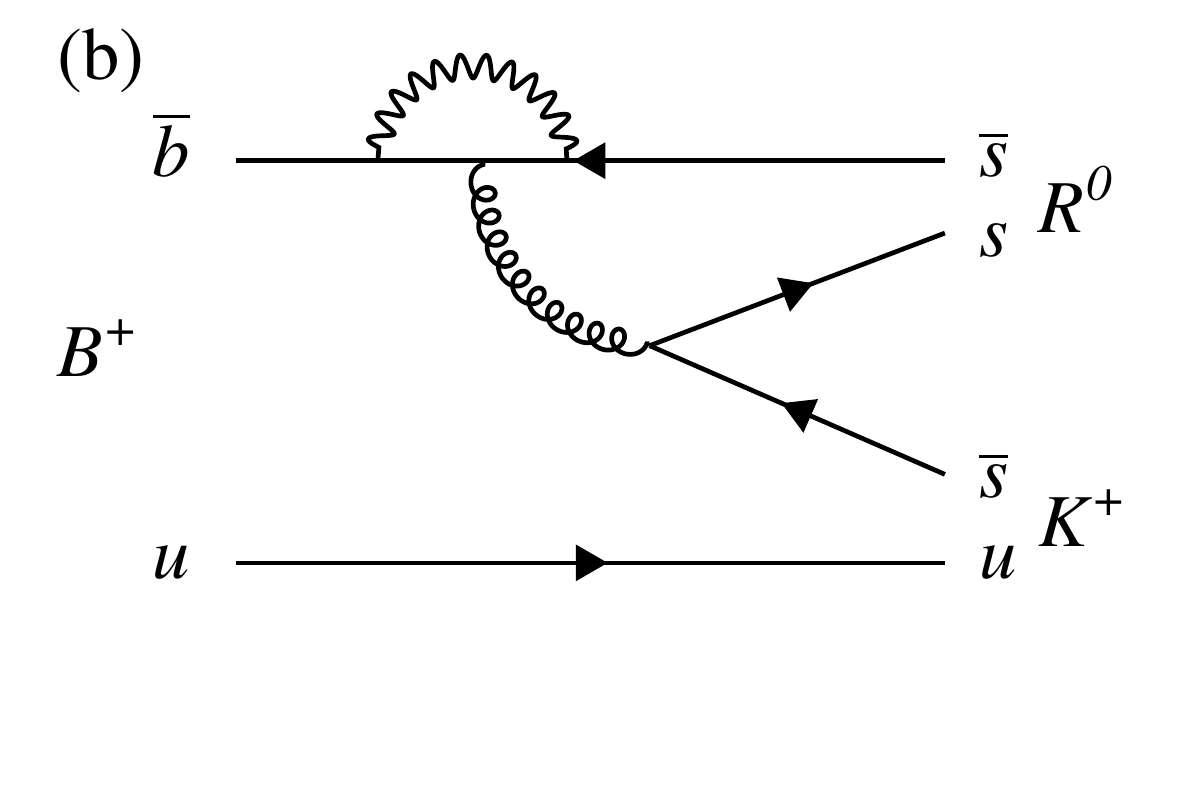}
\includegraphics[width=0.45\textwidth]{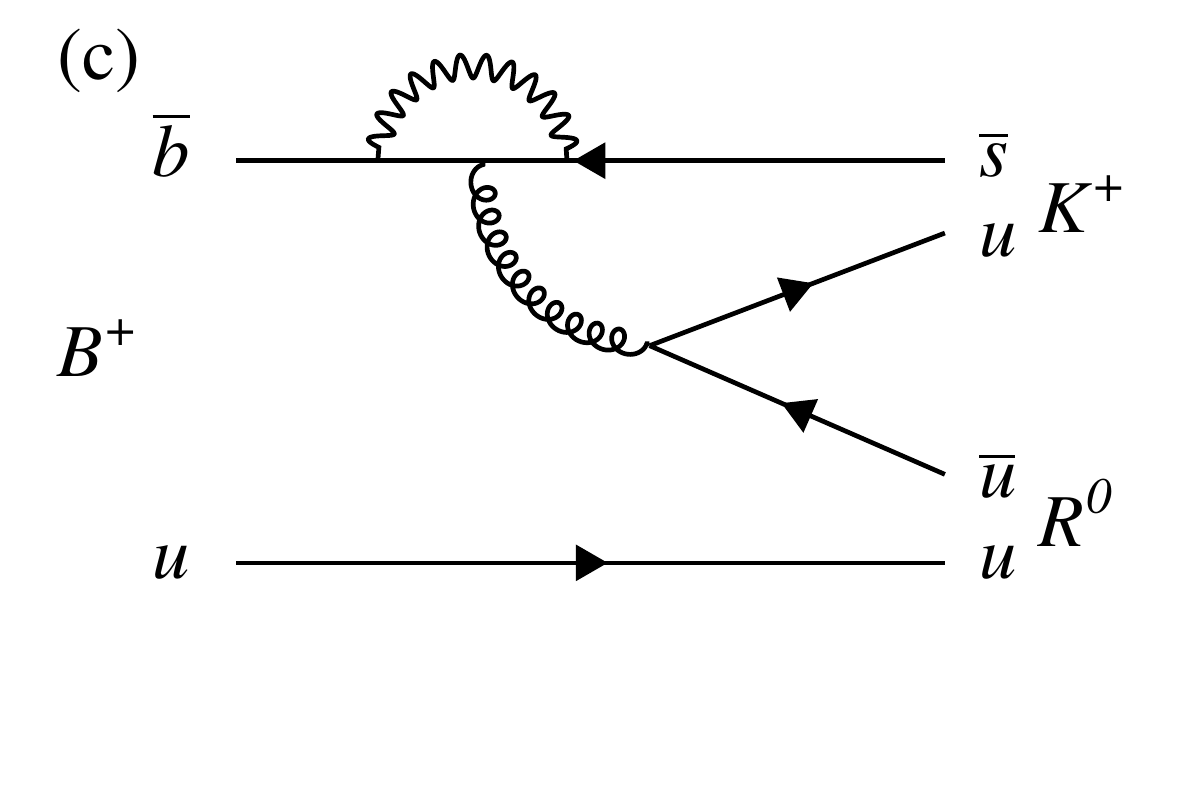}
\includegraphics[width=0.45\textwidth]{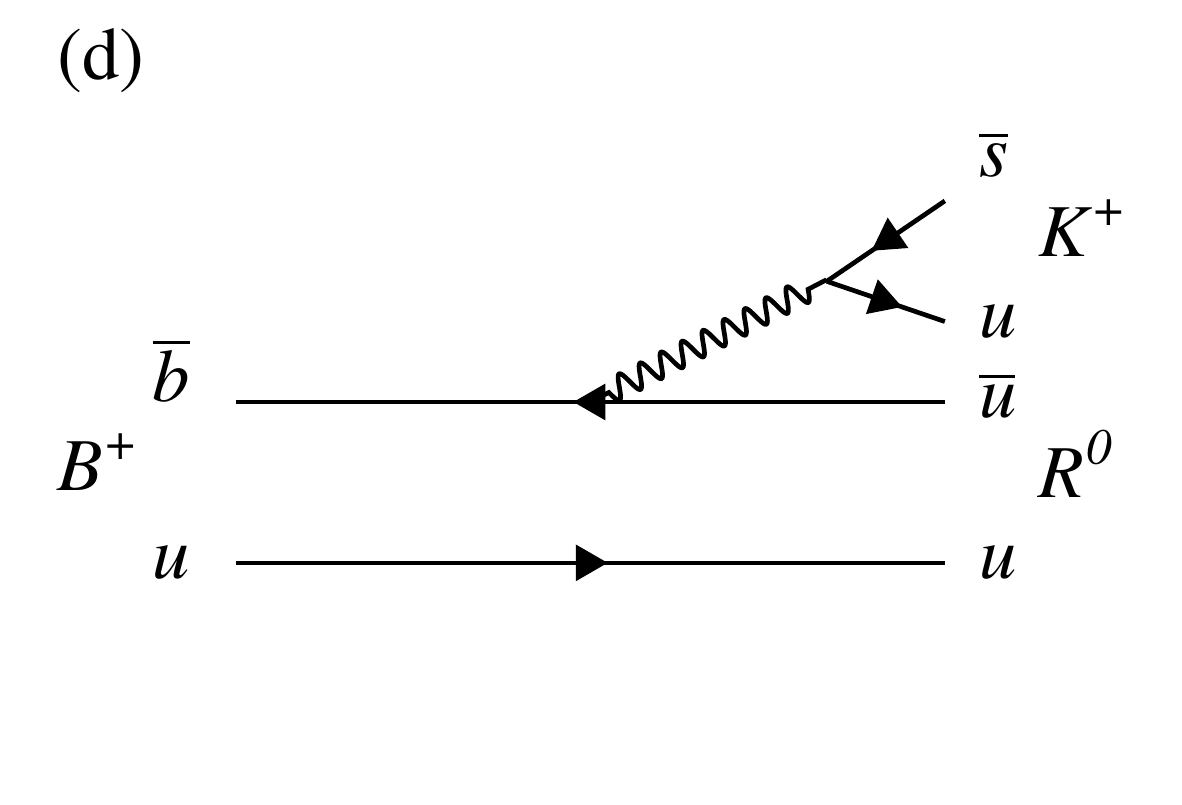}
\caption{\small\label{fig:Fig1} Possible diagrams for \Bu decays involving the production of a resonance $R^0$ being: (a) a gluonium state,
  (b) a $s \bar s$ meson, (c)--(d) an $u \bar u$ meson.
}
\end{figure}

\begin{figure}[htb]
\centering
\small
\includegraphics[width=0.45\textwidth]{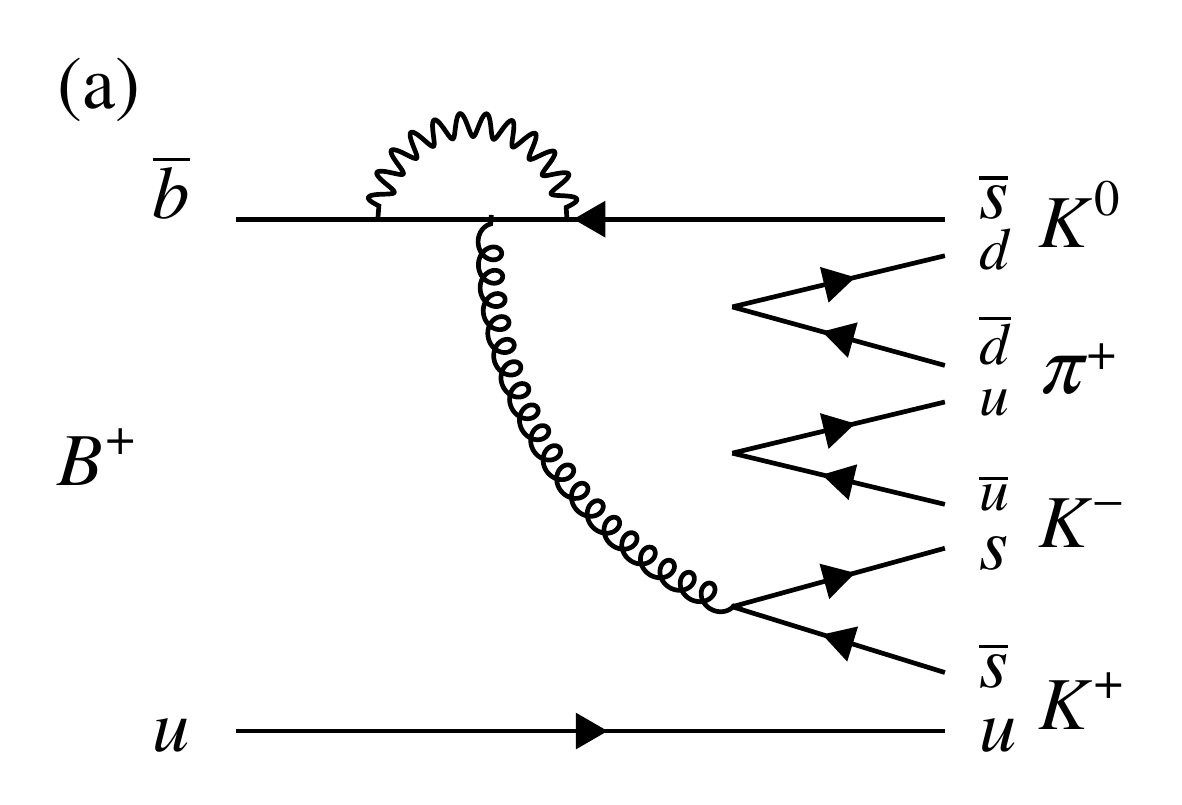}
\includegraphics[width=0.45\textwidth]{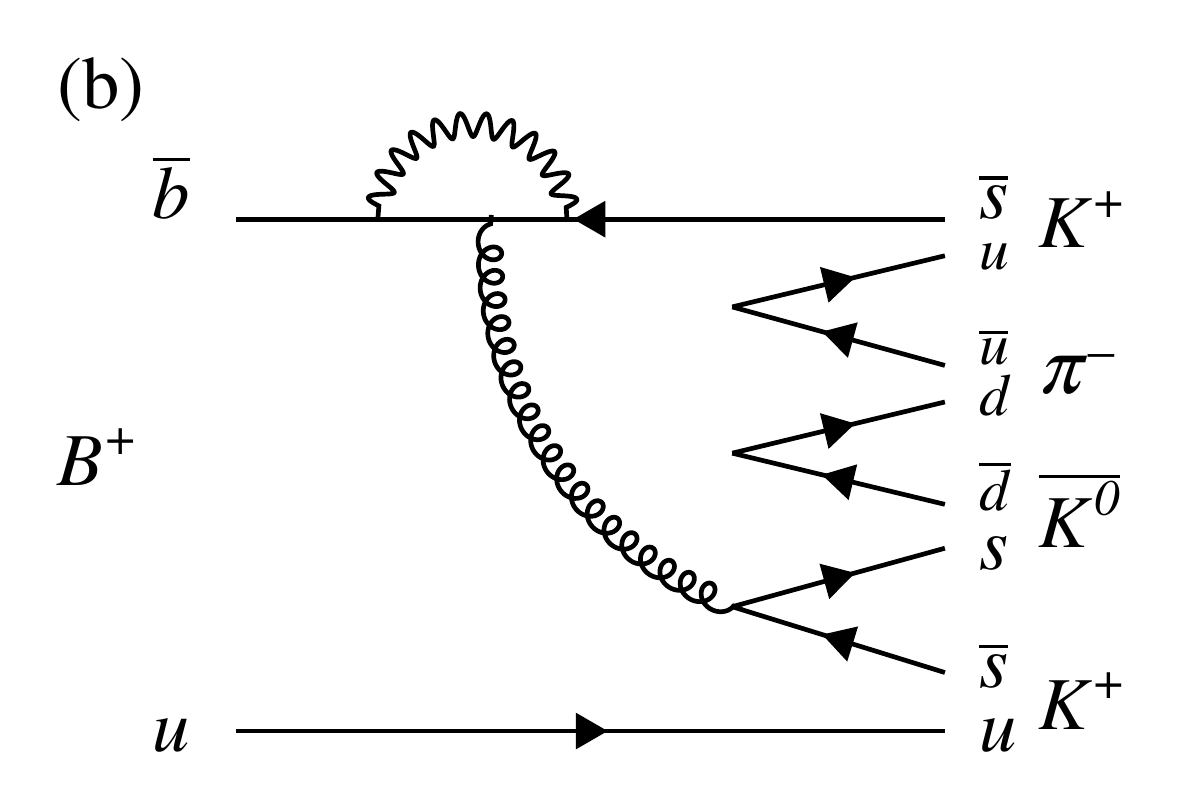}
\caption{\small\label{fig:Fig2} Possible diagrams describing (a) \bkzkkpip and (b) \bkzkkpim nonresonant decays.
}
\end{figure}

In the present paper, a study of the \bkskkpip and \bkskkpim final states is presented.\footnote{The inclusion of charge-conjugate processes is implied throughout the paper.} These decays can proceed through different diagrams, as shown in Fig.~\ref{fig:Fig1}. Decays of $B$ mesons to final states with strangeness have been proposed as 
potential channels for searching for gluonium states~\cite{Fritzsch:1997ps}. A possible diagram for the production of a gluonium state $R^0$ decaying to \kskpi is shown in Fig.~\ref{fig:Fig1}(a). However, in the same $B$ decays, contributions from  $s \bar s$ and $u \bar u$ resonances are also expected (see Fig.~\ref{fig:Fig1}(b)--(d)).
Possible diagrams describing the \bkskkpip and \bkskkpim nonresonant decays are shown in Fig.~\ref{fig:Fig2}. 
Note that the \kskpi system is charge-conjugated in the two \Bu decay modes.\footnote{Kaons and pions charges are not indicated when the sentence refers to both \Bu decay modes.}

Very little is known at present about charmless $B$ decays to pseudoscalar and axial meson resonances~\cite{PDG2024}. 
The \babar collaboration~\cite{Aubert:2008bk} has studied neutral resonances, referred to $R^0$ in the following, in the \mbox{$B^+ \to R^0 K^+$} decays, where the $R^0$ mass spectra were interpreted as signals from \mbox{$\eta(1475) \to K^*\Kb$} and $\eta(1295) \to \eta \pi^+ \pi^-$.

\section{Detector, data and simulation}
\label{sec:lhcb}
The \lhcb
detector~\cite{LHCb-DP-2008-001,LHCb-DP-2014-002} is a
single-arm forward spectrometer covering the pseudorapidity range $2 < \eta < 5$, designed for
the study of particles containing \bquark\ or \cquark\ quarks. The detector elements particularly
relevant to this analysis are: a silicon-strip vertex detector~(VELO)~\cite{Aaij:2014zzy} surrounding the $pp$ interaction
region that allows \cquark\ and \bquark\ hadrons to be identified by exploiting their characteristically long
flight distance; a tracking system that provides a measurement of the momentum, $p$, of charged
particles; and two ring-imaging Cherenkov detectors that are able to discriminate between
different species of charged hadrons. 
Photons, electrons and hadrons are identified by a calorimeter system consisting of
scintillating-pad and preshower detectors, an electromagnetic
and a hadronic calorimeter. Muons are
identified by a system composed of alternating layers of iron and multiwire proportional
chambers.

The analysis is performed on $pp$ collisions data at centre-of-mass
energies 7, 8 and $13\tev$ collected by the \lhcb experiment during Run1 and 2 and corresponding to an integrated luminosity of 9 \invfb.
The online event selection is performed by a trigger~\cite{Aaij:2012me}, which consists of a hardware stage, based on information from the calorimeter and muon systems, followed by a software stage, which applies a full event reconstruction.
During offline selection, trigger signatures
are associated with reconstructed particles. Since the trigger system uses the transverse momentum of the
charged particles with respect to the beam axis, \pt, the phase-space and time acceptance is different for events where signal
tracks were involved in the trigger decision 
and those where the trigger decision was made using information from the rest of the
event only.
Data from both trigger conditions are used but studied separately for consistency tests and for the evaluation of systematic uncertainties.

Simulation is required to model the effects of the detector acceptance and the
  applied selection requirements.
  In the simulation, $pp$ collisions are generated using
  \pythia~\cite{Sjostrand:2007gs,*Sjostrand:2006za} 
  with a specific \lhcb configuration~\cite{LHCb-PROC-2010-056}.
  Decays of unstable particles
  are described by \evtgen~\cite{Lange:2001uf}, in which final-state
  radiation is generated using \photos~\cite{davidson2015photos}.
  The interaction of the generated particles with the detector, and its response,
  are implemented using the \geant
  toolkit~\cite{Allison:2006ve, *Agostinelli:2002hh} as described in
  Ref.~\cite{LHCb-PROC-2011-006}.
  In the simulation, \Bu meson decays are modeled  according to a phase-space distribution.

\section{Event selection}
\label{sec:evsel}

This paper presents a study of the two \Bu decay modes, 
$\Bu \to (\KS \Km \pip) \Kp$ and 
\mbox{$\Bu \to (\KS \Kp \pim)\Kp$}~\cite{LHCb-PAPER-2022-051},
in the region of $m(\kskpi)<1.85\gev$.
Candidate \KS mesons are reconstructed via their decay into the $\pi^+\pi^-$ final state, classified into two categories. The first includes \KS
mesons that decay early enough for the pions to be reconstructed within the VELO, referred to as long \KS (\KSLL). The second category includes \KS mesons that decay later, resulting in pion track segments that lie outside the VELO, referred to as downstream \KS (\KSDD).
While the \KSLL category has better mass, momentum and vertex
resolution, the number of \KSDD candidates is approximately twice as large.

Candidate \Bu mesons are formed by combining the \KS candidate with three additional charged tracks.
A kinematic fit of the entire decay tree is performed~\cite{Hulsbergen:2005pu}, under the assumption that the \Bu candidate originates from a good-quality primary vertex. The selection of \KS and \Bu candidates requires appropriate particle-identification information for each track and imposes broad invariant-mass selections around the known \KS and \Bu masses~\cite{PDG2024}.

To suppress background contribution, in particular the combinatorial background formed from random combinations of unrelated tracks, the candidates satisfying the trigger requirements are filtered
by a loose selection, followed by a multivariate analysis optimized separately for each final state. The selection criteria are tuned to minimize correlation of the signal
efficiency with the kinematic variables, resulting a in better control of the corresponding systematic uncertainties. As a result, the selection relies minimally on the kinematics of
the final-state particles and instead focuses on the topological features associated with the detached vertex of the \Bu candidate.
These features include the impact parameters of the \Bu candidate and its decay products, the fit quality of the decay vertices of the \Bu and \KS candidates, and the separation of these vertices from each other and from the
primary vertex. 

The separation of signal from combinatorial background is achieved by means of a 
boosted decision tree~(BDT) classifier~\cite{Breiman,AdaBoost}, implemented within the TMVA
toolkit~\cite{Hocker:2007ht,*TMVA4}.
For this analysis a BDT with a gradient boosting algorithm~\cite{Roe:2004na} is used, with separate classifiers for the \KSLL and \KSDD categories.

\begin{figure}[!tb]
\centering
\small
\includegraphics[width=0.45\textwidth]{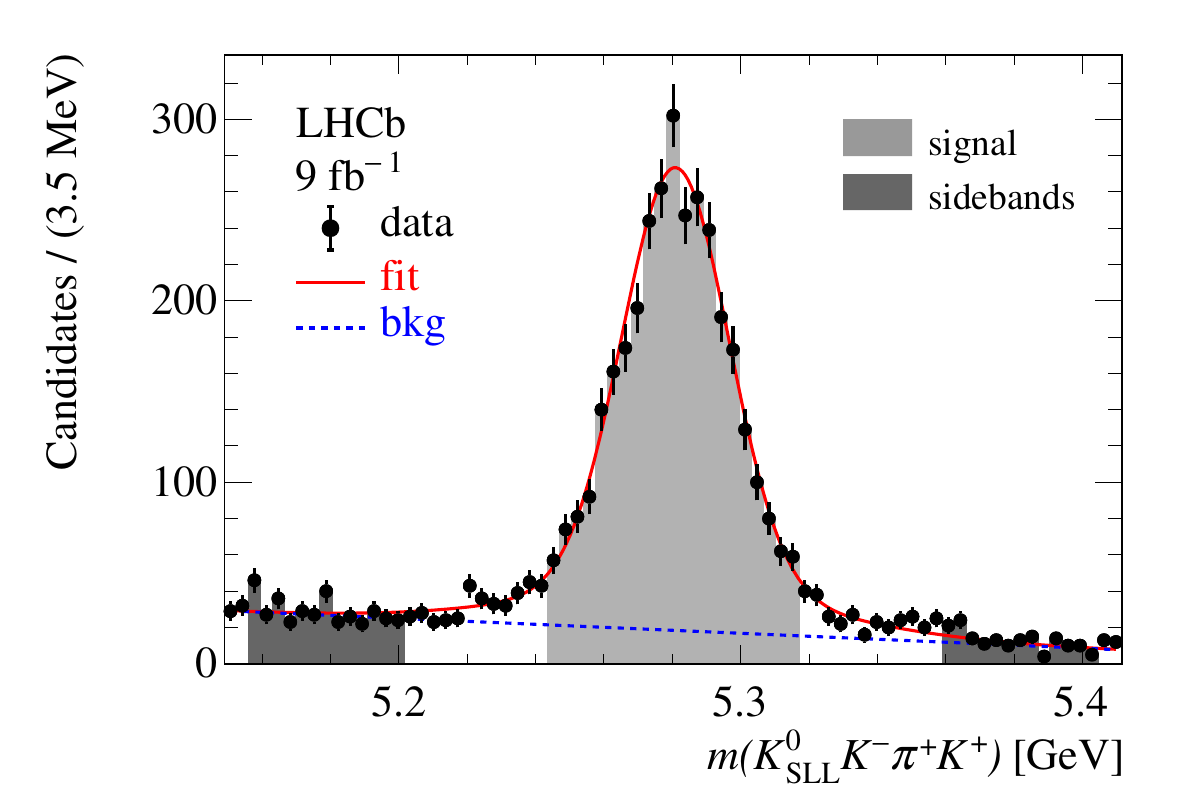}
\includegraphics[width=0.45\textwidth]{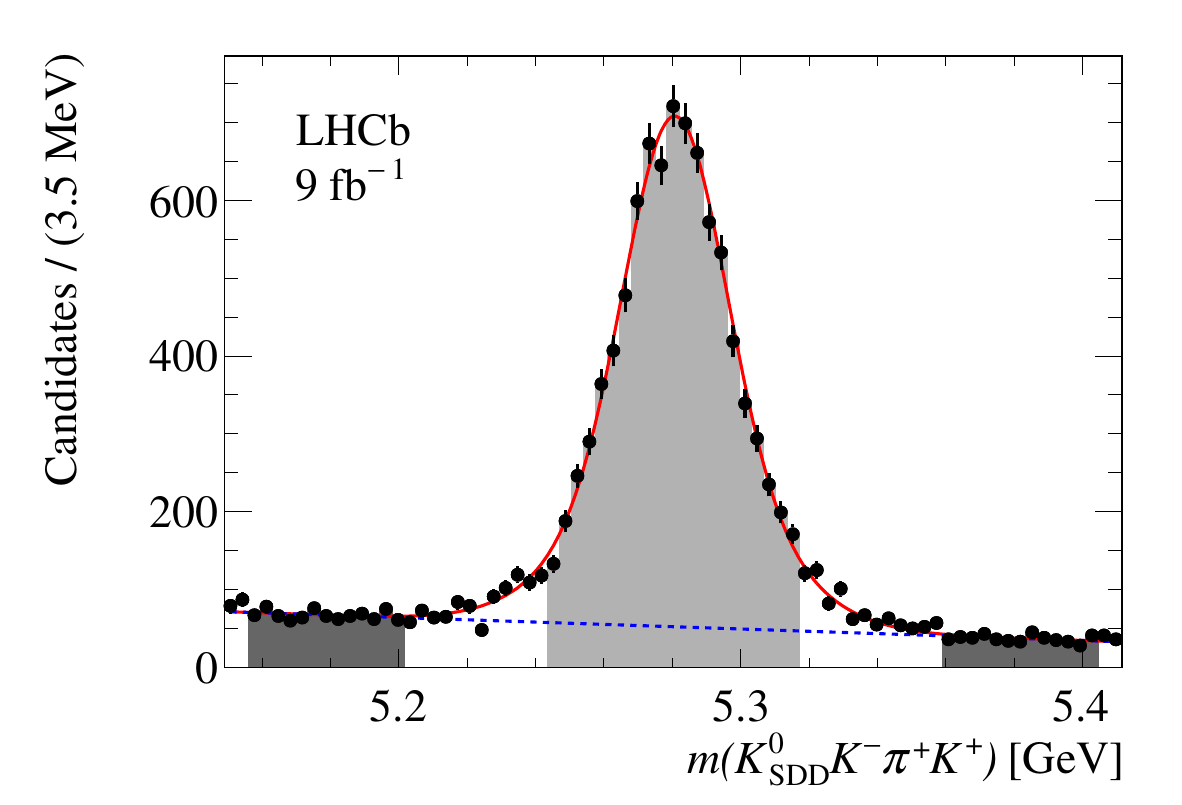}
\includegraphics[width=0.45\textwidth]{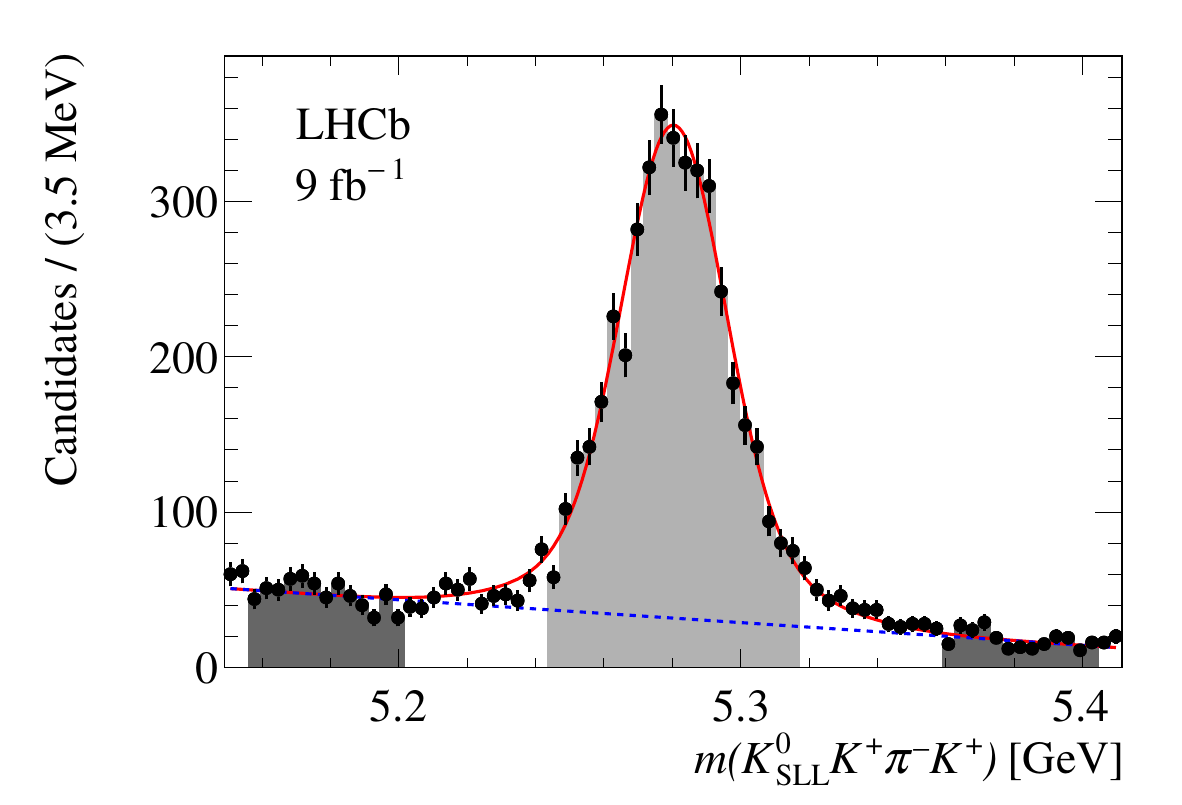}
\includegraphics[width=0.45\textwidth]{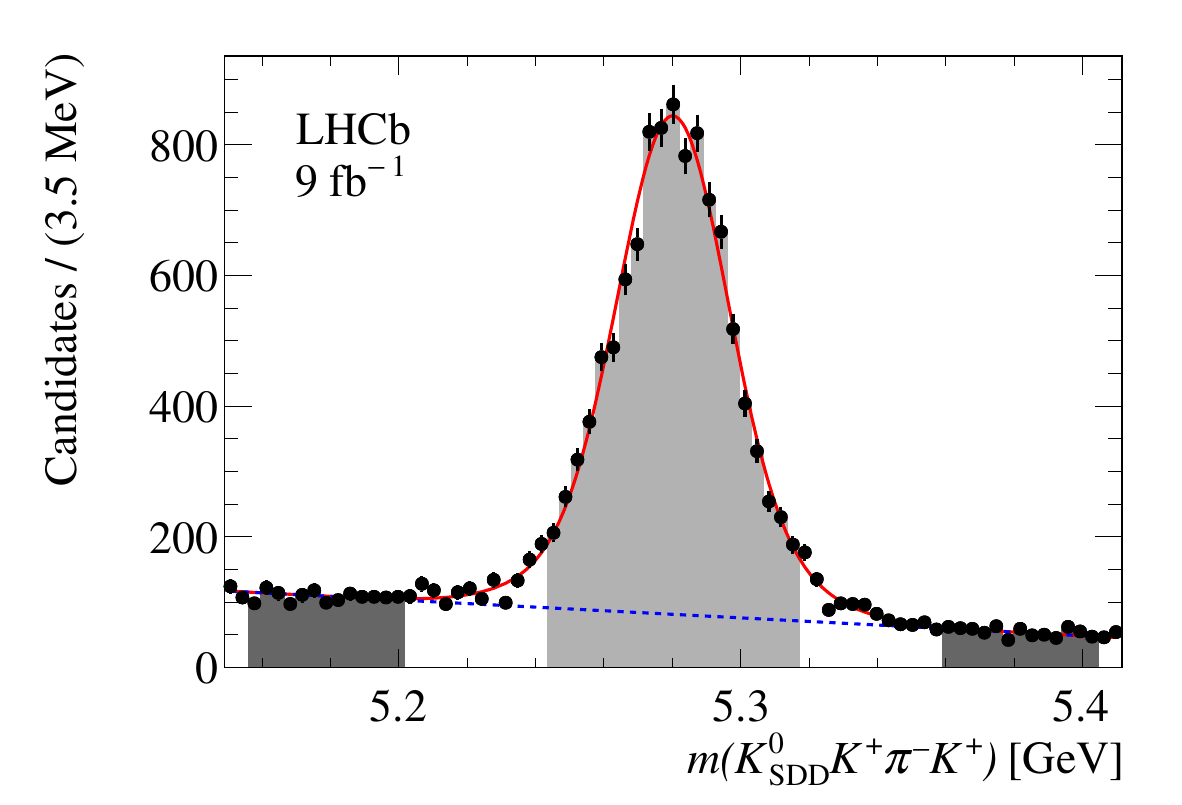}
\caption{\small\label{fig:Fig3} 
Distributions of the (top) \kskkpip and (bottom) \kskkpim invariant mass for (left) \KSLL and (right) \KSDD candidates.}
\end{figure}

Although the amplitude analysis is performed for $m(\kskpi)<1.85\gev$ (see Sec.~\ref{sec:data}), to obtain information on the resonant contribution in an extended mass region,
the classifiers are trained using data and simulated signal decays with $m(\kskpi)<2.5\gev$ from both \Bu decay modes. The simulation matches the relative yields of the dataset at the various center-of-mass energies. It is assumed that the efficiencies for the reconstruction of the \bkskkpip and \bkskkpim decays are the same.
Data from the lower and upper mass sidebands of the \Bu signal region are used as background proxy in the BDT training, as indicated in Fig.~\ref{fig:Fig3}.
The composition of the background sample reflects the data-taking conditions, and events from the sidebands from both \Bu decays are included in equal proportions. 
The optimization of the BDT-classifier working point is performed by considering the figure-of-merit
\begin{equation}
  S=\frac{N_{\rm sig}}{\sqrt{N_{\rm sig} + N_{\rm bkg}}},
 \label{eq:sign} 
  \end{equation}
where $N_{\rm sig}$ and $N_{\rm bkg}$ represent the \Bu signal, respectively, in the signal region and combinatorial background yield in the signal region evaluated by fitting the $m(\kskkpi)$ mass distributions.

In order to facilitate the extraction of the \KS and \Bu signal and combinatorial background components from these invariant-mass spectra, a kinematic fit of the decay without constraints on the masses of the \KS and \Bu is performed.
The binned $\pip \pim$ invariant-mass distributions for \KSLL and \KSDD candidates are fitted separately. The fit model uses the sum of two Gaussian functions sharing the same mean for the signal and a linear function for the background.
An effective resolution is computed as
\begin{equation}
  \sigma_{\rm eff}=f\sigma_1+(1-f)\sigma_2
  \label{eq:sigma}
  \end{equation}
where $\sigma_1$ and $\sigma_2$ indicate the width of the two Gaussian functions and $f$ is the fraction of the first Gaussian contribution.
The resulting effective resolutions are $\sigma_{\rm eff} = 2.5\mev$ and $\sigma_{\rm eff} =6.5\mev$ for the LL and DD categories, respectively.
The \KS signals are selected within $3.0 \, \sigma_{\rm eff}$ of the fitted \KS mass of $497.8\mev$.

To improve the resolution of the other invariant masses, the 
energy of the selected candidate \KS is calculated as
\begin{equation}
E_{\KS}=\sqrt{p^2_{\KS}+m^2_{\KS}},
\end{equation}
where $p_{\KS}$ is the reconstructed \KS momentum, and $m_{\KS}$ the known \KS mass~\cite{PDG2024}.
Compared to using the \KS mass constraint, this method achieves the same resolution for the $\kskkpi$
invariant mass but results in a slightly worse resolution, by  $\sim 6$\%, for the $\KS K \pi$ invariant mass. However, it retains the ability to extract both the \Bu and \KS signals and the combinatorial background from a fit to the \kskkpi and $\pip \pim$ invariant-mass spectra.

Particle identification of the three charged hadrons is performed using the output of a neural network (NN) trained on the information of all the relevant subdetectors. The figures of merit are expressed as \mbox{$P_{K} = \mathit{\rm NN}_K(1 - \mathit{\rm NN}_{\pi})$} for kaon identification and
\mbox{$P_{\pi} =\mathit{\rm NN}_{\pi}(1 - \mathit{\rm NN}_K)$} for pion identification, where $\mathit{\rm NN}_{\pi}$ and $\mathit{\rm NN}_K$ are the NN probabilities for pion and kaon identification, respectively.
Very loose selections are applied to these quantities to maximize the significance of the \Bu candidate invariant-mass peak as a function of $P_K$ or $P_{\pi}$. 
Tests performed using a large $\etac\to \kskpi$ sample~\cite{LHCb-PAPER-2022-051} show that ($0.35 \pm 0.05$)\% and ($3.0 \pm 0.1$)\%
of \etac signal decays are removed from both \Bu decays under the pion and kaon hypotheses, respectively. This procedure ensures that minimal bias is introduced into the angular distributions of the \Bu decays. With such particle-identification requirements it is found that data and  simulation agree, in fractional \etac losses, within two standard deviations~($\sigma$).

Figure~\ref{fig:Fig3} shows the \kskkpip and \kskkpim invariant-mass spectra for the selected candidates, separated by \KSLL and \KSDD  categories. 
The fits give a signal peak \Bu-mass value of $5280.0\mev$ and an effective width of \mbox{$\sigma_{\rm eff}= 17.7\mev$}.
Signal candidates are selected in a window of $\pm 2 \, \sigma_{\rm eff}$ (used in the amplitude analysis reported in Sec.~\ref{sec:data}) of the fitted \Bu mass, common to the four datasets.
Table~\ref{tab:Tab1} lists the fitted yields and purities~($P$)
in the \Bu signal region for the different datasets, where the purity is defined as 
\begin{equation}
P=\frac{N_{\rm sig}}{N_{\rm sig}+N_{\rm bkg}}.
\label{eq:purity}
\end{equation}
It is found that $P$, for the four datasets, does not depend on the collision energy nor the data-taking conditions, which simplifies the amplitude analysis and they are analysed together. 
Approximately 0.02\% of the events contain multiple \Bu decay candidates, all of which are retained for further analysis.
\begin{table} [!tb]
  \centering
  \caption{\small\label{tab:Tab1} Fitted \Bu signal yield and purity for \kskkpip and \kskkpim final states separated by \KS type.}
\begin{tabular}{lcc}
\hline
Final state & \Bu signal yield & \Bu purity  [\%]\cr
\hline\\ [-2.3ex]
\kskkpip &                &       \cr
\KSLL & $2911\pm68$ & $87.1 \pm 0.6$ \cr
\KSDD &  \al $7672\pm116$ & $87.7 \pm 0.4$ \cr
\hline\\ [-2.3ex]
\kskkpim &                &       \cr
\KSLL & $3497\pm86$ & $83.5\pm0.6$\cr
\KSDD & \al $8955\pm112$ & $83.7\pm0.4$\cr
\hline
\end{tabular}
\end{table}

\begin{figure}[!tb]
\centering
\small
\includegraphics[width=0.45\textwidth]{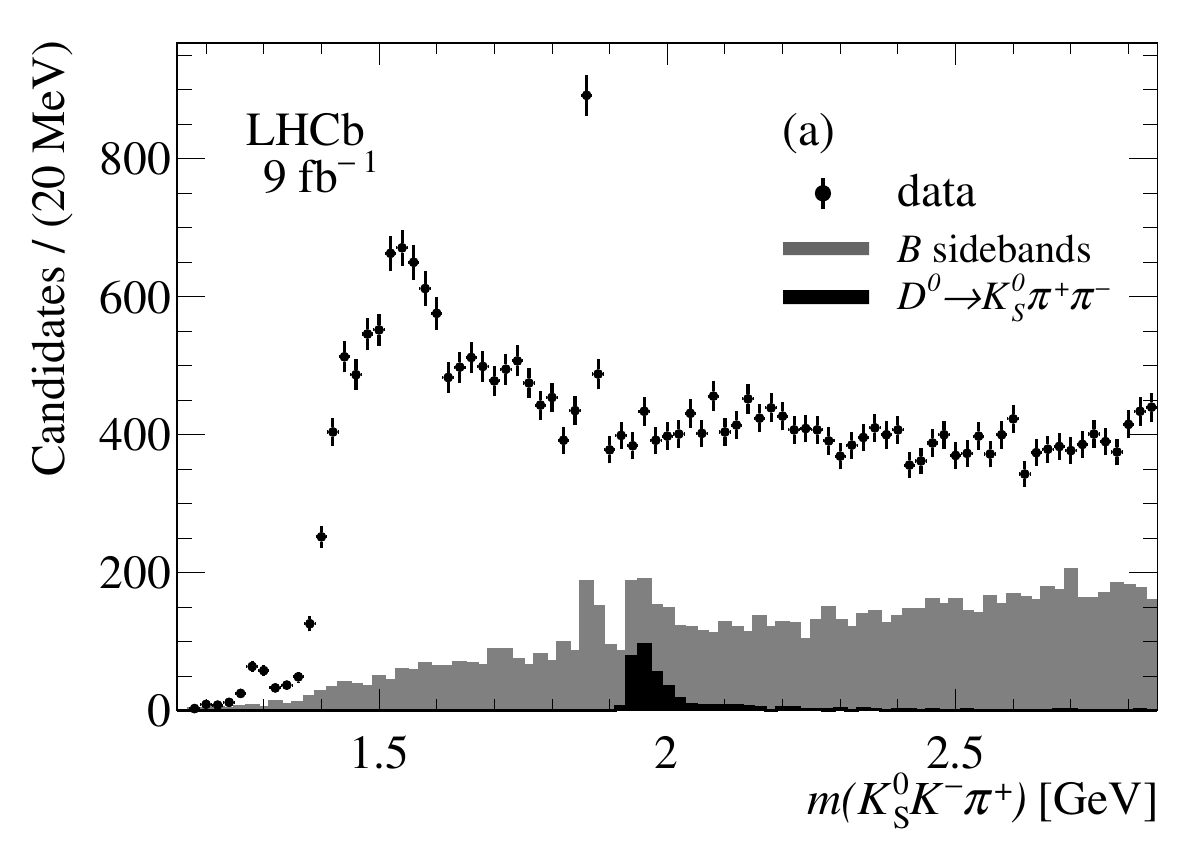}
\includegraphics[width=0.45\textwidth]{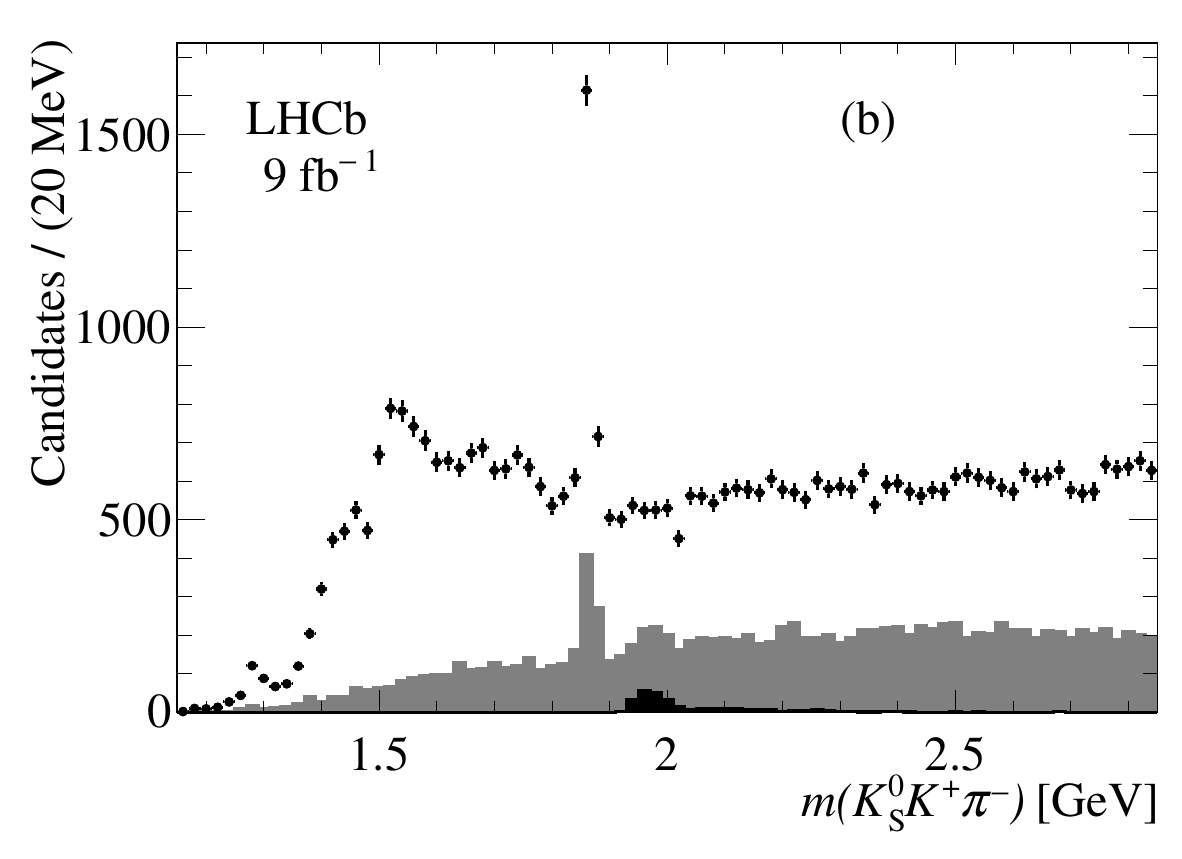}
\caption{\small\label{fig:Fig4} \kskpi invariant-mass distributions for (a) \bkskkpip and (b) \mbox{\bkskkpim} candidates in the \Bu signal region. The gray distributions are obtained from the \Bu mass sidebands normalized to the expected background in the signal region, the black distributions show the $\Dz \to \KS \pip \pim$ reflection from pions misidentified as kaons. 
}
\end{figure}

The \kskpi invariant-mass spectra for events in the \Bu signal region, summed over the \KSLL and \KSDD datasets, are shown in Fig.~\ref{fig:Fig4}.
The lower and upper mass sidebands around the \Bu signal peak, representing the background, defined in the ranges [$-6 \, \sigma_{\rm eff},-4 \, \sigma_{\rm eff}$] and [$4 \, \sigma_{\rm eff},6 \, \sigma_{\rm eff}$], are superimposed onto the
\kskpi invariant-mass spectrum from the \Bu signal region. 
For the \bkskkpim final state, which has two kaons with the same charge, the smallest of the two possible mass combinations is plotted.
Notably, for $m(\kskpi)<2.3\gev$, there is only one combination possible. 

The \kskpi mass spectra show a peak at the position of the $f_1(1285)$ resonance, followed by a broad enhancement suggesting the presence of several states. A $\Dz \to \kskpi$ peak is visible, originating from the open-charm decay $\Bu \to \Dzb \Kp$.
The structure above $1.9\gev$, which appears only in the background samples, is due to reflections from $\Dz \to \KS \pip \pim$ decays, where one pion is misidentified as a kaon. This \Dz background contribution has two different sources: (a)  $\Bu \to \Dzb \pip$ and (b) \Dz open-charm production in the background. The \Bu contribution is reconstructed by assigning the pion mass to both kaon candidates and selecting the candidate if the recalculated masses
fall within mass windows around the \Dz and \Bu masses.
The contribution (b) is reconstructed after having enhanced the \Dz signal by requiring the $\KS \pi^{\pm}$ mass to be in the $K^{*\pm}$ mass region.
The combination of these two selections is illustrated by the black distributions shown in Fig.~\ref{fig:Fig4}.
This contribution is subsequently removed from the sidebands sample.

\section{Mass resolution, efficiency and background}
\label{sec:reso}

The \kskpi mass resolution is obtained from simulation as the difference between true and reconstructed mass in slices of the \kskpi mass. The resolution is described by
the sum of two Gaussian functions, with $\sigma_{\rm eff}$ (see Eq.~\ref{eq:sigma}) 
varying across the \kskpi invariant-mass values range $1.3 < m(\KS K \pi) < 1.9\gev$. For \KSLL candidates $\sigma_{\rm eff}$ spans from 4.5 to $7.0\mev$, while for \KSDD candidates it ranges from 5.5 to $9.0\mev$.
Since the width of the resonances present in this mass range is much larger than the experimental resolution, its effect is ignored. An exception is the $f_1(1285)$ resonance, whose description is discussed in Sec.~\ref{sec:f1}.

Several angular variables are used to show projected distributions and determine the efficiency profiles.
The \Bu decay can be described by the process
\begin{equation}
  B \to (\KS K \pi) K_4
  \label{eq:k1k2}
\end{equation}
where $K$ indicates the kaon participating in resonant decay to the $\KS K \pi$ system, with $m(\KS K \pi)<1.85\gev$, and
$K_4$ is the spectator kaon.

\begin{figure}[!tb]
\centering
\includegraphics[width=5.50cm]{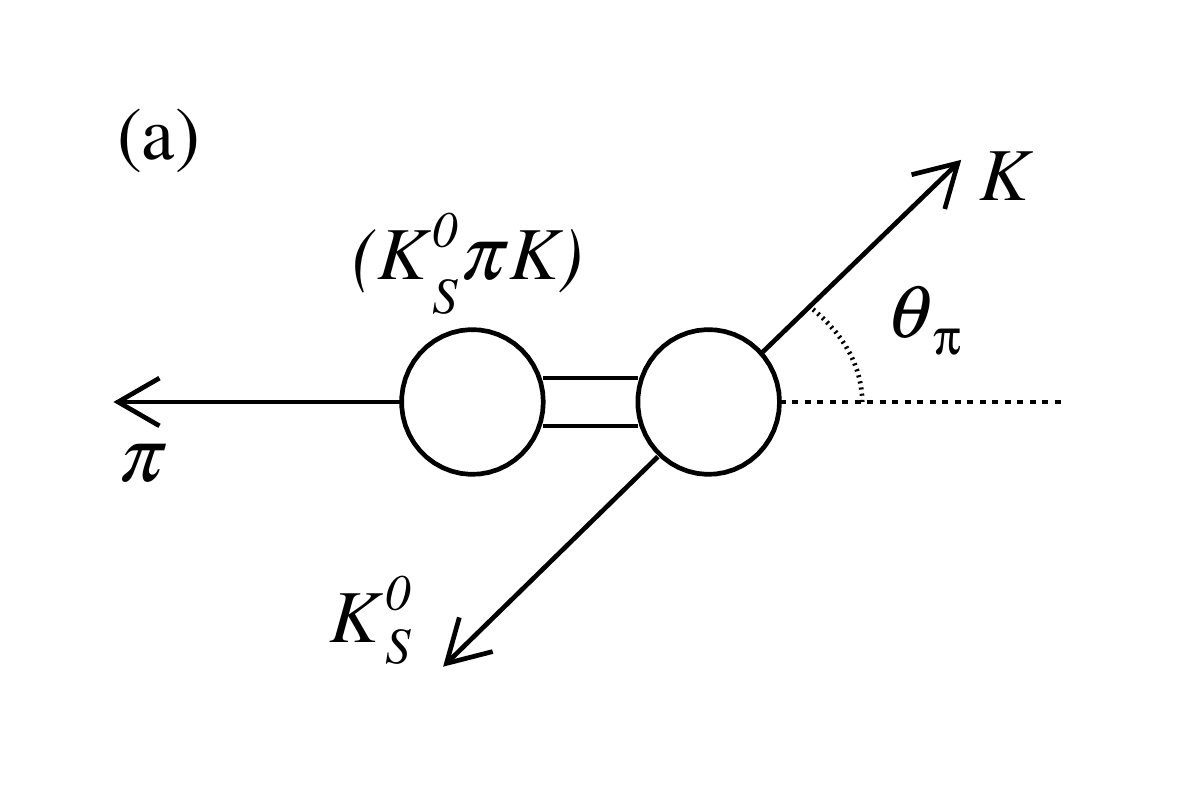}
\includegraphics[width=5.50cm]{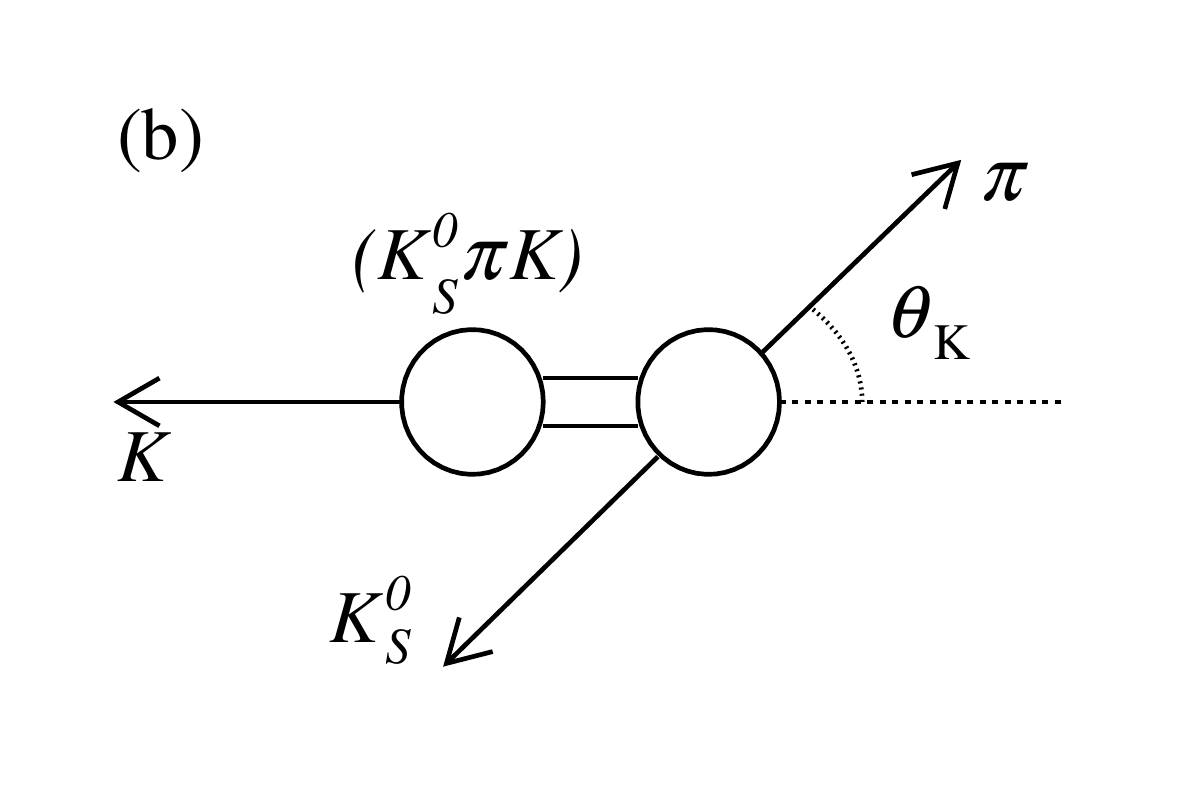}\\
\includegraphics[width=5.50cm]{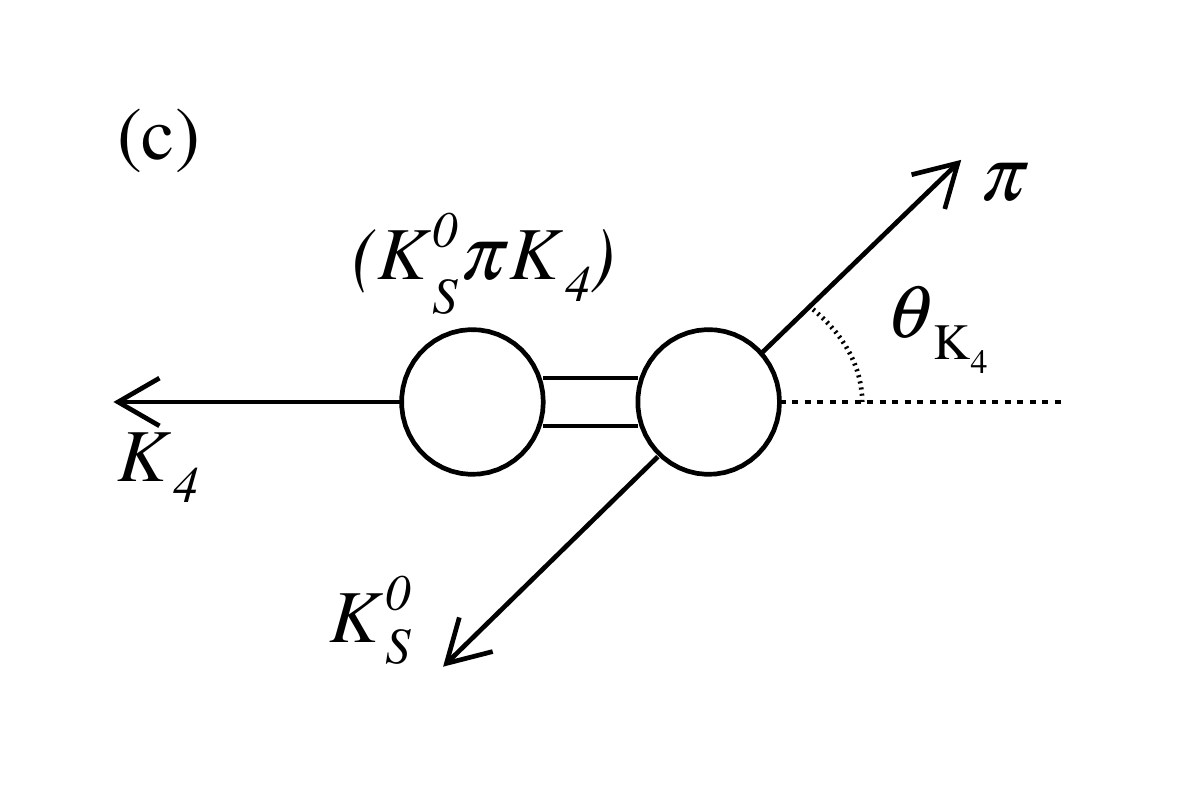}
\includegraphics[width=5.50cm]{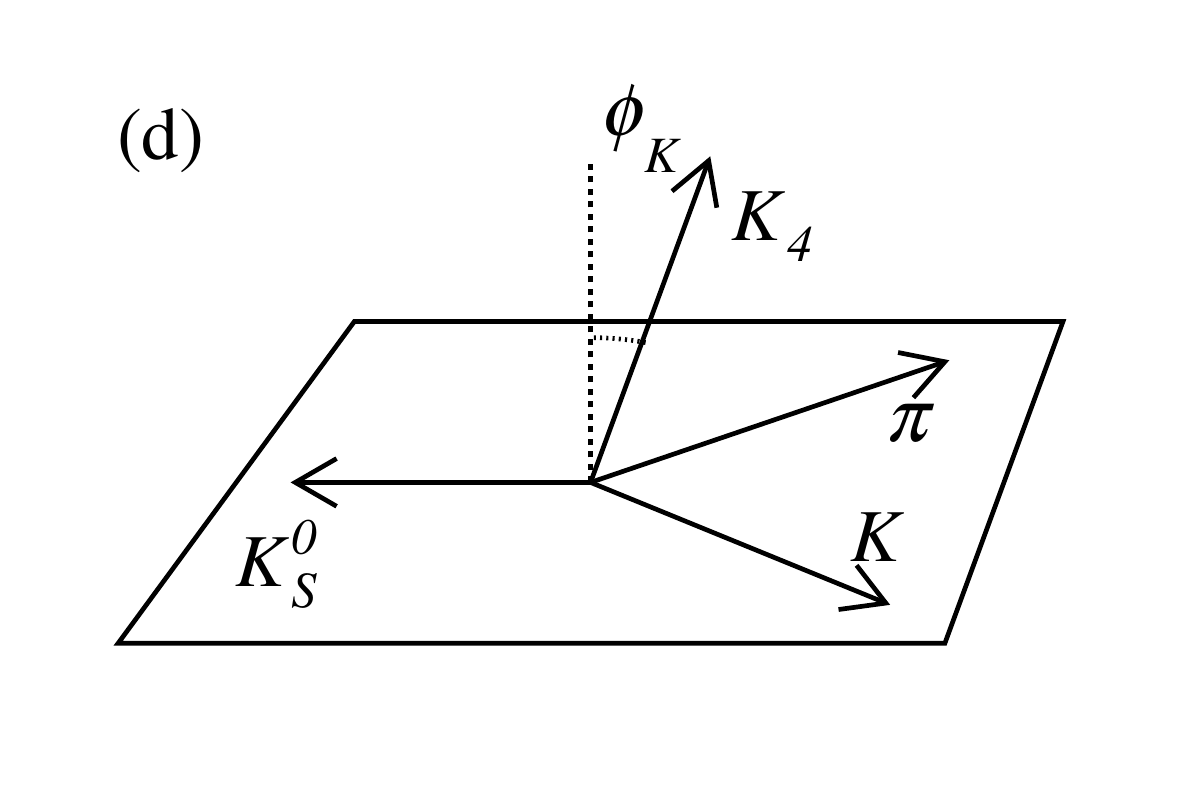}
\caption{\small\label{fig:Fig5} Diagrams illustrating the angular variables $\theta_{\pi}$, $\theta_K$, $\theta_{K_4}$ and $\phi_K$.}
       \end{figure}

The angular distributions used in this analysis are defined as follows and illustrated in Fig.~\ref{fig:Fig5}.
The angle $\theta_{\pi}$($\theta_{K}$) is defined as the angle between the $K$($\pi$) in the $\KS K$($\KS \pi$) rest frame and the $\KS K$($\KS \pi$) direction in the $\KS K \pi$ rest frame.
Similarly, the angle $\theta_{K_4}$ is defined by exchanging $K$ with $K_4$. Finally,
$\phi_K$ is the angle formed by the spectator $K_4$ momentum with the normal to the $\KS K \pi$ decay plane.

\subsection{Efficiency}
\label{sec:effy}

The kinematics of a four-body decay are fully described by five independent variables. A mixture of invariant-mass combinations and decay angles is used as variables in this analysis.
The various possible invariant-mass combinations have different kinematic bounds, therefore mass-reduced variables are used instead as they always range between 0 and 1. They are defined as~\cite{LHCb-PAPER-2018-045}

\begin{equation}
   m_x = \frac{1}{\pi} {\arccos}\left(2\frac{m - m_{\rm min}}{m_{\rm max}-m_{\rm min}}-1\right),
\end{equation}
where $m$, $m_{\rm min}$ and $m_{\rm max}$ indicate the invariant-mass and its minimum and maximum kinematically allowed values, respectively. Note that the $m_x$ ranges are reduced by the request $m(\kskpi)<1.85\gev$.

Two types of efficiencies are evaluated, total and local.
The total efficiency describes the effects of the reconstruction on the full phase space of the \Bu decay to the four-body final state.
Local efficiencies are evaluated in the specific \kskpi mass region considered in the present analysis.
The efficiencies are evaluated using simulated samples that undergo the same reconstruction and selection criteria as the data. Efficiency distributions are determined by taking the ratio of selected to generated events, projected onto the relevant kinematic variables.
A comparison of the \pt distributions of \Bu candidates between simulation and data shows a small disagreement, which is corrected by weighting the former to match the latter.

The total efficiency for the two \Bu decay modes is evaluated in an iterative manner as described in Ref.~\cite{LHCb-PAPER-2022-051}.
The local efficiency is evaluated separately for \KSLL and \KSDD simulations using the same method.

\begin{figure}[!tb]
\centering
\small
\includegraphics[width=0.45\textwidth]{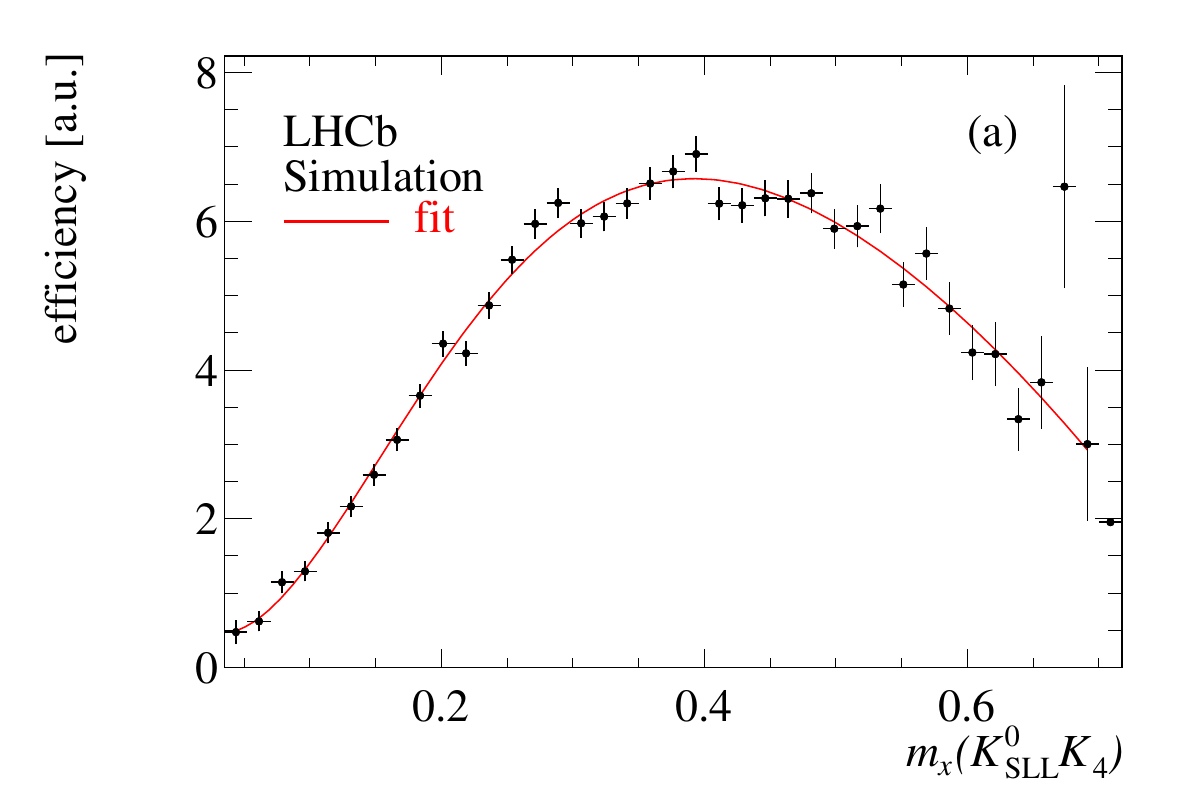}
\includegraphics[width=0.45\textwidth]{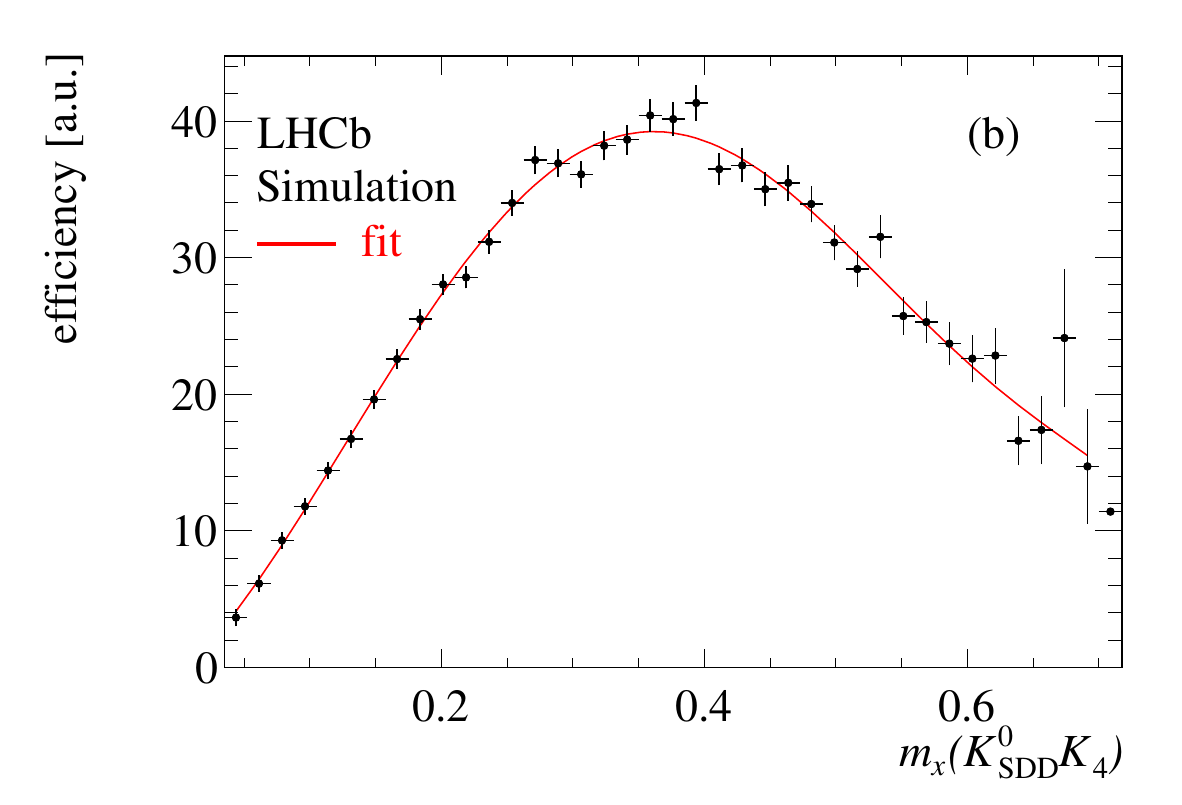}
\caption{\small\label{fig:Fig6} Efficiency projections (in arbitrary units) on $m_x(\KS K_4)$ for the (a) \KSLL and (b) \KSDD samples with $m(\kskpi)<1.85\gev$.}
\end{figure}

First, the variable whose efficiency distribution shows the most significant deviation from uniformity, $m_x(\KS K_4)$, is identified. Figure~\ref{fig:Fig6} shows the efficiency projected onto $m_x(\KS K_4)$ separately for the \KSLL and \KSDD samples and the result of a fit with a seventh-order polynomial function, labeled as $\epsilon_1(m_x(\KS K_4))$.
The simulated candidates are then weighted by the inverse of the efficiency $1/\epsilon_1(m_x(\KS K_4))$ and a second variable ($m_x(\KS K)$) is chosen and fitted with a fifth-order polynomial function, labeled as $\epsilon_2(m_x(\KS K))$. 
The events are then weighted by the factor $1/(\epsilon_1(m_x(\KS K_4))\cdot\epsilon_2(m_x(\KS K)))$. The process continues in this fashion, terminating when the efficiency is consistent with being uniform across all nine of the considered variables ($m(\kskpi)$, five two-body $m_x(\KS \pi (K))$  combinations and the three angular variables, $\cos \theta_K$, $\cos \theta_{K^0_S}$ and $\cos \phi$) both in their one- and two-dimensional projections.
The total efficiency for each \KS category, $\epsilon_{\rm LL}$ and $\epsilon_{\rm DD}$, is found to be well described by the following functions
\begin{equation}
  \begin{split}
    \epsilon_{\rm LL} & = \epsilon_1(m_x(\KS K_4))\cdot \epsilon_2(m_x(\KS K))\cdot \epsilon_3(\cos \phi) \cdot \epsilon_4(m_x(\KS \pi)) , \\
    \epsilon_{\rm DD} & = \epsilon_1(m_x(\KS K_4))\cdot \epsilon_2(m_x(\KS K))\cdot \epsilon_3(\cos \phi).
  \end{split}
  \label{eq:effy}
\end{equation}

\begin{figure}[!tb]
\centering
\small
\includegraphics[width=0.48\textwidth]{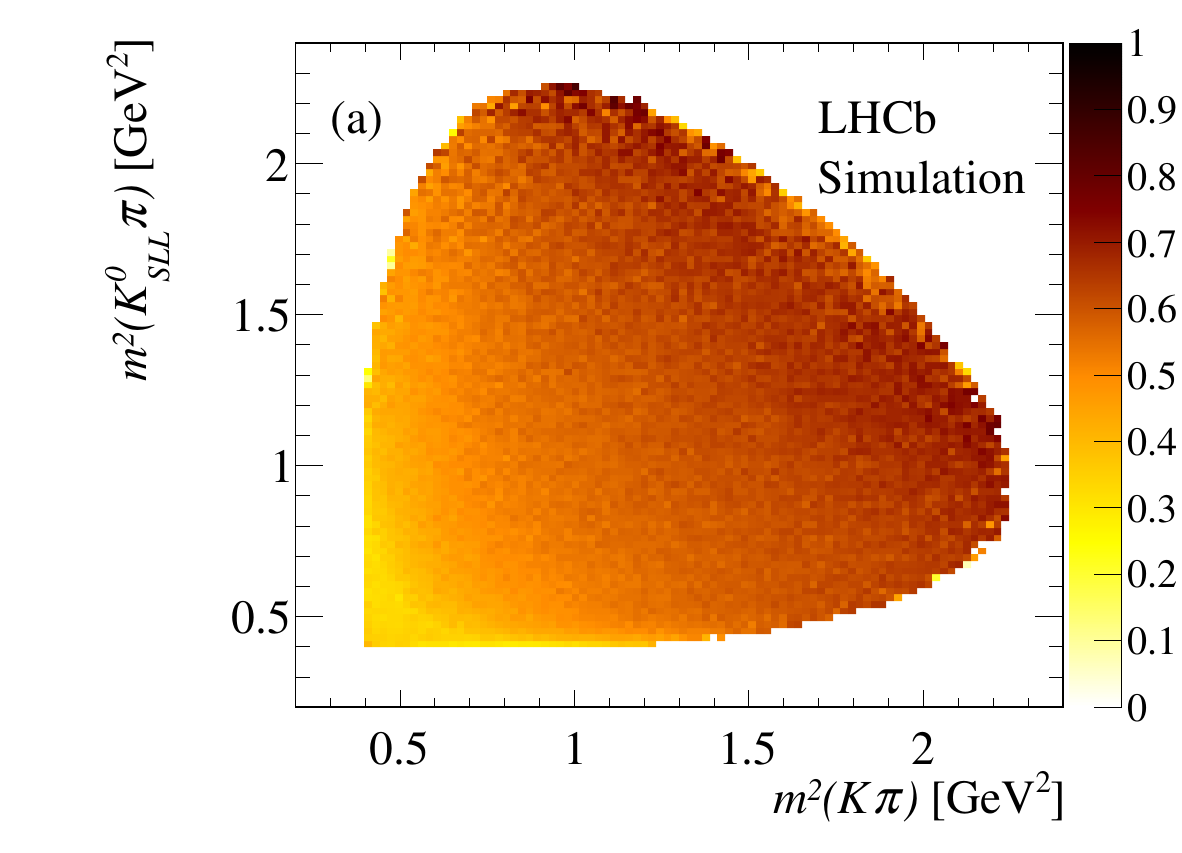}
\includegraphics[width=0.48\textwidth]{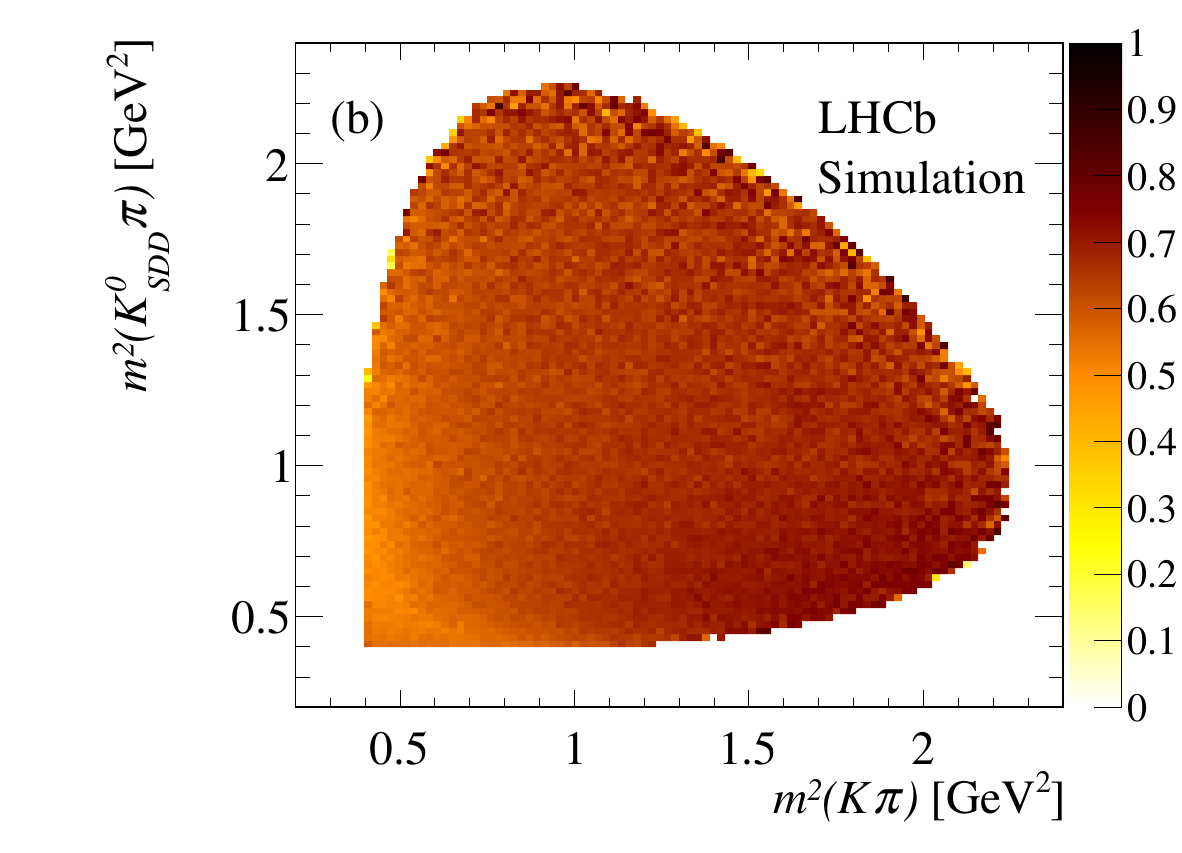}
\caption{\small\label{fig:Fig7} Two-dimensional efficiency distributions given in arbitrary units in the \mbox{$m(\kskpi)<1.85\gev$} mass region for the (a) \KSLL and (b) \KSDD samples.}
\end{figure}

\begin{figure}[!tb]
\centering
\small
\includegraphics[width=0.45\textwidth]{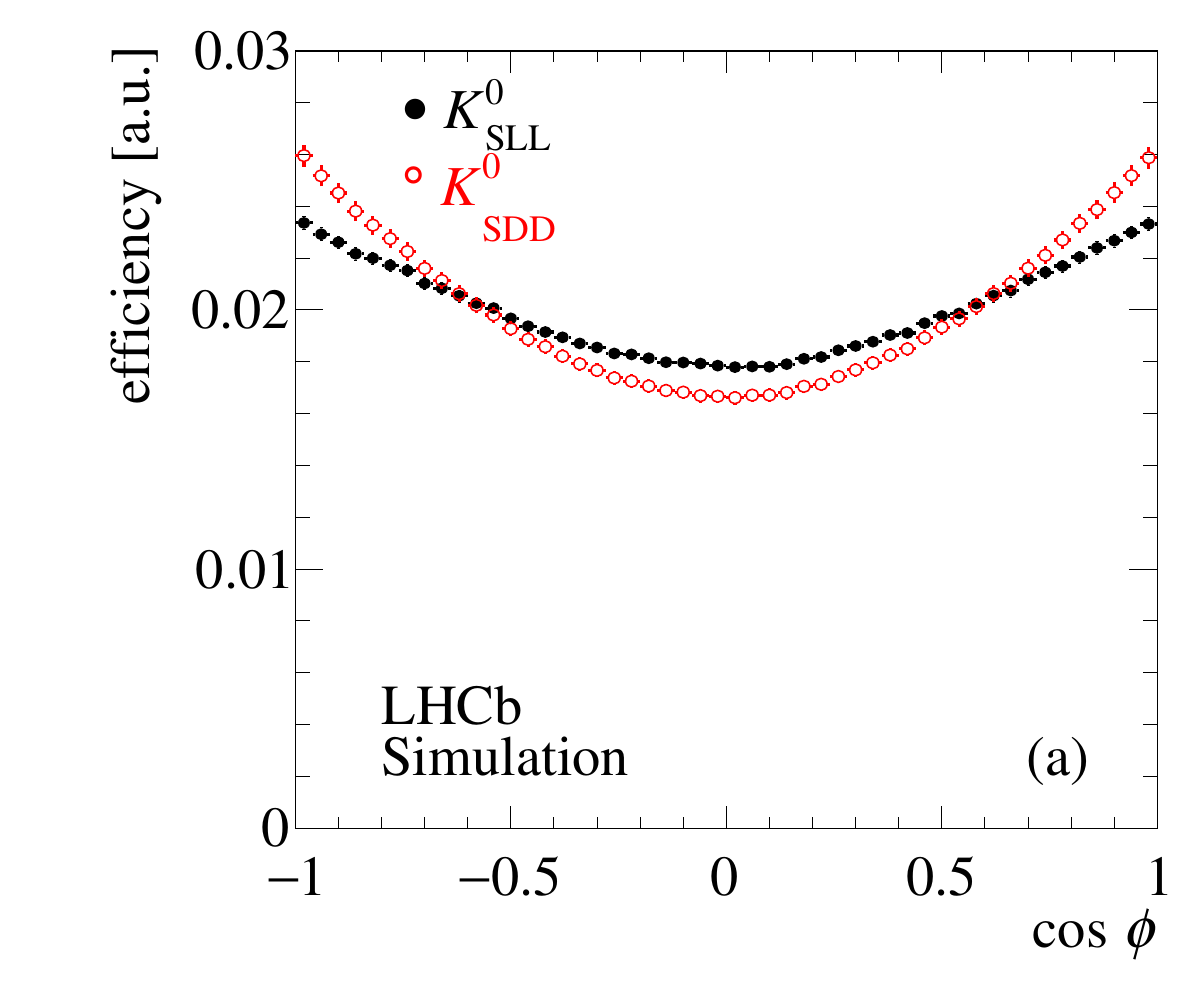}
\includegraphics[width=0.45\textwidth]{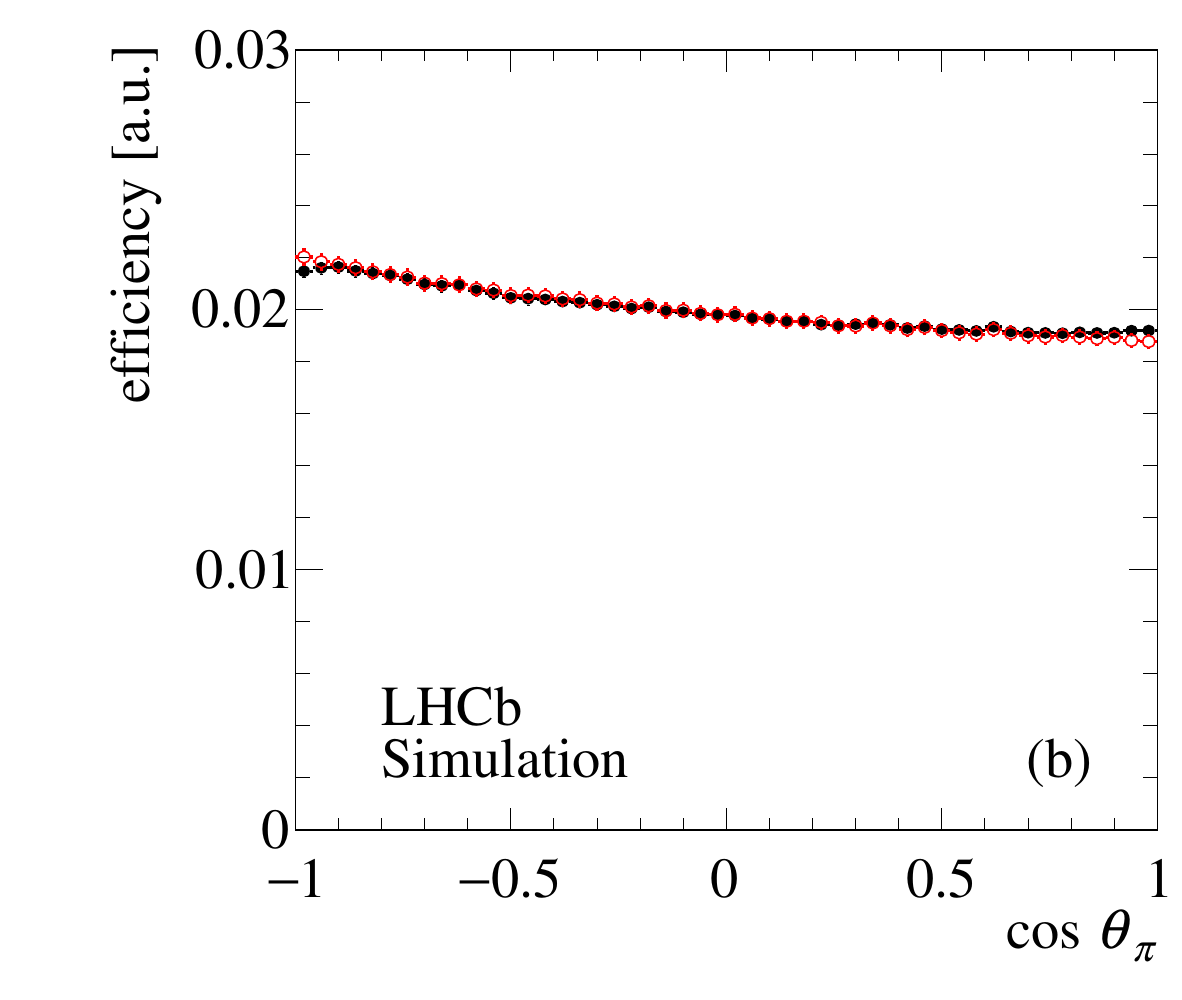}
\caption{\small\label{fig:Fig8} Normalized efficiency distributions from simulation in the $m(\kskpi)<1.85\gev$ mass region projected onto (a) $\cos \phi$ and (b) $\cos \theta_{\pi}$ for the \KSLL and \KSDD samples.}
\end{figure}

Figure~\ref{fig:Fig7} shows the resulting efficiency distributions across the Dalitz plot for \mbox{$m(\kskpi)<1.85\gev$} while Fig.~\ref{fig:Fig8} shows the normalized efficiency projected onto $\cos \phi$ and $\cos \theta_{\pi}$.

\subsection{Description of the background}
\label{sec:back}
The background distribution is obtained by inspecting the \Bu mass sidebands where the \KSLL and \KSDD data are combined and performing an unbinned maximum-likelihood fit to the \kskpi and two-body mass distributions.
The \kskpi mass distribution is described by a fourth-order polynomial function, while the two-particle mass distributions are modeled including the contributions from charged and neutral $K^*(892)$ resonances with parameters fixed to known values~\cite{PDG2024}, and a nonresonant contribution. The former are modeled with relativistic Breit--Wigner (BW) functions, while the latter is represented by a constant value.
In the following $K^{*}_{\rm{ne}}(892)$ and $K^{*}_{\rm{ch}}(892)$ indicate the neutral and charged $K^*(892)$ contributions, respectively.

\begin{figure}[!tb]
\centering
\small
\includegraphics[width=0.80\textwidth]{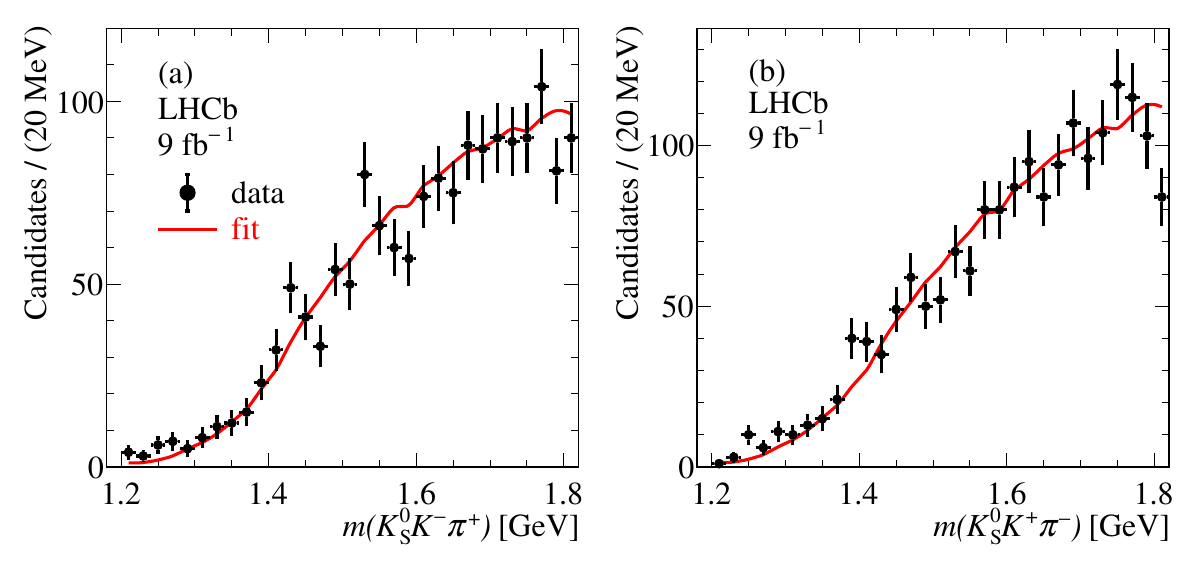}
\caption{\small\label{fig:Fig9} Distribution of the \kskpi mass of background candidates from the \Bu mass sidebands for (a) \bkskkpip and (b) \bkskkpim decays.}
\end{figure}

\begin{figure}[htb]
\centering
\small
\includegraphics[width=0.95\textwidth]{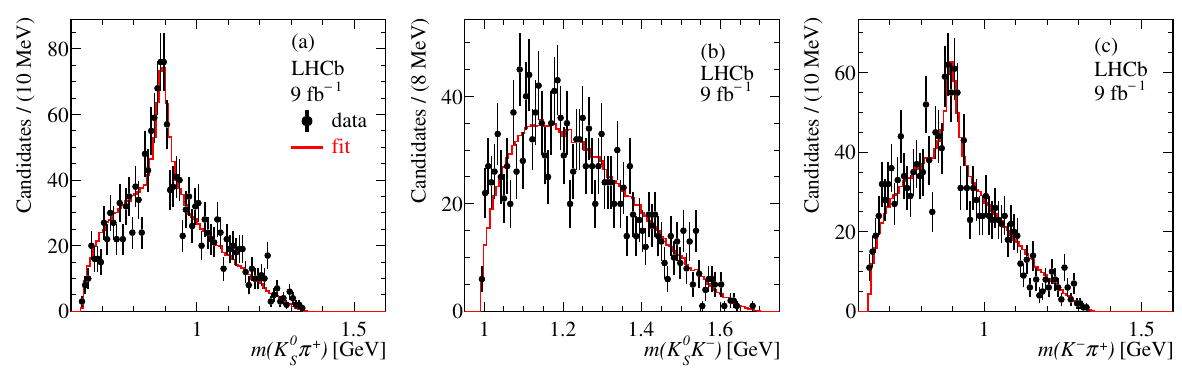}
\includegraphics[width=0.95\textwidth]{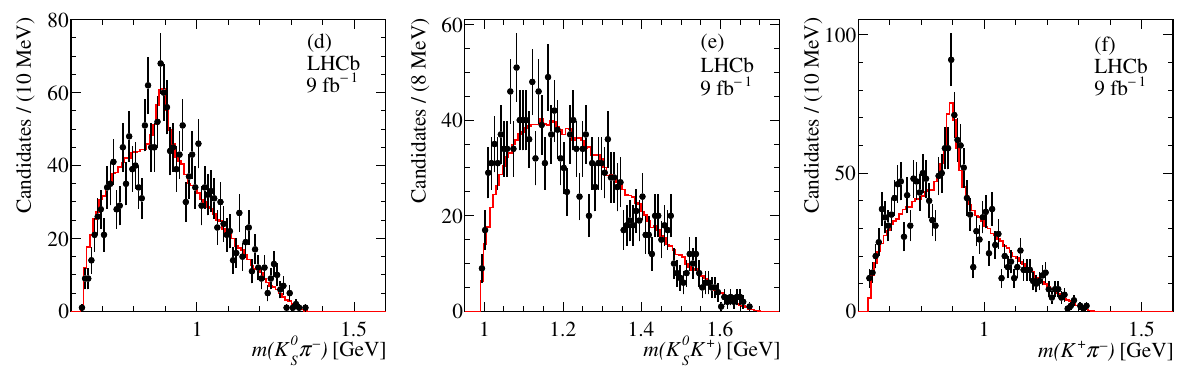}
\caption{\small\label{fig:Fig10} Two-particle mass distributions of the background candidates in the \Bu mass sidebands with the result of the fit also shown for (a)--(c) \kskkpip and (d)--(f) \kskkpim decays.}
\end{figure}

Figure~\ref{fig:Fig9} shows the \kskpi mass spectra of the \Bu sidebands along with the result from the fits.
The two-particle mass distributions and fits results are shown in Fig.~\ref{fig:Fig10}, with the fit results listed in Table~\ref{tab:Tab2}.

\begin{table} [!tb]
  \centering
  \caption{\small\label{tab:Tab2} Yields and fractional composition of the background from a fit to the candidates in the \Bu mass sidebands for \kskkpip and \kskkpim decays.}
  {\small
\begin{tabular}{lcccc}
\hline\\ [-2.3ex]
Final state & Candidates & $K^*_{\rm{ne}}(892)$ & $K^*_{\rm{ch}}(892)$ & Nonresonant\cr
\hline\\ [-2.3ex]
\kskkpip & 1702 & $0.110 \pm 0.016$ & $0.177 \pm 0.018$ & $0.713\pm  0.025$ \cr
\kskkpim & 1933 & $0.126 \pm 0.016$ & $0.073 \pm 0.016$ & $0.801\pm  0.022$ \cr
\hline
\end{tabular}
}
\end{table}

\section{Amplitude analysis}
\label{sec:data}

\begin{figure}[!tb]
\centering
\small
\includegraphics[width=0.45\textwidth]{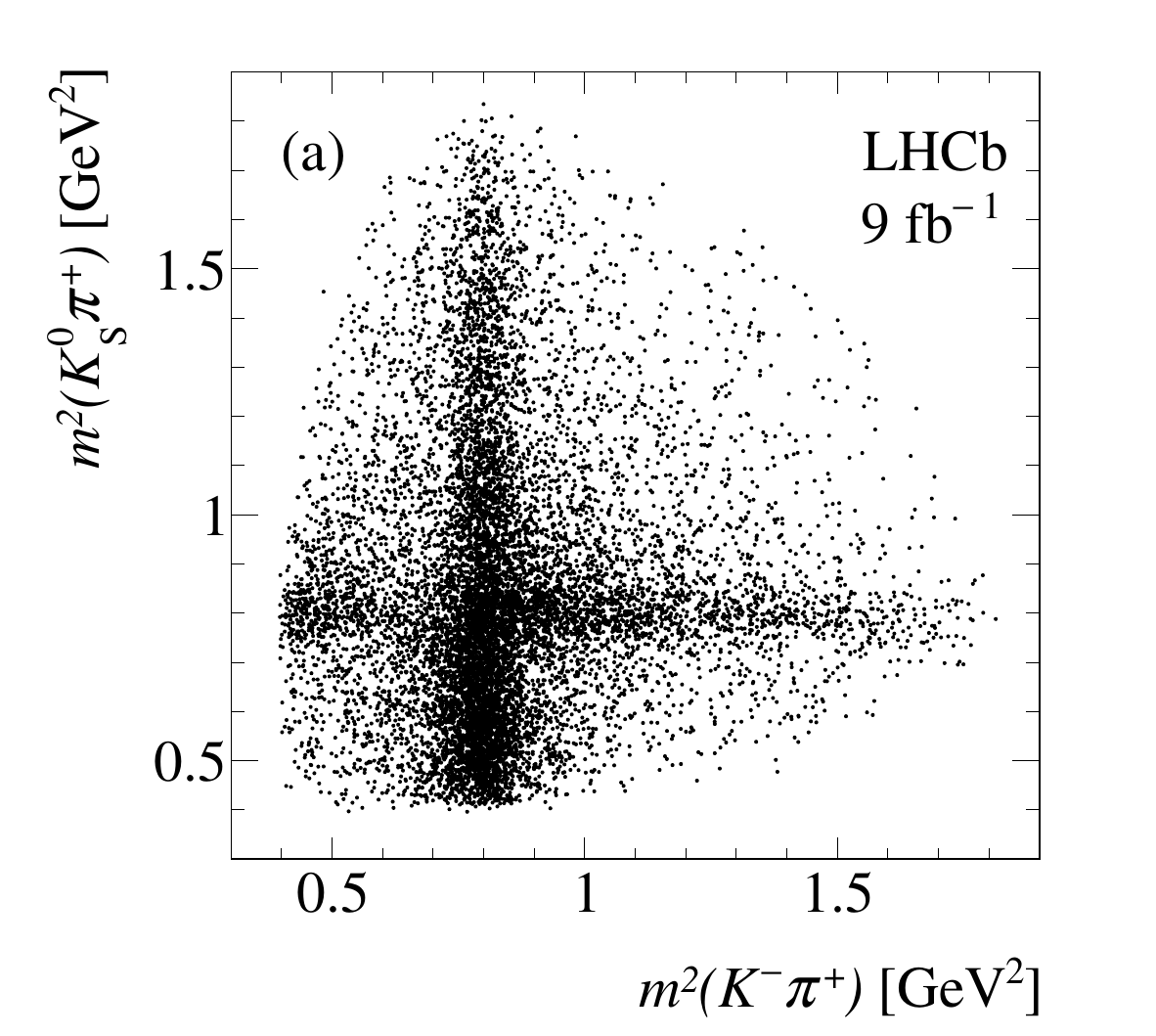}
\includegraphics[width=0.45\textwidth]{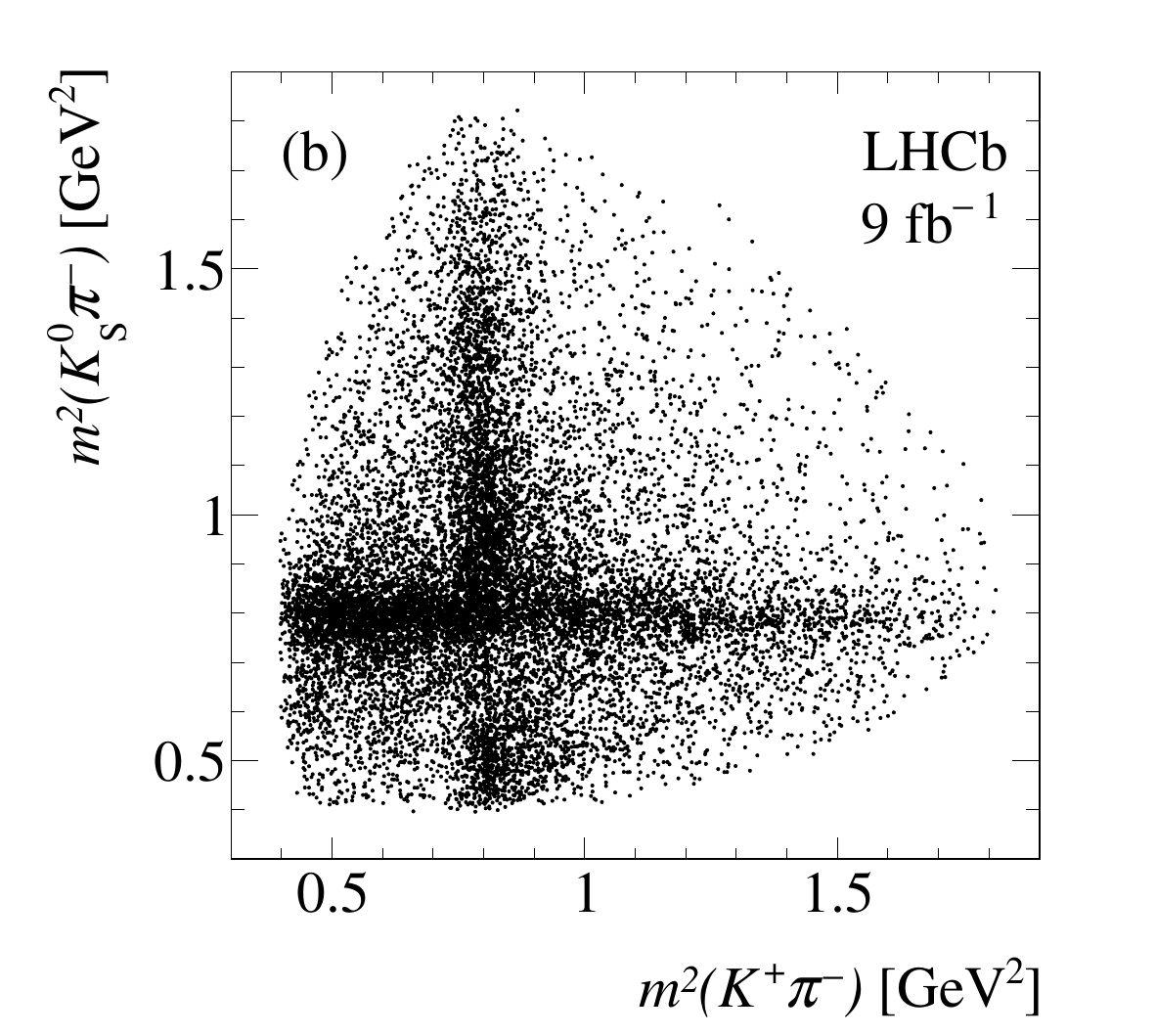}
\caption{\small\label{fig:Fig11} Dalitz plot distribution of the \kskpi system for candidates with \mbox{$1.30<m(\kskpi)<1.85\gev$} for (a) \bkskkpip and (b) \bkskkpim decays.}
\end{figure}

Figure~\ref{fig:Fig11} shows the \kskpi Dalitz plot in the $1.30<m(\kskpi)<1.85\gev$ mass region, separately for \bkskkpip (10830 candidates) and \bkskkpim (12930 candidates).
The distributions are dominated by two intersecting bands associated with the $K^{*}_{\rm{ne}}(892)$ and $K^{*}_{\rm{ch}}(892)$ resonances. Notably, the event distribution along the bands is not uniform, due to the spin of the contributing resonances and their interference. 
A comparison  between the  \bkskkpip and \bkskkpim data reveals opposite behavior in the interference pattern between the $K^{*}_{\rm{ne}}(892)$ and $K^{*}_{\rm{ch}}(892)$ bands.

\subsection{Fitting method}
\label{sec:fitmeth}

An amplitude analysis of \bkskkpip and \bkskkpim decays is performed with two unbinned maximum-likelihood fits, one for each decay channel.
The likelihood function is defined as
\begin{equation}
  \mathcal{L} = \prod_{n=1}^N \bigg[P\epsilon(\vec{y}_n)\frac{\sum_{i,j} c_i c_j^* A_i(\vec{x}_n) A_j^*(\vec{x}_n)}{\sum_{i,j} c_i c_j^* I_{A_i A_j^*}}+(1-P)\frac{\sum_{k}f_kB_k(\vec{z}_n)}{\sum_{k} f_kI_{B_k}}\bigg],
  \label{eq:like}
\end{equation}
\noindent where
$N$ is the number of events in the \Bu signal region and $P$ is the signal purity listed in Table~\ref{tab:Tab1}; $\epsilon(\vec{y}_n)$ is the efficiency parameterized as described in Sec.~\ref{sec:effy} in terms of the list of mass-reduced variables, here indicated by $\vec{y}_n$; the $A_i(\vec{x}_n)$ function, modeled by the nonrelativistic Zemach-tensor formalism~\cite{Zemach:1963bc,Dionisi:1980hi,Filippini:1995yc}, described in Appendix~\ref{sec:app1} and listed in Table~\ref{tab:Tab15}, describes the complex signal-amplitude contribution parameterized as a function of the list of parameters $\vec{x}_n$. The parameter $c_i$ is the complex coefficient for the $i$-th signal component, which is allowed to vary in the fit. One amplitude, the largest, is taken as the reference by setting $|c_i|=1$ with zero phase.  
The term $B_k(\vec{z}_n)$ represents the background probability-density function, described in terms of the parameters $\vec{z}_n$  as discussed in Sec.~\ref{sec:back}. It is assumed that interference between signal and background amplitudes can be ignored. 
The parameter $f_k$ is the magnitude of the $k$-th background component, obtained from the fit to the candidates in the sideband regions as described in Sec.~\ref{sec:back}. The terms $I_{A_i A_j^*}=\int A_i (\vec{x})A_j^*(\vec{x}) \epsilon(\vec{y}) {\rm d}\vec{x}\,{\rm d}\vec{y}$ and 
$I_{B_k}=\int B_k(\vec{z}) \, {\rm d}\vec{z}$ are normalization integrals. 
They are determined through numerical integration on phase-space-generated events, with initial-state masses for the signal and background samples set according to their respective distributions measured from the \Bu-candidate mass.
The \KSLL and \KSDD datasets enter in the likelihood function according to their efficiency and purity.

For each contribution, resonant or nonresonant, the fraction is defined as
\begin{equation}
f_i = \frac {|c_i|^2 \int |A_i(\vec{x})|^2 \, {\rm d}\vec{x}}
{\int |\sum_j c_j A_j(\vec{x})|^2 \, {\rm d}\vec{x}}.
\label{eq:frac}
\end{equation}
The fractions $f_i$ do not necessarily sum to 100\% because of interference effects. The uncertainty of each $f_i$ is evaluated by propagating the covariance matrix obtained from the fit.
Interference fractions are evaluated as

\begin{equation}
f_{ij} = \frac {\int 2 \Real[c_ic^*_j A_i(\vec{x}) A^*_j(\vec{x})]\, {\rm d}\vec{x}} 
{\int |\sum_j c_j A_j(\vec{x})|^2 \, {\rm d}\vec{x}}.
\label{eq:int}
\end{equation}

To evaluate the quality of the fit, a large simulated sample is prepared, where events are generated uniformly in the phase space~\cite{James:1968gu}. These events are weighted by the fitted likelihood function, normalized to the yields in data and compared to the data distribution on several invariant-mass and angular projections.
Several two-dimensional distributions are divided into a grid of $n \times n$
cells. In total $N_{\rm cells}$ cells are considered whose weighted yield is at least two.  A $\chi^2$ estimator is used, defined as
\begin{equation}
\chi^2 = \sum_{i=1}^{N_{\rm cells}} (N^i_{\rm obs}-N^i_{\rm exp})^2/\sigma^2,
\label{eq:chi2}
\end{equation}
where $N^i_{\rm obs}$ and $N^i_{\rm exp}$ are event yields from data and simulation, respectively. Here $\sigma\equiv\sqrt{N^i_{\rm exp}}$ for cells containing more than nine entries, while it is approximated as the average of the lower and upper Poisson uncertainties at the 68\% confidence level for lower statistics cells. The figure of merit for the fit quality is defined as \chisqndf, with ${\rm ndf} =N_{\rm cells}-n_{\rm par}-1$, where  $n_{\rm par}$ is the number of free parameters and 
 one degree of freedom is removed because of the normalization constraint.

\section{Amplitude fit}
\label{sec:fits}
  
\subsection{Study of the \boldmath{$f_1(1285)$} mass region}
\label{sec:f1}

\begin{figure}[!tb]
\centering
\small
\includegraphics[width=0.5\textwidth]{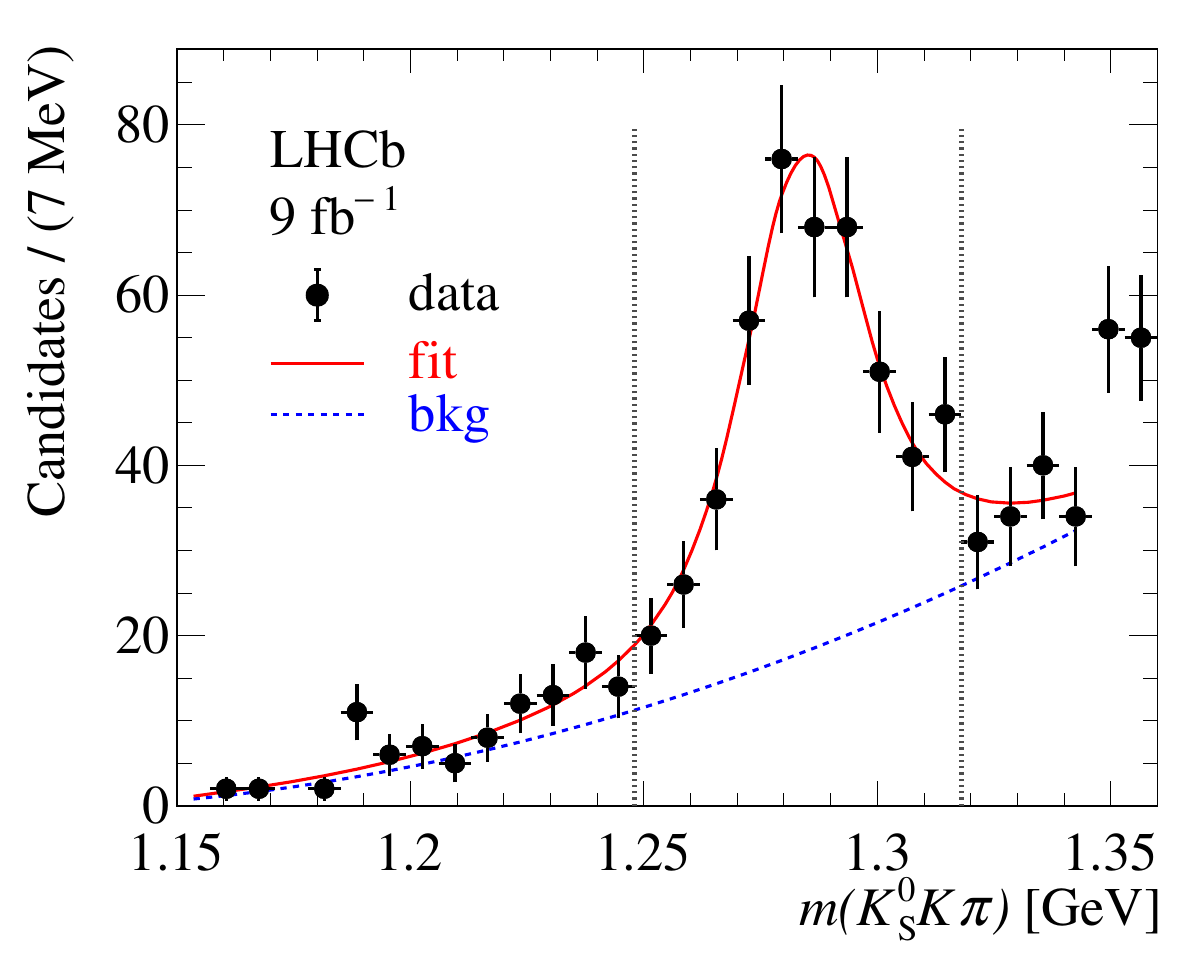}
\caption{\small\label{fig:Fig12} Fit to the $m(\kskpi)$ mass distribution in the $f_1(1285)$ mass region with the result of the fit also shown. The dashed line indicates the fitted background, the vertical dotted lines indicate the region used for the amplitude analysis. 
}
\end{figure}

The $f_1(1285)$ state is a well known resonance with established $J^P=1^+$ quantum numbers. The resonance is narrow and located in the threshold region of the \kskpi phase space, therefore little interference with other resonant amplitudes is expected. Therefore a study of this resonance allows to perform a simple test of the amplitude analysis model to correctly obtain a good description of the data and confirm the $f_1(1285)$ quantum numbers. The data from both \Bu decay modes are added as we do not expect significant differences.
The $f_1(1285)$ decays to the \kskpi final state mainly through the $a_0(980) \pi$ intermediate state~\cite{PDG2024}, hereafter denoted as $a_0\pi$.
Figure~\ref{fig:Fig12} shows the low-mass region of the \kskpi spectrum, summed over the \KSLL, \KSDD, \bkskkpip and \bkskkpim data samples, where a significant
$f_1(1285)$ signal is observed.

The data is fitted using the modulus squared of the BW function
\begin{equation}
 {\rm BW}(m) = \frac{1}{(m_0 - m) - i \Gamma/2},  
 \label{eq:bw}
\end{equation}
where $m$ indicates the \kskpi mass, multiplied by a function representing the \kskpi phase space for the signal and a second-order polynomial function for the background. Since
the known width of the $f_1(1285)$ resonance, $\Gamma=23.0 \pm 1.1 \mev$~\cite{PDG2024}, is comparable with the average mass resolution in this region, $\sigma=5.3\mev$, the signal is alternatively modeled using a BW convolved with the experimental mass resolution fixed to this value. 
This fit is shown in Fig.~\ref{fig:Fig12} with the results from both fits summarized in Table~\ref{tab:Tab3}.

Both approaches yield similar-quality fits and results, though the width is larger without the BW convolution (see Table~\ref{tab:Tab3}). 
However, for simplicity, the BW model with no convolution is used to describe the $f_1(1285)$ lineshape in the amplitude analysis described in Sec.~\ref{sec:fit_bw}. 

\begin{table} [!tb]
  \centering
  \caption{\small\label{tab:Tab3} Results from the fits in the $f_1(1285)$ mass region.}
  {\small
\begin{tabular}{lcccc}
\hline\\ [-2.3ex]
Fitting method & \chisqndf & $m_0$ [\mev] & $\Gamma$ [\mev]& Yield \cr 
\hline\\ [-2.3ex]
BW with resolution & 13.3/22 & $1283.5 \pm 1.5$ & $27.4 \pm 5.6$ & $360\pm50$ \cr
BW & 13.6/22 & $1283.5 \pm 1.6$ & $32.3 \pm 5.4$& $381\pm51$ \cr
\hline
\end{tabular}
}
\end{table}

An amplitude analysis of the data in the $f_1(1285)$ mass region is performed by selecting candidates in the $1.248<m(\kskpi)< 1.318\gev$ interval, and combining the \bkskkpip and \bkskkpim data samples. This sample corresponds to 497 events and a \Bu purity of $P=0.890 \pm 0.015$.

\begin{figure}[!ht]
\centering
\small
\includegraphics[width=0.9\textwidth]{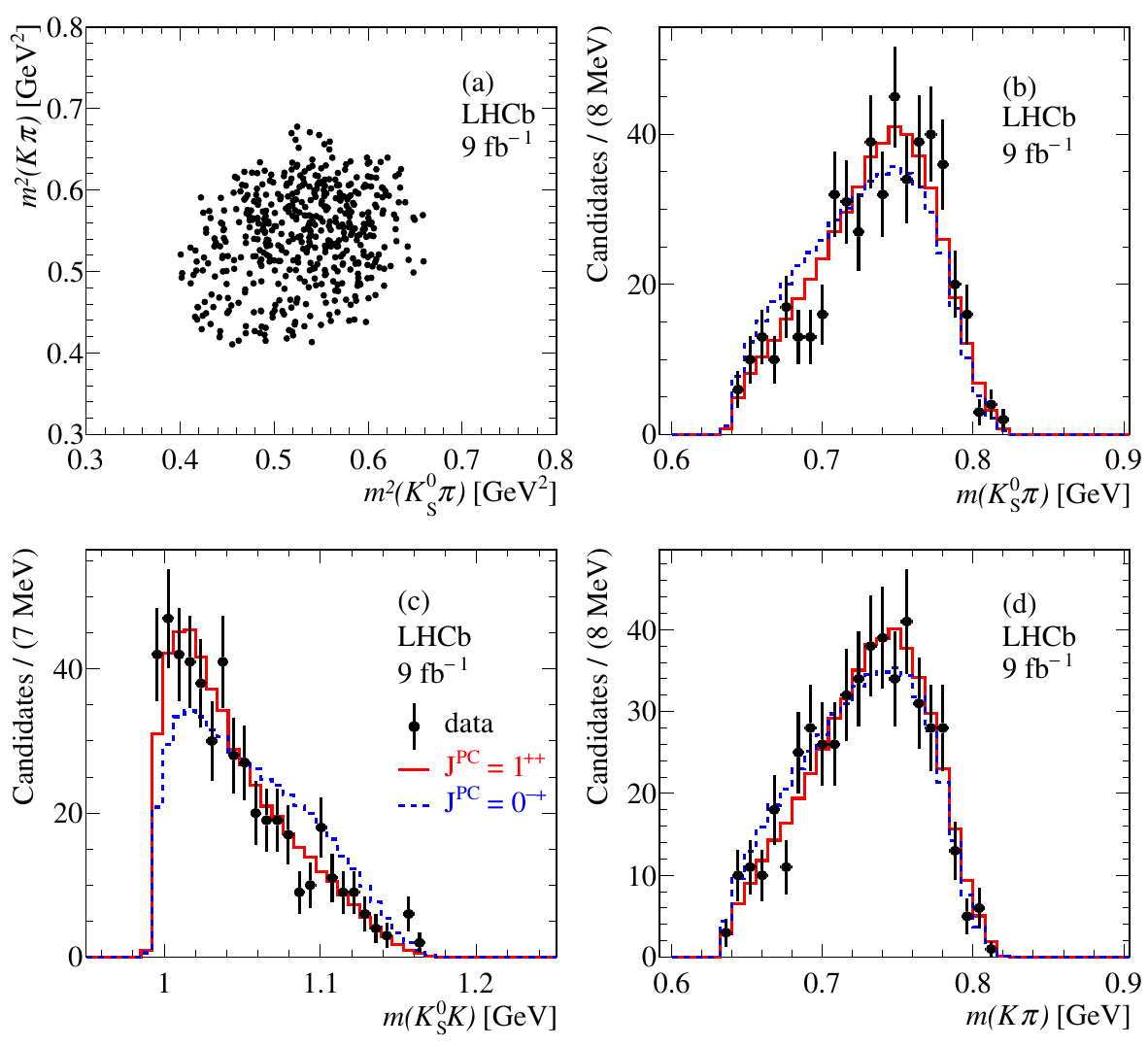}
\caption{\small\label{fig:Fig13} (a) Dalitz plot of candidates in the $f_1(1285)$ mass region and (b)--(d) projections with the fit result also shown.}
 \end{figure}

Three different hypotheses, listed in Table~\ref{tab:Tab4}, are used to fit the data; in all cases the interference between the different contributions is included.
In fit (a), the fit model consists of two contributions: $J^P=1^+$ $f_1(1285)\to a_0(980) \pi$ and a phase-space ($PS$) contribution.
In fit (b), the $f_1(1285)$ resonance is assumed to have quantum numbers $J^P=0^-$. This hypothesis is discarded by the fit, having a much worse likelihood and \chisqndf values.
For (c) an additional $\eta(1295) \to a_0(980) \pi$ contribution is included.
This gives a similar quality to fit (a), but returns large interference terms. The fraction of this additional contribution is $0.129 \pm 0.050$, consistent with zero within $2.6 \ \sigma$. 
Given the low significance and the large interference effects, fit (c) is discarded while fit (a) is considered as the baseline solution.
These results are confirmed by the full amplitude analysis of the \kskpi mass spectrum (see Sec.~\ref{sec:fit_bw}).
The fit projections, with comparisons between $J^P=1^+$ and $J^P=0^-$ $f_1(1285)$ assignments, are shown in Fig.~\ref{fig:Fig13}.

\begin{figure}[!ht]
\centering
\small
\includegraphics[width=0.9\textwidth]{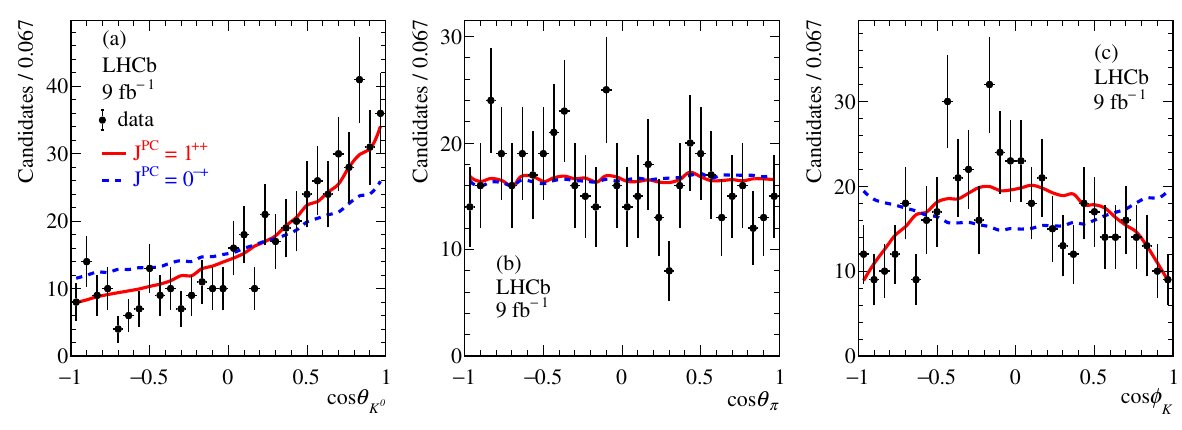}
\caption{\small\label{fig:Fig13a} Angular distributions in the $f_1(1285)$ mass region with fit projections corresponding to the two $J^{PC}$ hypotheses.}
 \end{figure}

To better visualize the difference between the two $J^{PC}$ hypotheses, Fig.~\ref{fig:Fig13a} shows the fit projections
to the $\cos\theta_{K^0}$, $\cos\theta_{\pi}$ and $\cos \phi_K$ distributions. A significant discriminant power can be observed in the
distribution of $\cos \phi_K$ due to the $f_1(1285)$ resonance originating from a \Bu decay. 

The quoted \chisqndf value listed in Table~\ref{tab:Tab4} for the three different scenarios is obtained by dividing the Dalitz plot into a $10\times10$ grid and using the method described in Sec.~\ref{sec:fitmeth}

\begin{table} [!tb]
  \centering
  \caption{\small\label{tab:Tab4} Results from the amplitude analysis of the $f_1(1285)$ mass region. Likelihood variations are evaluated with respect to the baseline fit (a). Positive changes in the $\Delta(-2 \log \calL)$ means better results.}
  {\small
\begin{tabular}{lccccccc}
\hline\\ [-2.3ex]
Amplitudes & $\Delta(-2 \log \calL)$ & Fractions  & Sum of           & \chisqndf \cr
           &                         &            & fractions        &           \cr
\hline\\ [-2.3ex]
(a) $1^+ \ f_1(1285),\ PS$ & -- & $0.601\pm0.042$, $0.392 \pm 0.042$ & 0.993 & 62/66=0.94 \cr
(b) $0^- \ f_1(1285), \ PS$ & \alm\alm$-405.3$ & $0.164\pm0.041$, $0.784 \pm 0.104$ & 0.948 & 85/64=1.32 \cr
(c) $1^+ \ f_1(1285), \ PS $, & 19.7 & $0.577 \pm 0.043$, $0.766 \pm 0.101$ & 1.471 & 64/66=0.97\cr
\al\al\al$0^- \ \eta(1295)$ &     & $0.129 \pm 0.050$ & \cr
\hline
\end{tabular}
}
\end{table}

\subsection{Amplitude analysis of the full low-mass \boldmath{\kskpi} region}
\label{sec:fit_bw}

It is well known that the BW function provides a simplified description, strictly valid only in cases of isolated, single resonances, far away from thresholds. In the present analysis a particularly complex problem is faced, with the presence, in a limited mass region, of several interfering amplitudes with different quantum numbers. This approach, taking into account its limitations, is employed to describe the data in the present analysis.

An unbinned maximum-likelihood fit is performed to the  data in the low-mass region \mbox{$m(\kskpi)<1.85$\gev} inserting all possible resonances which decay into the \kskpi final state~\cite{PDG2024}.
The \bkskkpip and \bkskkpim data are fitted separately.
A search for the best solution is performed by adding resonances one-by-one and considering as figures of merit the significance of their fit fractions, which relates to the increase of the likelihood and the decreasing \chisqndf.
Defining the significance of a given contribution $\sigma_R\equiv f/\delta_f$, where $f$ and $\delta_f$ are the fitted fraction and the corresponding statistical error, contributions with $\sigma_R<3$ are discarded.
The list of resonances contributing to the two \Bu decay modes is given in Table~\ref{tab:Tab5}.
In addition, a nonresonant phase-space contribution, and incoherent $K^*_{\rm ne}(892)$ and $K^*_{\rm ch}(892)$ contributions are included, modeled by relativistic Breit--Wigner functions with no angular dependence.
According to the results of Ref.~\cite{BESIII:2022chl}, $R^0$ resonances are also allowed to decay directly to the \kskpi final state, here described by a constant term,  neglecting a possible intermediate contribution from the $K^*_0(700)$ whose parameters are affected by large uncertainties~\cite{PDG2024}.

Different fitting sequences are tried, taking as reference amplitude the $J^{PC}=0^{-+}$ $\eta(1475) \to K^*\Kb$ or the $J^{PC}=1^{++}$ $f_1(1420)\to K^*\Kb$, which are found to be significant in all the fits. Additional contributions are included and removed to test their significance and the effects on the fitting sequence. After finding the best solution,
 additional tests are performed to evaluate the significance of lower level contributions. It is concluded that the baseline solution described below is unique and the list of resonances that describe the data is the same for both \Bu decay modes.

\begin{table} [!tb]
   \caption{\small\label{tab:Tab5} List of the light-meson $R^0$ resonances which contribute to the \Bu decays studied.  For the $\eta(1475)$ resonance both the PDG values~\cite{PDG2024} (in parentheses) and values from Ref.~\cite{BESIII:2022chl} are listed. For other resonances PDG values are shown. The $f_1(1285)$ width has been increased to $32.3\mev$  to take into account the experimental resolution (see Sec.~\ref{sec:f1}). The decay products of the $h_1$ resonances can be in either a relative $S$ or $D$ wave (see Table~\ref{tab:Tab15}).}
  \centering
\begin{tabular}{lccccl}
  \hline\\ [-2.3ex]
Resonance &  $J^{PC}$  & $m_0 \ [\mev]$ & $\Gamma \ [\mev]$ & Decay mode \cr
\hline\\ [-2.3ex]
$\eta(1295)$ & $0^{-+}$ & $1294.4 \pm 4$ & $55 \pm 5$ & $a_0 \pi$ \cr
$\eta(1405)$ & $0^{-+}$ & $1408.8\pm2$ & $50.1\pm 2.6$ & $a_0 \pi$  \cr
             &         &              &               & $K^*\Kb$    \cr
             &         &              &               & $PS  $     \cr
$\eta(1475)$ & $0^{-+}$ & $(1475\pm4)$   & $(90\pm 9)$     & $a_0 \pi$  \cr
             &         & $1507.6\pm1.6$&\alm $115.8\pm2.4$ & $K^*\Kb$    \cr
             &         &              &               & $PS  $    \cr
$\eta(1760)$ & $0^{-+}$ & $1751\pm15$  & $240\pm30$    & $a_0 \pi$   \cr
             &         &              &               & $K^*\Kb$     \cr
             &         &              &               & $PS  $    \cr
$f_1(1285)$  & $1^{++}$ & $1281.9\pm0.5$& $22.7\pm 1.1$ & $a_0 \pi$ \cr
  & & & \alm\alm\alm\alm\alm(32.3) & \cr
$f_1(1420)$  & $1^{++}$ &  $1426.3\pm 0.9$ & $54.5\pm 2.6$& $K^*\Kb$    \cr
$f_1(1510)$  & $1^{++}$ & $1518\pm5$ & $73\pm 25$ & $K^*\Kb$         \cr
$h_1(1415)$ & $1^{+-}$ & $1416\pm8$ & $90\pm 15$ & $K^*\Kb [S]$          \cr
            &         &            &            & $K^*\Kb [D]$       \cr
$h_1(1595)$ & $1^{+-}$ & $1594\pm15$ & $384\pm 60$ & $K^*\Kb [S]$   \cr
            &         &             &             & $K^*\Kb [D]$    \cr
$\eta_2(1645)$ & $2^{-+}$ & $1617\pm5$ & $181\pm11$ & $a_0 \pi$   \cr
&         &            &            & $K^*\Kb$     \cr
\hline
\end{tabular}
\end{table}

In the baseline fit, the amplitude for the $\eta(1475) \to K^*\Kb$ decay is used as a reference, with parameters fixed to the values of Ref.~\cite{BESIII:2022chl}.
The fit-fraction and relative-phase results from the fits to the \bkskkpip and \bkskkpim data are summarized in Table~\ref{tab:Tab6}.
\begin{table} [!tb]
  \centering
  \caption{\small\label{tab:Tab6} Results of the amplitude analysis for the two $\Bu \to R^0(\to \kskpi) \Kp$ final states. The parameters of the listed contributions are given in Table~\ref{tab:Tab5}.}
  \resizebox{1.0\columnwidth}{!}{
\begin{tabular}{ll| rr|rr}

                &       & \multicolumn{2}{c|}{\bkskkpip}  & \multicolumn{2}{c}{\bkskkpim}  \cr
\hline\\ [-2.3ex]
Contribution & Decay & Fraction [\%]& Phase  [rad]& Fraction [\%] & Phase [rad] \cr \hline\\ [-2.3ex]
$\eta(1475)$& $K^*\Kb$ & $10.7\pm 1.1 \pm 1.1$ & 0\alp\alp\alp& $10.3\pm 1.1\pm1.4$ & 0 \alp\alp\alp\cr
& $a_0 \pi$ & $ 1.4\pm 0.4 \pm 0.4$& $ 3.18\pm 0.19\pm0.15$ & $ 1.8\pm 0.4\pm0.4$ &$ 2.92\pm 0.14\pm0.13$\cr
& $PS$ &$15.2\pm 2.1 \pm 2.1$ & $ 3.33\pm 0.10\pm0.12$&$ 8.9\pm 1.4\pm2.7$ &$ 3.57\pm 0.11\pm0.13$\cr
\cline{2-6} \\ [-2.3ex]
& Total & $27.4 \pm 2.4 \pm 2.4$ & -- \alp\alp\alp& $21.0 \pm 1.8 \pm 3.1$ & --\alp\alp\alp \cr
\hline\\ [-2.3ex]
$\eta(1760)$ & $K^*\Kb$ & $ 1.9\pm 0.4 \pm 0.3$ & $-1.53\pm 0.16\pm0.28$ & $ 3.1\pm 0.4\pm0.5$ & $-1.16\pm 0.13\pm0.20$ \cr
& $a_0 \pi$& $ 2.0\pm 0.4 \pm 0.3$ & $ 2.11\pm 0.15\pm0.20$ &$ 1.7\pm 0.4\pm0.3$ & $ 3.06\pm 0.12\pm0.24$\cr
& $PS$ & $11.9\pm 1.8 \pm 2.7$ & $ 1.60\pm 0.10\pm0.19$ & $23.2\pm 2.4\pm5.1$ &$ 1.92\pm 0.07\pm0.22$ \cr
\cline{2-6}\\ [-2.3ex]
& Total & $15.8 \pm 1.9 \pm 2.7$ & -- \alp\alp\alp& $27.9 \pm 2.5 \pm 5.1$ & -- \alp\alp\alp\cr
\hline\\ [-2.3ex]
$\eta(1405)$ & $K^*\Kb$ & $ 3.5\pm 0.6 \pm 1.9$ & $-0.10\pm 0.10\pm0.20$ &$ 2.3\pm 0.5\pm0.7$ &$-0.01\pm 0.11 \pm 0.18$\cr
& $PS$ & $ 5.2\pm 0.5\pm 0.8$ & $ 1.77\pm 0.11\pm0.28$ &$ 6.4\pm 0.5\pm 0.9$ &$1.96\pm 0.13\pm0.20$\cr
\cline{2-6}\\ [-2.3ex]
& Total & $8.7 \pm 0.8 \pm 2.0$ & -- \alp\alp\alp& $8.6 \pm 0.7 \pm 1.1$ & -- \alp\alp\alp\cr
\hline\\ [-2.3ex]
$f_1(1285)$ & $a_0 \pi$ & $ 2.0\pm 0.2\pm0.2$ & $-0.35\pm 0.13\pm0.26$ &$ 2.0\pm 0.2\pm0.2$ &$-0.47\pm 0.11\pm0.12$\cr
$f_1(1420)$ & $K^*\Kb$& $11.4\pm 0.7\pm2.1$ & $ 4.25\pm 0.08\pm0.23$ &$ 6.6\pm 0.5\pm1.7$ &$ 4.67\pm 0.10\pm0.27$ \cr
\hline\\ [-2.3ex]
$h_1(1415)$ & $K^*\Kb [S]$& $10.0\pm 0.9\pm2.0$ & $ 4.59\pm 0.08\pm0.20$ &$18.6\pm 1.2\pm3.5$ &$ 1.57\pm 0.08\pm0.55$\cr
& $K^*\Kb [D]$& $ 3.3\pm 0.3\pm0.2$ & $-0.13\pm 0.08\pm0.16$ & $ 2.4\pm 0.3\pm0.2$ &$-2.57\pm 0.08\pm0.42$\cr
\cline{2-6}\\ [-2.3ex]
& Total & $13.3 \pm 1.0 \pm 2.0$ & -- \alp\alp\alp& $21.0 \pm 1.2 \pm 3.6$ & --\alp\alp\alp \cr
\hline\\ [-2.3ex]
$f_1(1510)$ & $K^*\Kb$ & $ 2.9\pm 0.4\pm1.2$ & $-3.23\pm 0.09\pm0.47$ & $ 2.6\pm 0.3\pm2.7$ &$-2.88\pm 0.09\pm0.23$\cr
$h_1(1595)$ & $K^*\Kb [S]$ &$5.3 \pm 0.8\pm1.4$ & $ 1.44\pm 0.08\pm0.19$ &  $14.8\pm 1.4\pm2.8$ &$ 4.20\pm 0.06\pm0.48$\cr
$\eta_2(1645)$ &$K^*\Kb$ & $ 1.6\pm 0.2\pm0.8$ & $ 2.28\pm 0.10\pm0.09$ &$ 0.8\pm 0.2\pm0.2$ & $ 2.15\pm 0.12\pm0.09$ \cr
$PS$ & & $18.2\pm 2.4\pm2.9$ & $ 3.13\pm 0.09\pm0.15$ & $25.1\pm 2.6\pm5.0$ &$ 2.85\pm 0.07\pm0.16$\cr
$K^*_{\rm ne}$ & & $ 3.2\pm 0.4\pm0.3$ & $-0.87\pm 0.08\pm0.22$ &$ 2.0\pm 0.3\pm0.3$ &$-1.36\pm 0.08\pm0.16$\cr
$K^*_{\rm ch}$ & & $ 8.4\pm 0.7\pm 0.9$ & $-0.86\pm 0.07\pm0.19$ &$ 2.8\pm 0.4\pm1.0$ &$-0.82\pm 0.07\pm0.10$\cr
\hline\\ [-2.3ex]
Sum & & $118.0 \pm 4.3 \pm 6.2$ & -- \alp\alp\alp & $135.5 \pm 4.6 \pm 9.7$& -- \alp\alp\alp \cr
\hline
\end{tabular}
}
\end{table}

The fits find the same resonance composition for both \bkskkpip and \mbox{\bkskkpim} decays and the results can be summarized as follows.
The decay \mbox{$\Bu \to R^0(\to \kskpi) \Kp$} is dominated by pseudoscalar $\eta(1405)$, $\eta(1475)$ and $\eta(1760)$ resonances,
which contribute via the $\Kstar \Kb$, $a_0 \pi$ and direct \kskpi decay modes.
Contributions from $a_0(1450)\pi$ are tested but found consistent with zero.
Significant contributions from the $J^{PC}=1^{++}$ $f_1(1420)$ and $f_1(1510)$ resonances are also observed. Concerning the negative $C$-parity states, the $J^{PC}=1^{+-}$ $h_1(1415)$ resonance is present in both $S$ and $D$-waves (see Table~\ref{tab:Tab15}) as well as the $h_1(1595)$ resonance, which contributes only in the $S$-wave decay mode.

An attempt to include the $\eta(1295)\to a_0\pi$ contribution is made, but the significance $\sigma_R$ is found to be 
below the threshold defined above.
An $f_2'(1525)$ component is also included, but its fit fraction is consistent with zero as expected due to the very small branching fraction reported in Ref.~\cite{BESIII:2022chl}. Similarly, the presence of a $\phi(1680)$ resonance is tested, but rejected by the fits to the data.

The sum of the fractions exceeds 100\%, indicating the presence of interference effects.
The majority of the interference terms are found to be very small.
However, contributions with fractions greater than 5\% are present and are mostly due to resonances decaying directly to \kskpi or via global $PS$ contributions.

A comparison is made between the results obtained from the amplitude analyses of the two \Bu decays. This is achieved by introducing the significance $n\sigma_1$, defined as $\Delta f/\delta$, where $\Delta f$ indicates the fraction difference and  $\delta$ is the associated statistical uncertainty. A significance $n\sigma_2$ is also evaluated by replacing $\epsilon$ as the sum in quadrature of the statistical and systematic uncertainties.
Similar quantities are defined for the evaluation of the significance of the phase difference $\Delta \phi$.

The results, summarized in Table~\ref{tab:Tab7}, can be outlined as follows. Most of the fractional contributions are consistent between the two \Bu decay modes. A few fractional contributions have $n\sigma_1>5$ but with lower significances when $n\sigma_2$ is considered. In addition it can be noted that relative phases of positive $C$-parity contributions are consistent, while significant differences are observed for negative $C$-parity resonance contributions. This effect is observed here for the first time.

\begin{table} [!t]
  \centering
  \caption{\small\label{tab:Tab7} Fit-fractions and relative-phase differences between the results from amplitude analyses of the \bkskkpip and \bkskkpim decay modes. The significances $n\sigma_1$ and $n\sigma_2$ defined in the text are also listed for each contribution.}
  {\small
\begin{tabular}{ll|  ccc| ccc}
  \hline\\ [-2.3ex]
  Contribution & Decay  & \ $\Delta f$ & $n\sigma_1$ & $n\sigma_2$ & $\Delta \phi$ & $n\sigma_1$ & $n\sigma_2$ \cr
\hline\\ [-2.3ex]
$\eta(1475)$ & $K^*\Kb$ & \alpp $0.5\pm 1.5\pm 1.8$ &   0.3 &   0.2 &  -- &   -- &   -- \cr
& $a_0 \pi$ & $ -0.4\pm 0.6\pm 0.5$ &   0.7 &   0.6 & \alpp$  0.26\pm 0.24\pm 0.20$ &   1.1 &   0.8 \cr
& $PS$ & \alpp $  6.2\pm 2.5\pm 3.5$ &   2.5 &   1.5 & $ -0.24\pm 0.15\pm 0.18$ &   1.6 &   1.0 \cr
\cline{2-8}\\ [-2.3ex]
& Total  & \alpp$  6.3\pm 3.0\pm 3.9$ &   2.1 &   1.3 &   --  &   -- &   -- \cr
\hline\\ [-2.3ex]
$\eta(1760)$ & $K^*\Kb$ & $ -1.1\pm 0.6\pm 0.6$ &   1.9 &   1.4 & $ -0.37\pm 0.21\pm 0.34$ &   1.8 &   0.9 \cr
& $a_0 \pi$& \alpp $  0.3\pm 0.5\pm 0.4$ &   0.5 &   0.4 & $ -0.95\pm 0.19\pm 0.31$ &   5.0 &   2.6 \cr
 & $PS$ & $-11.4\pm 3.0\pm 5.8$ &   3.9 &   1.8 & $ -0.32\pm 0.12\pm 0.29$ &   2.6 &   1.0 \cr
\cline{2-8}\\ [-2.3ex]
& Total  & $-12.2\pm 3.0\pm 5.8$ &  4.1 &  2.0 &  -- &   -- &   -- \cr
\hline\\ [-2.3ex]
$\eta(1405)$ & $K^*\Kb$ & \alpp$  1.2\pm 0.8\pm 2.0$ &   1.5 &   0.5 & $ -0.09\pm 0.15\pm 0.27$ &   0.6 &   0.3 \cr
&  $PS$ & $ -1.2\pm 0.8\pm 1.2$ &   1.5 &   0.8 & $ -0.19\pm 0.17\pm 0.34$ &   1.1 &   0.5 \cr
\cline{2-8}\\ [-2.3ex]
& Total  & \alpp $  0.0\pm 1.1\pm 2.3$ &   0.0 &   0.0 &   -- &   -- &   -- \cr
\hline\\ [-2.3ex]
$f_1(1285)$ & $a_0 \pi$ & $ -0.1\pm 0.3\pm 0.3$ &   0.2 &   0.2 & \al$  0.1\pm 0.2\pm 0.3$ &   0.7 &   0.4 \cr
\hline\\ [-2.3ex] $f_1(1420)$ &$K^*\Kb$ & \alpp$  4.8\pm 0.9\pm 2.7$ &   5.6 &   1.7 &  $ -0.4\pm 0.1\pm 0.4$ &   3.3 &   1.1 \cr
 \hline\\ [-2.3ex]
$h_1(1415)$ & $K^*\Kb [S]$&  $ -8.6\pm 1.5\pm 4.1$ &   5.8 &   2.0 & \al $  3.0\pm 0.1\pm 0.6$ &  \alm 26.7 &   5.1 \cr
& $K^*\Kb [D]$& \alpp $  0.9\pm 0.4\pm 0.3$ &   2.4 &   1.9 &\al $  2.4\pm 0.11\pm 0.5$ & \alm 21.6 &   5.3 \cr
\cline{2-8}\\ [-2.3ex]
& Total  &  $ -7.7\pm 1.5\pm 4.1$ &   5.0 &   1.8 &  -- &   -- &   -- \cr
\hline\\ [-2.3ex]
$f_1(1510)$ & $K^*\Kb$ & \alpp $  0.3\pm 0.5\pm 3.0$ &   0.6 &   0.1 & $ -0.35\pm 0.13\pm 0.52$ &   2.8 &   0.7 \cr
$h_1(1595)$ & $K^*\Kb [S]$ &$ -9.5\pm 1.7\pm 3.1$ &   5.8 &   2.7 & $ -2.76\pm 0.10\pm 0.52$ & \alm 27.6 &   5.3 \cr
$\eta_2(1645)$ &$K^*\Kb$ & \alpp $  0.8\pm 0.3\pm 0.8$ &   3.0 &   0.9 & \al $  0.13\pm 0.16\pm 0.13$ &   0.8 &   0.7 \cr
$PS$ & & $ -6.8\pm 3.5\pm 5.8$ &   2.0 &   1.0 & \al $  0.28\pm 0.11\pm 0.22$ &   2.5 &   1.1 \cr
\hline
\end{tabular}
}
\end{table}  

\begin{figure}[!tb]
\centering
\small\includegraphics[width=0.45\textwidth]{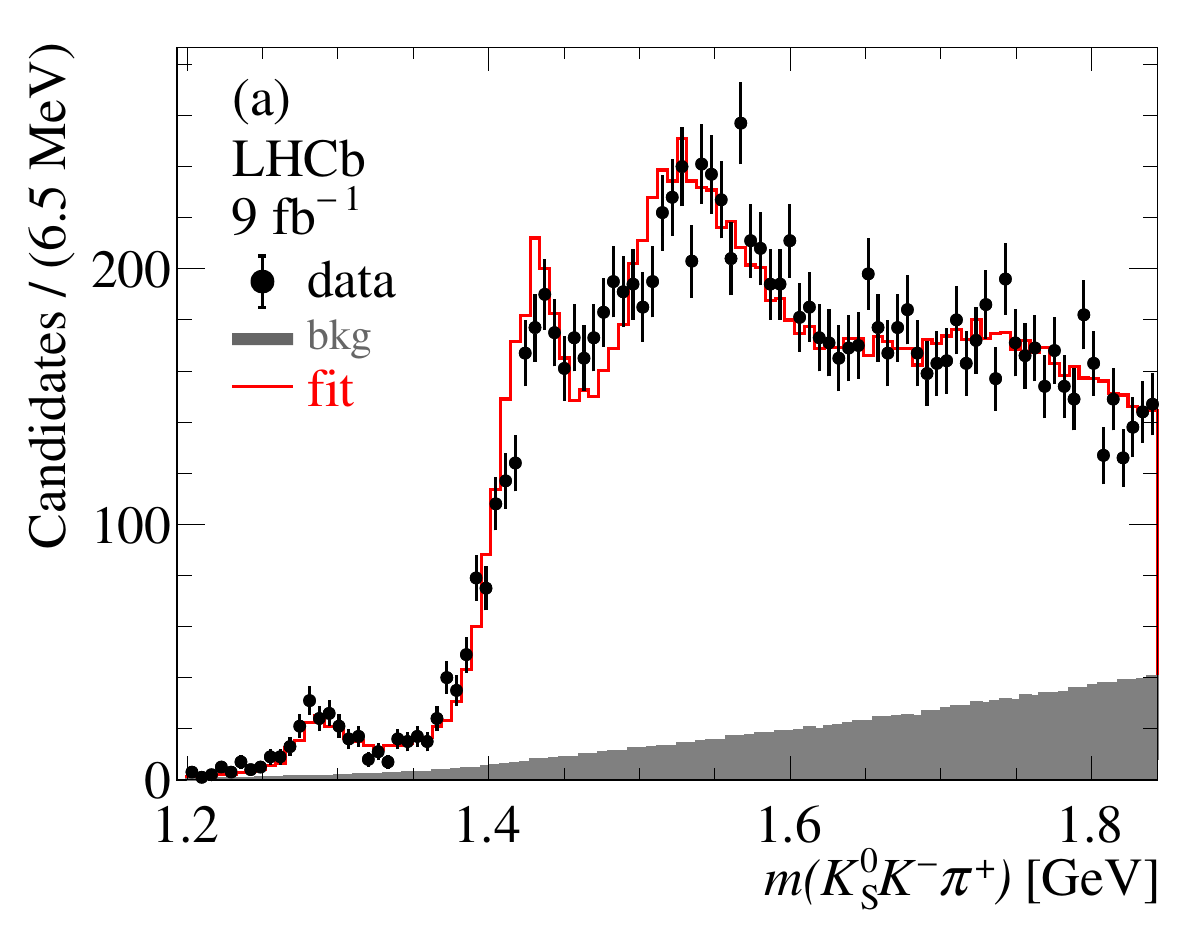}
\includegraphics[width=0.45\textwidth]{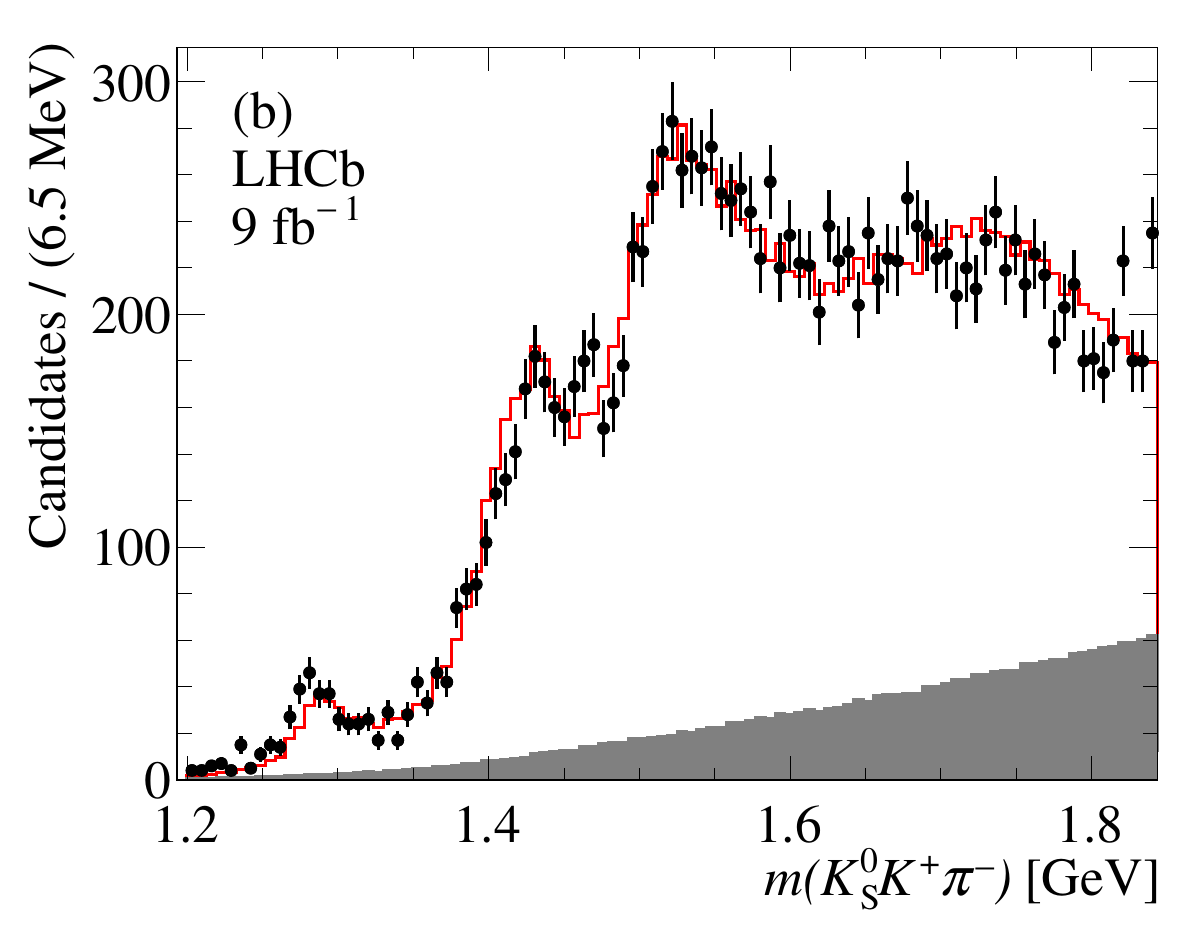}
\caption{\small\label{fig:Fig14} \kskpi invariant-mass for (a) \bkskkpip and (b) \mbox{\bkskkpim} candidates with the amplitude analysis fit projections also shown.
}
\end{figure}

\begin{figure}[!tb]
\centering
\small
\includegraphics[width=0.95\textwidth]{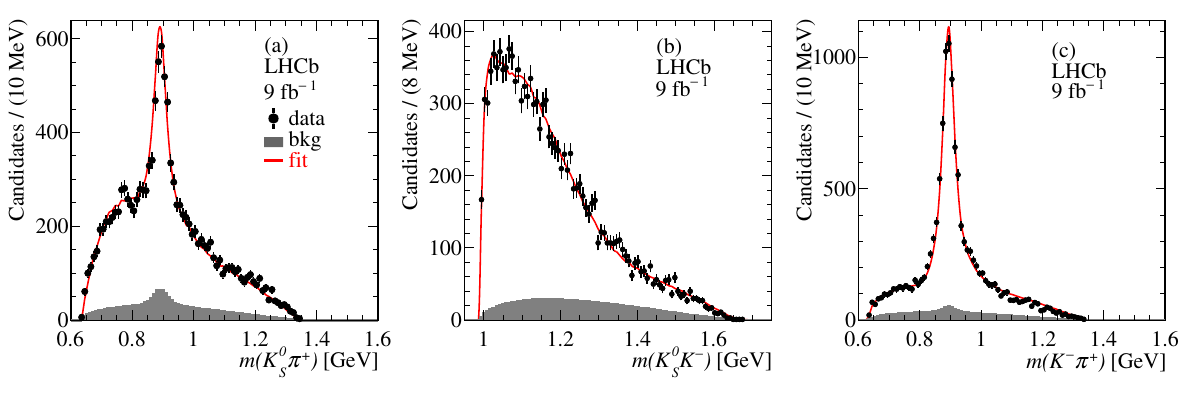}
\includegraphics[width=0.95\textwidth]{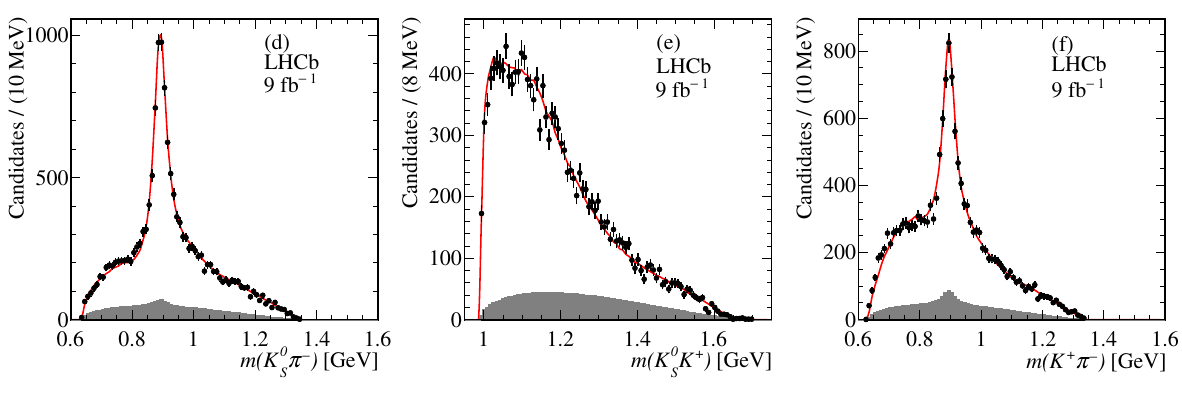}
\caption{\small\label{fig:Fig15} Two-body invariant-mass distributions for (a)--(c) \bkskkpip and (d)--(f) \mbox{\bkskkpim} candidates with the projections from the amplitude analysis.
}
\end{figure}

In Sec.~\ref{sec:data} a significant difference is found between the \kskpi Dalitz plots for \mbox{\bkskkpip} and \bkskkpim data which are explained by the observed phase differences in the two decay modes. 
Figure~\ref{fig:Fig14} shows the \kskpi mass spectrum and Fig.~\ref{fig:Fig15} the two-body mass projections for \bkskkpip and \mbox{\bkskkpim} data together with the projections of the fit. 
Note the different behavior in the $K \pi$ mass distributions between the two decay channels.
A reasonable agreement between the fit and the data is observed in all the distributions.

\begin{figure}[!tb]
\centering
\small
\includegraphics[width=0.950\textwidth]{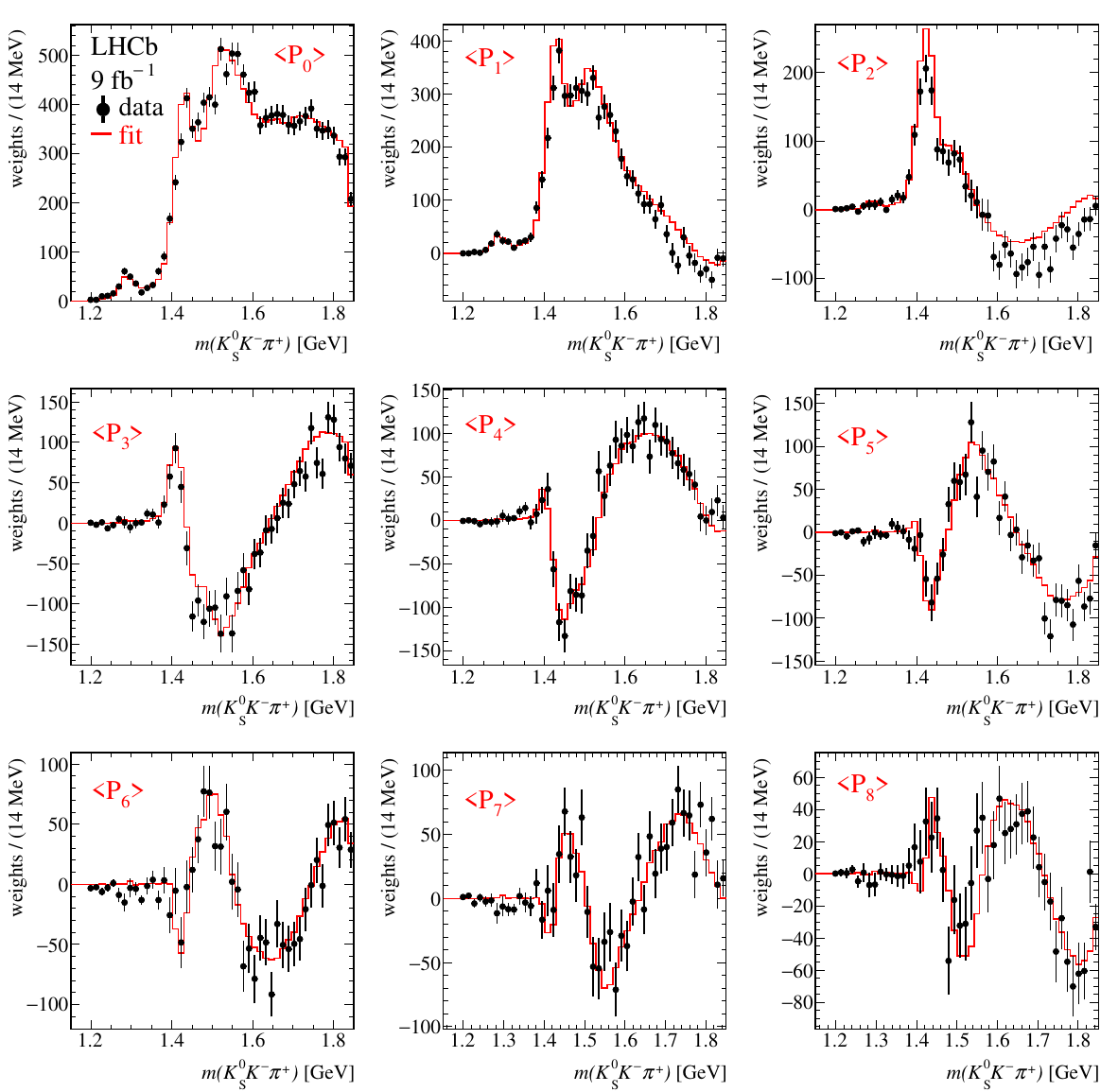}
\caption{\small\label{fig:Fig16} Fit projections of the \kskpi mass spectra where both the fit and the data are weighted by Legendre-polynomial moments as functions of $\cos \theta_K$ for \bkskkpip decays.
}
\end{figure}

\begin{figure}[!tb]
\centering
\small
\includegraphics[width=0.950\textwidth]{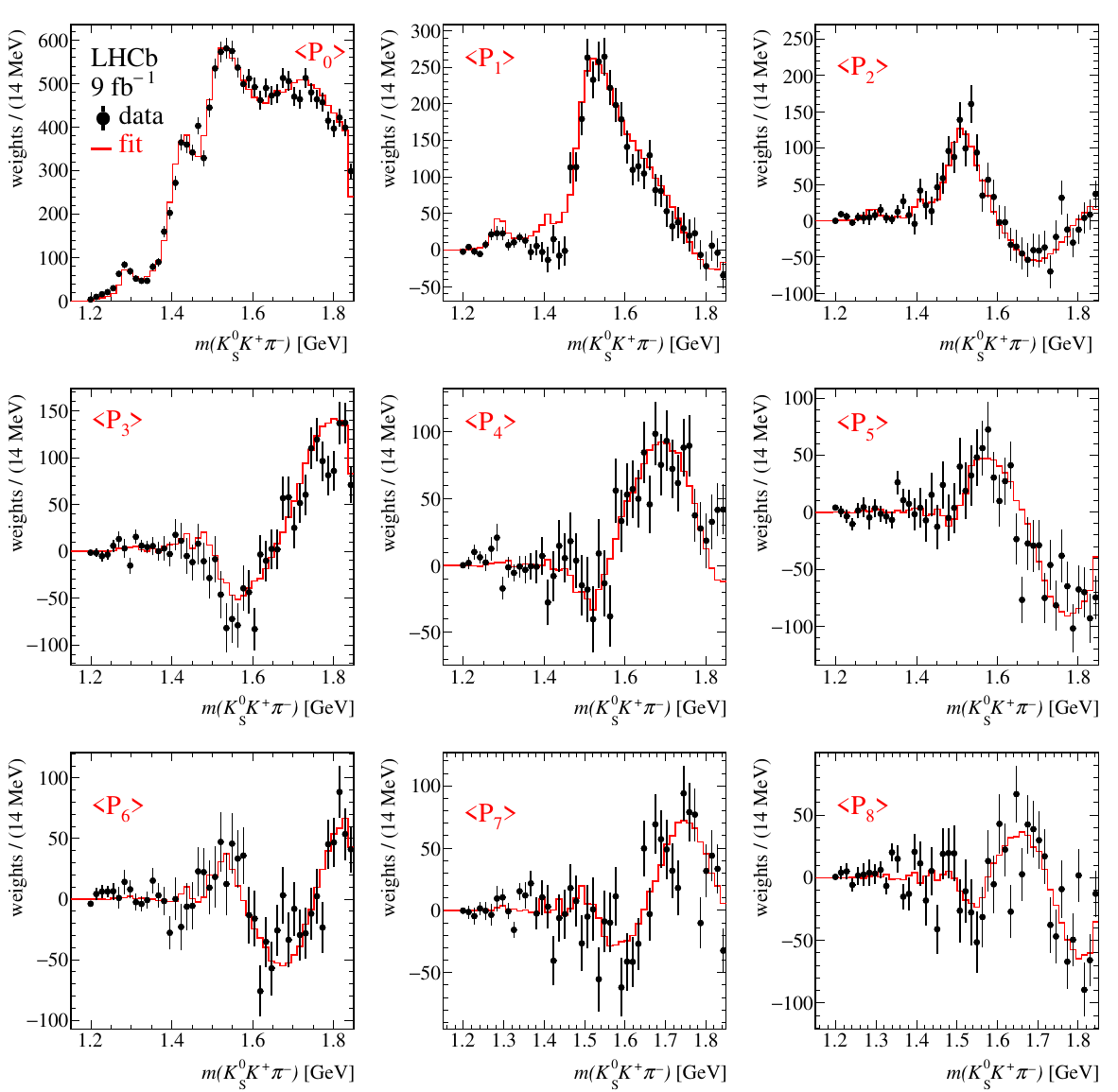}
\caption{\small\label{fig:Fig17} Fit projections on \kskpi mass spectra where both the fit and the data are  weighted by Legendre polynomial moments as functions of $\cos \theta_K$ for \bkskkpim decays.
}
\end{figure}

Figures~\ref{fig:Fig16} and~\ref{fig:Fig17} show the invariant-mass distributions and the fitted functions weighted
by Legendre-polynomial moments up to order eight, for \bkskkpip and \mbox{\bkskkpim} data, respectively.
 A good agreement between the data and the fit is observed for all the distributions, 
indicating that the fit is capable of reproducing local structure observed in the Dalitz plot.
A comparison between the two figures emphasizes once more the difference in the angular distributions between the two \Bu decay modes.

An additional test on the fit quality is performed by evaluating the \chisqndf value of different two-dimensional distributions, divided into $40 \times 40$ cells, following the method described in Sec.~\ref{sec:fitmeth}. The results are listed in Table~\ref{tab:Tab8} and confirm the good description of the data in most of the distributions. A few higher \chisqndf values are present as systematic uncertainties are not included by construction. These are possibly related to imperfections in the fitting model or poor knowledge of the parameters of some resonance contributions~\cite{PDG2024}.

\begin{table} [!tb]
  \centering  \caption{\small\label{tab:Tab8} Results from the \chisqndf tests on various two-dimensional distributions for \mbox{\bkskkpip} and \bkskkpim data.}
  {\small
\begin{tabular}{lrr}
\hline
\Bu decay mode & Variables & \chisqndf \cr
\hline\\ [-2.3ex]
\bkskkpip & $(m(\kskpip),\cos\theta_{\pi})$ & $1328/1113=1.19$ \cr
 & $(m^2(\Km \pip),m^2(\KS \pip))$ & $920/809=1.14$ \cr
  & $(m_X(\Km \pip),\cos \theta_{K^0})$ & $1230/1102=1.12$ \cr
& $(m_X(\Kp \Km \pip),\cos\theta_{\Kp})$ & $994/964=1.03$ \cr
 & $(m_X(\KS \pip),\cos\theta_K)$ & $1316/1039=1.27$ \cr
\hline\\ [-2.3ex]  
\bkskkpim & $(m(\kskpim),\cos\theta_{\pi})$ & $1201/1185=1.01$ \cr
 & $(m^2(\Kp \pim),m^2(\KS \pim))$ & $1063/847=1.25$ \cr
  & $(m_X(\Kp \pim),\cos \theta_{K^0})$ & $1222/1142=1.07$ \cr
  & $(m_X(\Kp \Kp \pim),\cos\theta_{\Kp})$ & $1286/1064=1.21$ \cr
 & $(m_X(\KS \pip),\cos\theta_K)$ & $1117/1084=1.03$ \cr
\hline
\end{tabular}
}
\end{table}

\subsection{Partial-waves decomposition}

\begin{figure}[!tb]
\centering
\small
\includegraphics[width=0.45\textwidth]{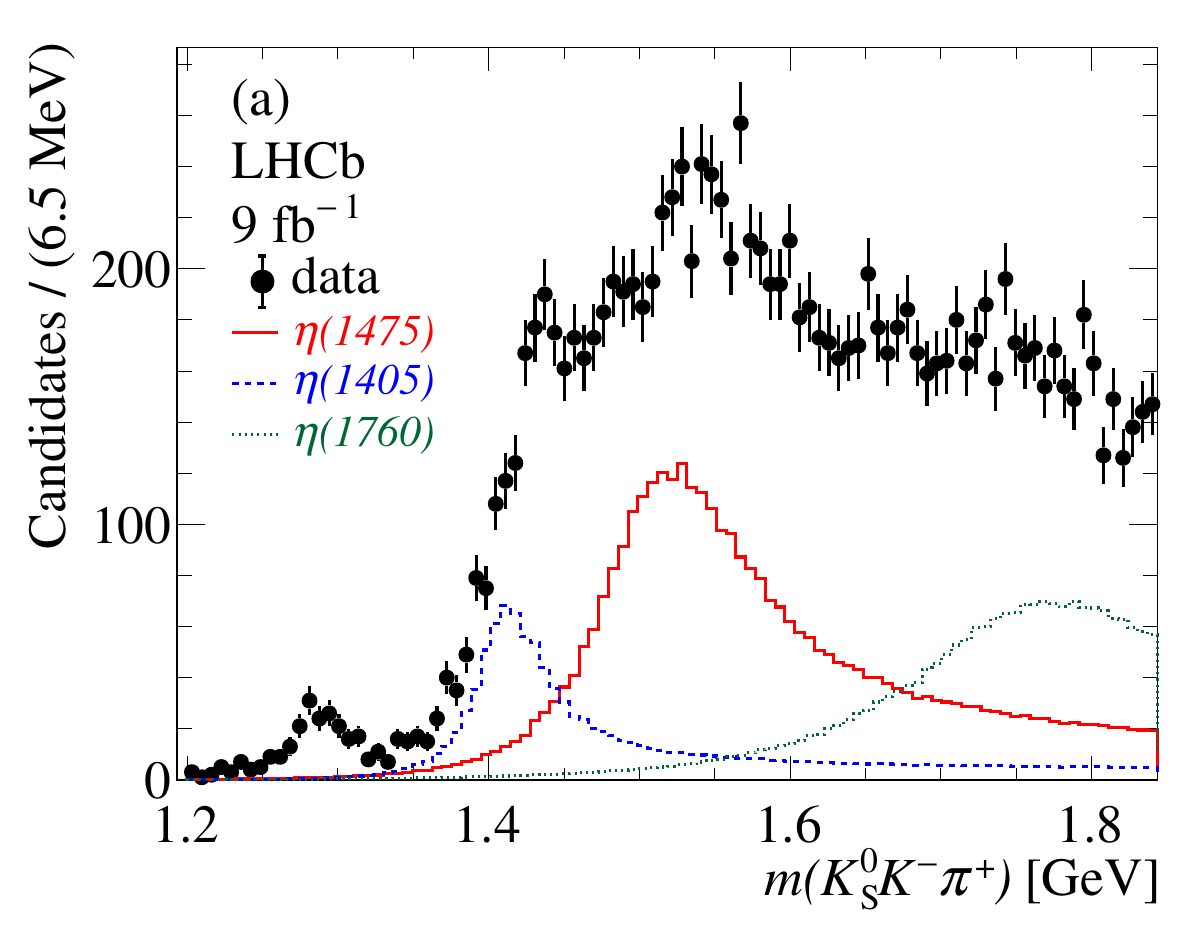}
\includegraphics[width=0.45\textwidth]{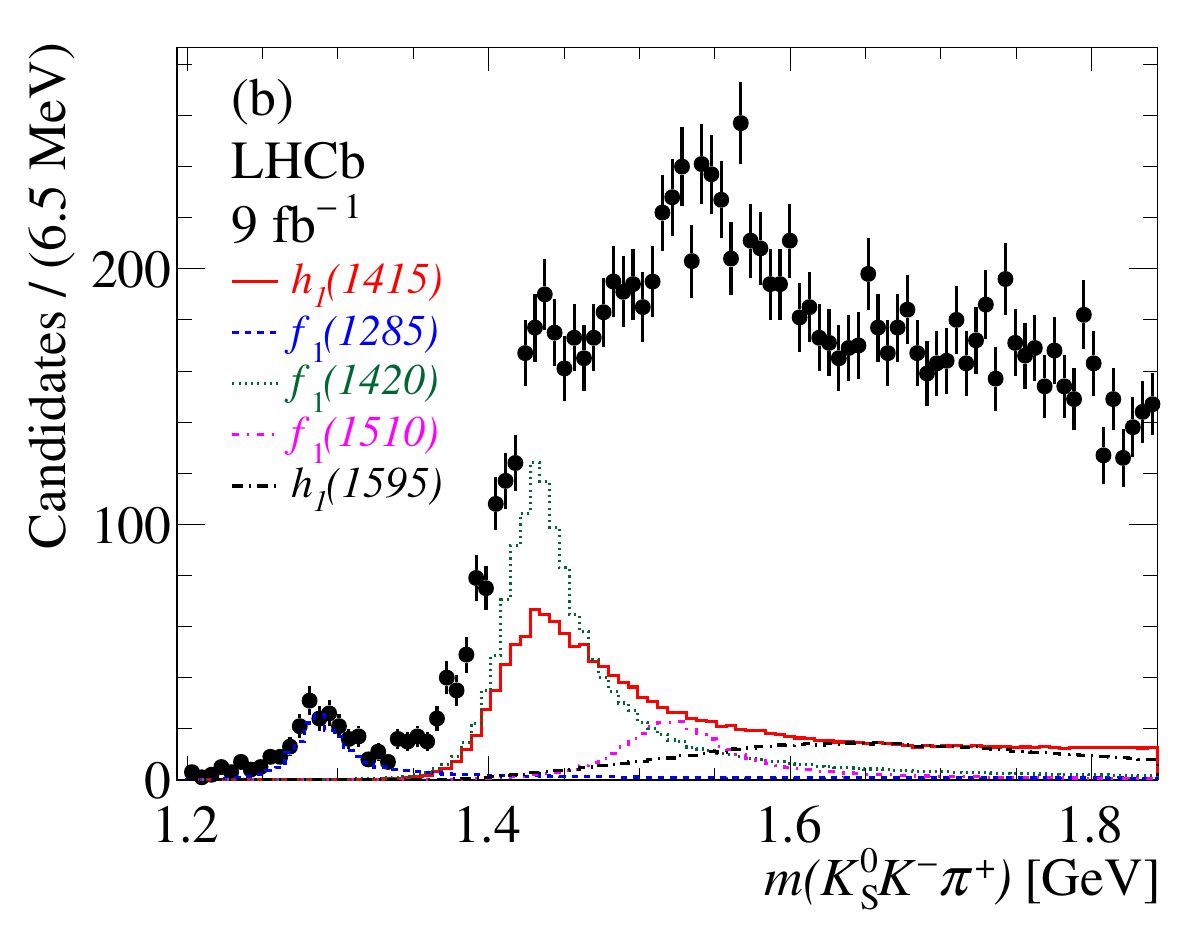}
\includegraphics[width=0.45\textwidth]{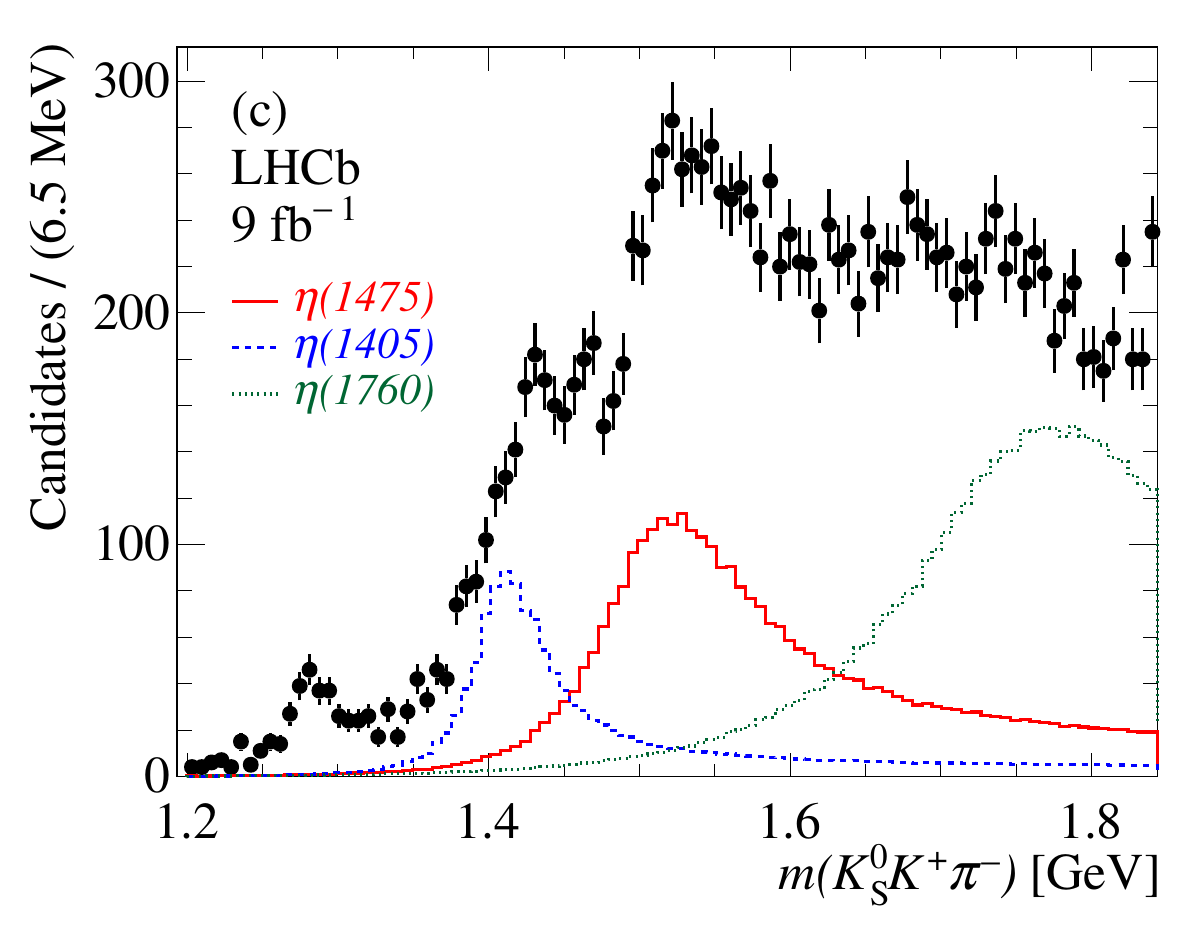}
\includegraphics[width=0.45\textwidth]{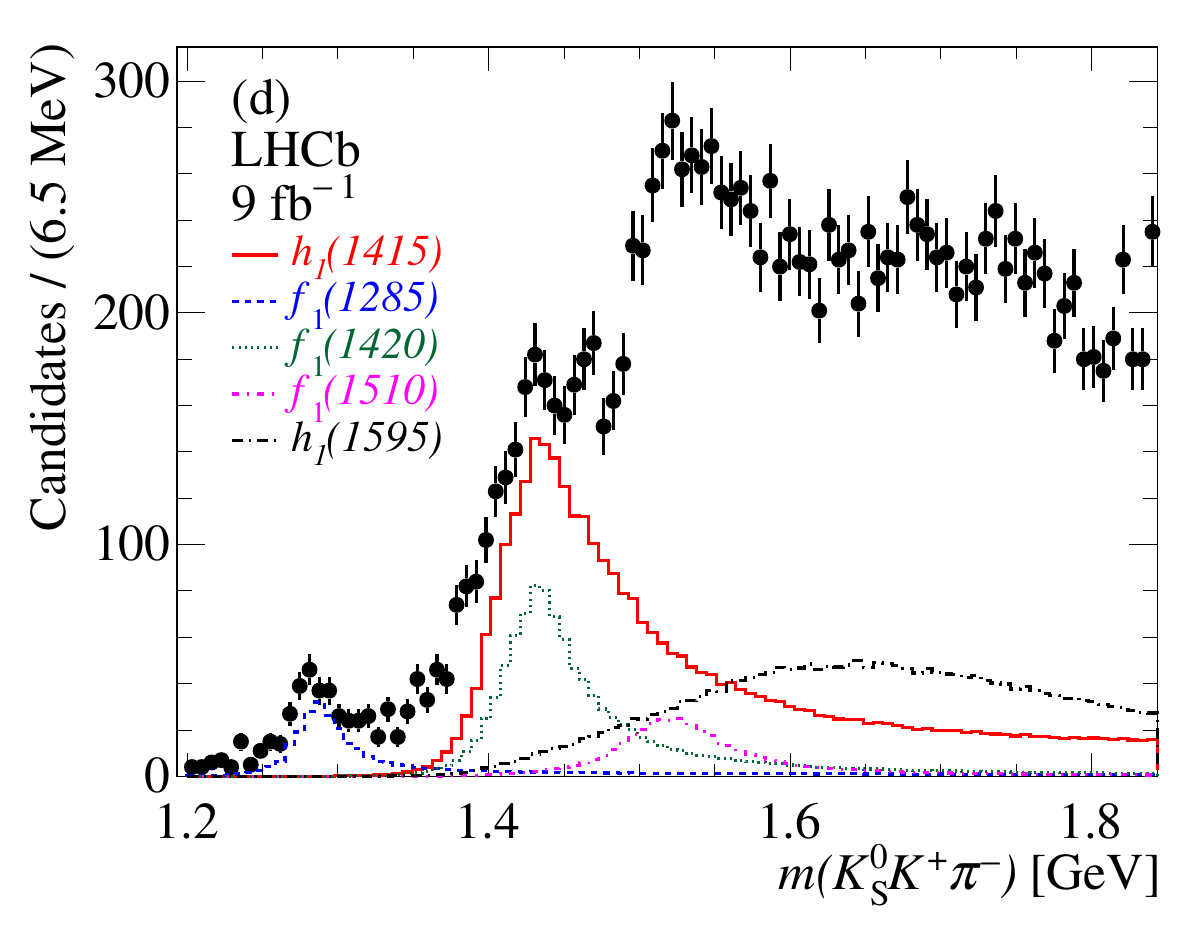}
\caption{\small\label{fig:Fig18} \kskpi invariant-mass distribution of (a)--(b) \bkskkpip and (c)--(d) \bkskkpim decays. 
}
\end{figure}

Figure~\ref{fig:Fig18} shows the \kskpi mass spectra for the two decay modes with contributions of the different resonances superimposed. Here, the squared moduli of the amplitudes are summed over all their partial decay modes.

The composition of the structure around 1.5\gev resulting from this analysis is rather complex. A superposition of $\eta(1405)$, $f_1(1420)$ and $\eta(1475)$ is present, with similar fractional composition to that obtained from radiative $\jpsi$ decays~\cite{Bai:1990hs,BESIII:2022chl}. In addition, significant contributions from $h_1(1415)$, $h_1(1595)$ and $f_1(1510)$ resonances are found. The ratios of fit fractions for the $f_1(1420)$ and $h_1(1415)$ resonances are found to be rather different, $0.86\pm0.08\pm0.20$ and $0.32\pm0.03\pm0.10$ for \bkskkpip and \bkskkpim decays, respectively (Fig.~\ref{fig:Fig18}(b) and (d)). 

Other experiments also studied the \kskpi system.
The $f_1(1510)$ resonance was first observed in $\Km p$ interactions by the ACNO collaboration~\cite{Gavillet:1982tv} where it is found that this state fits well the hypothesis of being the $s \bar s$ member of the $J^{PC}=1^{++}$ nonet. Furthermore, the $h_1(1415)$ was first observed in $\Km p$ interactions~\cite{Aston:1987ak} through a significant interference with the $f_1(1510)$ resonance.
On the other hand, the $f_1(1420)$ resonance dominates the \kskpi mass spectrum in central production~\cite{WA76:1986sfi} and a molecular assignment for this resonance has been proposed~\cite{Longacre:1990uc}.


\section{Systematic uncertainties and tests}
\label{sec:sys}

Several systematic uncertainties are evaluated and listed in Tables~\ref{tab:Tab16} and \ref{tab:Tab17} of Appendix~\ref{sec:app2} for \bkskkpip and \bkskkpim data, respectively.
In the baseline fits the radius $r$, which enters the Blatt--Weisskopf form factors~\cite{Blatt:1952ije} used by the relativistic BW function describing the $K^*(892)$ resonances, is fixed to $2.5\gev^{-1}$~\cite{LHCb-PAPER-2022-051}. This value is varied to 1.5 and $3.5\gev^{-1}$ and the average of the variations with respect to the baseline fit results is taken as systematic uncertainties (listed as $r$ in Tables~\ref{tab:Tab16} and \ref{tab:Tab17}).

The mass and width of the resonances included in this analysis are fixed to their known values, as listed in Table~\ref{tab:Tab5}.
Attempts to vary these parameters freely during the fits were unsuccessful due to
fit instabilities introduced by the large number of free parameters.
The uncertainty associated to this effect is evaluated by performing 100 alternative fits to data where, in each fit,
the mass and width of each resonance is randomly sampled from Gaussian functions based on their known values and uncertainties~\cite{PDG2024}, also listed in Table~\ref{tab:Tab5}. The root-mean-squares of the differences with respect to the baseline fit results are taken as systematic uncertainties (indicated as BW).

The $a_0(980)$ resonance is described by a coupled-channel Breit--Wigner described by Eq.~\ref{eq:a0} in Appendix~\ref{sec:app1}. The measured parameters describing the resonance are varied within their known statistical uncertainties, and the average of the absolute deviations from the baseline fit results are taken as systematic uncertainties (indicated as $a_0$).

The effect of the uncertainty on the efficiency model is evaluated by repeating the fit using a modified model where the last correction term in the functional expressions describing the efficiency in Eq.~\ref{eq:effy} is removed for both \KSLL and \KSDD datasets. The deviations of the fit results from the baseline values are small (indicated as eff).

The uncertainty due to the background model 
is evaluated by varying within uncertainties the purity of the \Bu signal, listed in Table~\ref{tab:Tab1}, and the fitted fractions of the $K^*$ contributions describing the background model and listed in Table~\ref{tab:Tab2}.
The averages of the absolute deviations of the fitted fractions and phases from the baseline fit results are assigned as systematic uncertainties (indicated as bkg).

The \kskpi mass spectrum has limited phase space near the threshold and a large number of simulation samples are performed to evaluate normalization integrals. To assess the impact on the results, the number of the simulated candidates is doubled and halved and the average of the resulting variations with respect the baseline fit results is taken as systematic uncertainty (indicated as int).

The effect of the trigger (trig) on the composition of the dataset is evaluated by first recomputing separate efficiencies for the two types of trigger samples (described in Sec.~\ref{sec:lhcb}) and separate \KSLL and \KSDD simulations.
The \kskkpip and \kskkpim mass spectra are also fitted separately to evaluate the purities in each subsample. Then, the likelihood function given by Eq.~\ref{eq:like} is modified in order to include separate contributions for each dataset. The increase of the likelihood value determines a small improvement in the description of the data.
The absolute deviations from the baseline fit results are included as systematic uncertainties.
The different sources of systematic uncertainties are added in quadrature assuming no correlations among them, with the total uncertainties listed in Tables~\ref{tab:Tab16} and~\ref{tab:Tab17} in Appendix~\ref{sec:app2}.

Further tests of the amplitude model are performed for which no 
sizable impact on the results is found.
Possible fit biases are evaluated by generating, from the baseline fit solution, large samples of pseudoexperiments having the same size as the datasets. These samples are then fitted with the same baseline model, with deviations on the fractions and relative phases from the baseline fit results evaluated. It is found that the average values of these deviations are found to be consistent with zero, and the root-mean-square values of the distributions agree well with the statistical uncertainties resulting from the fits to the data.

The effect of a small discrepancy between data and simulation in the calculation of the tracking efficiency is evaluated by applying corrections to the simulation based on high yield control samples. It is found that the kinematic variables used in the description of the data and simulation are not affected by these corrections.
Charge conjugation on \bkskkpip and \bkskkpim decays is tested by fitting separately \Bu and \Bub data. The differences in the fractions and phases are found consistent with zero within uncertainties.

In Sec.~\ref{sec:reso} (Eq.~\ref{eq:k1k2}) it is assumed that the \kskpi system is ``decoupled'' from the kaon ($K_4$) which is assumed to be a spectator.  This assumption is reasonable given the strong kinematic constraint of $m(\kskpi)<1.85\gev$, which removes all possible resonances involving particles from the \kskpi system and the spectator kaon. Therefore
two-body and three-body invariant-mass combinations involving the spectator kaon should behave just as phase space (see Appendix~\ref{sec:app3}). It is found that the fit reproduces all of the invariant-mass distributions reasonably well.

\section{Measurement of the branching fractions}
\label{sec:br}

Branching fractions for the process $\Bu \to R^0 \Kp$ channels are evaluated.
Table~\ref{tab:Tab9} reports the branching fractions measured in Ref.~\cite{LHCb-PAPER-2022-051} for \bkskkpip and \mbox{\bkskkpim} decays using as reference
the known $\Bu \to\etac\Kp$  and $\Bu \to\jpsi\Kp$  branching fractions~\cite{PDG2024}. 

\begin{table} [!tb]
  \centering
\caption{\small\label{tab:Tab9} Measured branching fractions \calB \ for \bkskkpip and \bkskkpim from Ref.~\cite{LHCb-PAPER-2022-051}. 
The first uncertainty is statistical, the second systematic and the third is due to the uncertainties on the known $\Bu \to \etac \Kp$ and $\Bu \to \jpsi \Kp$ branching fractions.
Inverse-variance-weighted averages of the two measurements are also reported.}  
  {\small
\begin{tabular}{llc}
  \hline
  Final state & Reference & \calB\ $(\times 10^{-5})$\cr
  \hline\\ [-2.3ex]
      \bkzkkpip & \etac & $32.3 \pm  0.3 \pm 2.0 \pm 7.2$ \cr
       & \jpsi & $34.0 \pm 0.7 \pm 0.9 \pm 3.1$  \cr
      \hline\\ [-2.3ex]
      & average & $32.6 \pm 0.3 \pm 0.8 \pm 2.9$ \cr
      \hline\\ [-2.3ex]
      \bkzkkpim & \etac & $26.6 \pm 0.3 \pm 0.7 \pm 5.9$ \cr
      & \jpsi &$28.0 \pm 0.7 \pm 1.4 \pm 2.6$  \cr
      \hline
      & average & $26.8 \pm 0.3 \pm 0.6 \pm 2.3$ \cr
      \hline
\end{tabular}
}
\end{table}

The amplitude analysis discussed in Sec.~\ref{sec:fits} evaluates the fraction of events $f_R$ for resonance $R^0$ in the $m(\kskpi)<1.85$\gev region.
Using this information, it is possible to compute the branching fraction for resonance $R^0$ as
\begin{equation}
  \calB(\Bu \to R^0 (\to K^0 \Kpm \pimp)\Kp) = f_R \cdot \calR \cdot \calB \cdot f_{K^0},
  \label{eq:calr}
\end{equation}
where  $f_{K^0}$ indicates the correction for the unseen $K^0$ decay modes. This correction takes into account $\calB(\Kz \to \pip \pim)=0.6920 \pm 0.0005$~\cite{PDG2024} and a factor two for the \KL component in the \Kz meson, yielding a total factor $f_{K^0} = 2.890\pm 0.001$.
The $\calR$ factor is computed using efficiency-corrected data as
\begin{equation}
  \calR = \frac{N(m(\kskpi)<1.85  \gev)}{N_{\rm tot}}, 
  \label{eq:ratio}
\end{equation}
where $N(m(\kskpi))$ indicates the event yield in the indicated mass region and $N_{\rm tot}$ the total \bkskkpi event yield.

To evaluate the above ratios, the total efficiency correction for the four-body decay \bkskkpip is used, as described in Ref.~\cite{LHCb-PAPER-2022-051}. The candidates are weighted by the inverse of the total efficiency, and the resulting \kskkpi  mass spectra are fitted to obtain the total \Bu yields. The same method is used to obtain the efficiency-corrected yield for \mbox{$m(\kskpi)<1.85\gev$}.
Tables~\ref{tab:Tab10} and~\ref{tab:Tab11} give information on the uncorrected and corrected \Bu yields.
Before evaluating the 
 \bkskkpip and \bkskkpim yields, background contributions from open-charm
$b$-hadron decays $\Bu \to D_{(s)}X$, where $X$ indicates one or two additional particles, must be subtracted.

To account for possible charm resonances in the background, the two- and three-body invariant-mass distributions for \Bu candidates in the mass sideband regions are subtracted from the corresponding distributions of the signal region applying an appropriate normalization. 
In the mass region \mbox{$m(\kskpi)<1.85\gev$}, a $\Dzb \to \Kp \pim$ contribution is removed from the \bkskkpim candidates, while no open-charm contribution is present for \bkskkpip decays.

\begin{table} [!tb]
  \centering
  \caption{\small\label{tab:Tab10}  (Left)~uncorrected  and (right)~efficiency-corrected yields of \Bu candidates and charm contributions in \bkskkpip decays. The second uncertainty in the corrected yields is systematic.}
  {\small
\begin{tabular}{l|cc}
\hline
Contribution & Yield & Corrected yield\cr
\hline\\[-2.3ex]
\bkskkpip & $95200 \pm 480$  & $173600 \pm 800 \pm 240$ \cr
\hline \\[-2.3ex]
$D^-_s \to \KS \Km$ & $ 1490\pm 70$ & $ 2590\pm 80$\cr
$\Dz \to \Kp \Km$ & \al$280 \pm 30$ & \al $580 \pm 50$\cr
$\Dz \to \KS \Km \Kp$ & $17130 \pm 130$ & $28070 \pm 170$ \cr
$\Dz \to \KS \Km \pip$ & \al$490 \pm 40$ & \al $950 \pm 50$\cr
\hline\\[-2.3ex]
Sum of charm &  $19390 \pm 160$ &  $32190 \pm 200$ \cr
Charm fraction & & $0.180 \pm 0.002 \pm 0.006$ \cr
Charmless \Bu decays & $78090 \pm 430$ & $142390 \pm 720 \pm 970$\cr
\hline\hline\\[-2.3ex]
\bkskkpip $m(\kskpi)<1.85\gev$& $11310\pm 150$ & $20870 \pm 260 \pm 370$ \cr
\hline
\end{tabular}
}
\end{table}

\begin{table} [!tb]
  \centering
  \caption{\small\label{tab:Tab11} (Left)~uncorrected  and (right)~efficiency-corrected yields of \Bu candidates and charm contribution in \bkskkpim decays. The second uncertainty in the corrected yields is systematic.}
  {\small
\begin{tabular}{l|cc}
\hline
Contribution & Yield & Corrected yield \cr
\hline\\[-2.3ex]
\bkskkpim  & $68020\pm 120$ & $131130\pm 680 \pm 1270$ \cr
\hline\\[-2.3ex]
$D^+_s \to \KS \Kp$ & $ 1090\pm 60$ & $ 2060\pm 80$\cr
$\Dzb \to \Kp \pim$ & $4460 \pm 90$ & \al$8340\pm   120$\cr
$\Dzb \to \KS \Kp \pim$ & \al$820\pm  40$ & $1660\pm 60$ \cr
\hline\\[-2.3ex]
Sum of charm & $6370 \pm 110$ & $12050 \pm 150$ \cr
Charm fraction & & $0.090 \pm 0.002 \pm 0.002$ \cr
Charmless \Bu decays & $61650 \pm 160$ & $118520 \pm 650 \pm 1170$ \cr
\hline \hline\\[-2.3ex]
\bkskkpim $m(\kskpi)<1.85\gev$ & $12600 \pm 260$ & $24770\pm 280 \pm 20$  \cr
\hline\\ [-2.3ex]
$\Dzb \to \Kp \pim$ &\al $180\pm20$ & $410\pm30$ \cr
\hline\\ [-2.3ex]
Charm fraction &  & $0.014 \pm 0.002 \pm 0.002$ \cr
Charmless \Bu decays & $12420 \pm 250$ & $24410 \pm 280 \pm 60$ \cr
\hline
\end{tabular}
}
\end{table}
The resulting ratios are $\calR = 0.147 \pm 0.002 \pm 0.003$ for \bkskkpip and
$\calR = 0.206 \pm 0.003 \pm 0.002$ for \bkskkpim data.
Systematic uncertainties are due to variations of the fit models used for subtraction of the open-charm contributions~\cite{LHCb-PAPER-2022-051}.
The systematic uncertainty on the charm fraction is obtained by comparing results with and without efficiency corrections.

The fractional contributions resulting from the amplitude analysis given in Table~\ref{tab:Tab6} for \bkskkpip decays are converted to branching fractions, listed in Table~\ref{tab:Tab12}. This is achieved by multiplying by the total scaling factor \mbox{$s = \calR\cdot\calB\cdot f_{K^0}=(13.84 \pm 0.23 \pm 0.45 \pm 1.19) \times 10^{-5}$}, where \calB\ indicates the \mbox{\bkskkpip} branching fraction and the third uncertainty is due to the branching fraction \calB\ listed in Table~\ref{tab:Tab9}.
\begin{table} [ht]
  \centering
  \caption{\small\label{tab:Tab12} Measured branching fractions for $\Bu \to R^0 \Kp$ from \bkzkkpip decays.}
  {\small
\begin{tabular}{lr}
\hline
Contribution & $ \calB(\Bu \to R^0\Kp)\times 10^{-5}$\cr
\hline\\ [-2.3ex]
 $\Bu \to \eta(1475) \Kp \to (K^*\Kb) \Kp  $ &   $  1.49\pm  0.15\pm  0.16\pm  0.13$\cr
 $\Bu \to \eta(1475) \Kp \to (a_0(980)^- \pip) \Kp $ &  $  0.19\pm  0.05\pm  0.05\pm  0.02$\cr
 $\Bu \to \eta(1475) \Kp \to (\Kz \Km \pip) \Kp$ &   $  2.09\pm  0.29\pm  0.31\pm  0.18$\cr
\hline\\ [-2.3ex]
 $\Bu \to \eta(1760) \Kp \to (K^*\Kb) \Kp $ & $  0.27\pm  0.05\pm  0.05\pm  0.02$\cr  
 $\Bu \to \eta(1760) \Kp \to (a_0(980)^- \pip) \Kp $ &  $  0.28\pm  0.05\pm  0.04\pm  0.02$ \cr
 $\Bu \to \eta(1760) \Kp \to (\Kz \Km \pip) \Kp$ &  $  1.64\pm  0.25\pm  0.37\pm  0.14$ \cr
\hline\\ [-2.3ex]
 $\Bu \to \eta(1405) \Kp \to (K^*\Kb) \Kp $ &  $ 0.48\pm  0.08\pm  0.26\pm  0.04$ \cr
 $\Bu \to \eta(1405) \Kp \to (\Kz \Km \pip) \Kp$ &  $ 0.72\pm  0.08\pm  0.12\pm  0.06$\cr
\hline\\ [-2.3ex]
 $\Bu \to f_1(1285) \Kp \to (a_0(980)^- \pip) \Kp $ &   $0.27\pm  0.03\pm  0.02\pm  0.02$\cr
\hline\\ [-2.3ex]
 $\Bu \to f_1(1420) \Kp \to (K^*\Kb) \Kp  $ &  $ 1.58\pm  0.10\pm  0.30\pm  0.14$\cr
\hline\\ [-2.3ex]
 $\Bu \to f_1(1510) \Kp \to (K^*\Kb) \Kp  $ & $0.40\pm  0.05\pm  0.17\pm  0.03$\cr
\hline\\ [-2.3ex]
 $\Bu \to h_1(1415) \Kp \to (K^*\Kb) \Kp  $ & $   1.85\pm  0.14\pm  0.28\pm  0.13$\cr
\hline\\ [-2.3ex]
 $\Bu \to h_1(1595) \Kp \to (K^*\Kb) \Kp $ &  $  0.74\pm  0.12\pm  0.19\pm  0.06$\cr
\hline\\ [-2.3ex]
 $\Bu \to \eta_2(1645) \Kp \to (K^*\Kb) \Kp  $ &  $ 0.22\pm  0.03\pm  0.11\pm  0.02$ \cr
\hline
\end{tabular}
}
\end{table}

Similarly, the fractional contributions given in  Table~\ref{tab:Tab6} for \bkskkpim are converted to branching fractions, listed in Table~\ref{tab:Tab13} by multiplying by the total scaling factor $s= \calR\cdot\calB\cdot f_{K^0}=(15.96\pm0.29\pm0.43\pm 1.39)\times 10^{-5}$. 
\begin{table} [ht]
  \centering  \caption{\small\label{tab:Tab13} Measured branching fractions for $\Bu \to R^0 \Kp$ from \bkzkkpim decays.
  }
  {\small
\begin{tabular}{lr}
\hline
Contribution & $ \calB(\Bu \to R^0\Kp)\times 10^{-5}$\cr
\hline\\ [-2.3ex]
 $\Bu \to \eta(1475) \Kp \to (K^*\Kb) \Kp  $ & $   1.42\pm  0.15\pm  0.20\pm  0.12$ \cr
 $\Bu \to \eta(1475) \Kp \to (a_0(980)^- \pip) \Kp $ &  $  0.25\pm  0.06\pm  0.05\pm  0.02$\cr
 $\Bu \to \eta(1475) \Kp \to (\Kzb \Kp \pim) \Kp$ &  $ 1.23\pm  0.20\pm  0.38\pm  0.11$ \cr
 \hline\\ [-2.3ex]
 $\Bu \to \eta(1760) \Kp \to (K^*\Kb) \Kp  $ & $0.42\pm  0.06\pm  0.07\pm  0.04$\cr
 $\Bu \to \eta(1760) \Kp \to (a_0(980)^- \pip) \Kp $ &  $  0.24\pm  0.06\pm  0.04\pm  0.02$\cr
 $\Bu \to \eta(1760) \Kp \to (\Kzb \Kp \pim) \Kp$ &  $ 3.21\pm  0.31\pm  0.71\pm  0.28$\cr
 \hline\\ [-2.3ex]
 $\Bu \to \eta(1405) \Kp \to (K^*\Kb) \Kp $ & $ 0.32\pm  0.07\pm  0.09\pm  0.03$\cr
 $\Bu \to \eta(1405) \Kp \to (\Kzb\Kp \pim) \Kp$ & $  0.89\pm  0.08\pm  0.13\pm  0.08$\cr
\hline\\ [-2.3ex]
 $\Bu \to f_1(1285) \Kp \to (a_0(980)^- \pip) \Kp $ & $  0.28\pm  0.03\pm  0.03\pm  0.02$\cr
\hline\\ [-2.3ex]
 $\Bu \to f_1(1420) \Kp \to (K^*\Kb) \Kp  $ & $  0.92\pm  0.07\pm  0.24\pm  0.08$\cr
\hline\\ [-2.3ex]
 $\Bu \to f_1(1510) \Kp \to (K^*\Kb) \Kp $ & $  0.36\pm  0.05\pm  0.38\pm  0.03$ \cr
\hline\\ [-2.3ex]
 $\Bu \to h_1(1415) \Kp \to (K^*\Kb) \Kp  $ & $ 2.91\pm  0.17\pm  0.50\pm  0.22$\cr
\hline\\ [-2.3ex]
 $\Bu \to h_1(1595) \Kp \to (K^*\Kb) \Kp  $ &  $  2.05\pm  0.20\pm  0.39\pm  0.18$ \cr
\hline\\ [-2.3ex]
 $\Bu \to \eta_2(1645) \Kp \to (K^*\Kb) \Kp  $ &  $ 0.11\pm  0.02\pm  0.03\pm  0.01 $\cr
\hline
\end{tabular}
}
\end{table}

As noted in Sec.~\ref{sec:fit_bw} the resulting fractions are consistent among the two channels when including the systematic uncertainties. 
Assuming compatibility between the two sets of measurements, it is possible to evaluate the inverse-variance-averages of the two sets of measurements, as given in Table~\ref{tab:Tab14}.

\begin{table} [ht]
  \caption{\small\label{tab:Tab14} Inverse-variance-averages of the branching fraction measurements for $\Bu \to R^0 \Kp$ from \bkzkkpip and \bkzkkpim data.}  
  \centering
\begin{tabular}{lr}
\hline\\ [-2.3ex]
Contribution & $\calB(\Bu \to R^0\Kp) \times 10^{-5}$\cr
\hline\\ [-2.3ex]
$\Bu \to \eta(1475) \Kp \to (K^*\Kb) \Kp  $ & $1.45\pm  0.11\pm  0.12\pm  0.09$\cr
$\Bu \to \eta(1475) \Kp \to (a_0(980) \pi) \Kp $ & $ 0.22\pm  0.04\pm  0.03\pm  0.01$ \cr 
$\Bu \to \eta(1475) \Kp \to (\Kz K \pi) \Kp$ & $1.51\pm  0.16\pm  0.24\pm  0.09$ \cr
\hline\\ [-2.3ex]
 $\Bu \to \eta(1760) \Kp \to (K^*\Kb) \Kp  $  & $ 0.34\pm  0.04\pm  0.04\pm  0.02$ \cr
$\Bu \to \eta(1760) \Kp \to (a_0(980) \pi) \Kp $ & $0.26\pm  0.04\pm  0.02\pm  0.02$ \cr
$\Bu \to \eta(1760) \Kp \to (\Kz K \pi) \Kp$ & $2.21\pm  0.20\pm  0.33\pm  0.13$ \cr
\hline\\ [-2.3ex]
$\Bu \to \eta(1405) \Kp \to (K^*\Kb) \Kp $ & $0.38\pm  0.05\pm  0.09\pm  0.02$ \cr
$\Bu \to \eta(1405) \Kp \to (\Kz K \pi) \Kp$ & $  0.80\pm  0.05\pm  0.09\pm  0.05$ \cr
\hline\\ [-2.3ex]
$\Bu \to f_1(1285) \Kp \to (a_0(980) \pi) \Kp $ & $  0.28\pm  0.02\pm  0.02\pm  0.02$ \cr
\hline\\ [-2.3ex]
 $\Bu \to f_1(1420) \Kp \to (K^*\Kb) \Kp  $ & $1.14\pm  0.06\pm  0.19\pm  0.07$ \cr
\hline\\ [-2.3ex]
 $\Bu \to f_1(1510) \Kp \to (K^*\Kb) \Kp $ & $0.38\pm  0.03\pm  0.15\pm  0.02$ \cr
\hline\\ [-2.3ex]
 $\Bu \to h_1(1415) \Kp \to (K^*\Kb) \Kp  $ & $2.22 \pm 0.10 \pm 0.23 \pm 0.10$ \cr
\hline\\ [-2.3ex]
$\Bu \to h_1(1595) \Kp \to (K^*\Kb) \Kp  $ & $  1.04\pm  0.10\pm  0.12\pm  0.06 $\cr
\hline\\ [-2.3ex]
$\Bu \to \eta_2(1645) \Kp \to (K^*\Kb) \Kp  $ & $0.15\pm  0.02\pm  0.02\pm  0.01$ \cr
\hline
\end{tabular}
\end{table}

\section{Summary}
\label{sec:summary}

A study is presented of \bkskkpip and \bkskkpim decays at proton-proton collision energies of 7, 8 and $13\tev$ using the \lhcb detector with an integrated luminosity of $9\invfb$. The \kskpi invariant-mass spectra, in the \mbox{$m(\kskpi)< 1.85\gev$} mass region, show a large activity which is resolved using an amplitude analysis.
 A simple model, were $J^{PC}$ amplitudes are described by multiple Breit--Wigner functions with appropriate angular distributions provides a good description of the experimental data.

It is found that the \kskpi mass spectrum is dominated by $J^{PC}=0^{-+}$, $1^{++}$ and $1^{-+}$ amplitudes.
Strong interference is observed between the
$J^{PC}=1^{++}$ and $J^{PC}=1^{-+}$ contributions with different patterns in the two decay modes. 
Associating the measured amplitudes to known resonances~\cite{PDG2024}, these interference patterns are similar to those observed in $\Km p$ interactions~\cite{Aston:1987ak} between the $h_1(1415)$ and $f_1(1510)$ resonances. In the present analysis, the contribution from the $f_1(1510)$ is found to be small, with the interference instead generated between the $h_1(1415)$ and $f_1(1420)$ resonances. The presence of two $I=0$, 
$J^{PC}=1^{++}$ resonances that are close in mass, namely the $f_1(1420)$ and $f_1(1510)$, complicates the interpretation as both states compete to be interpreted as the $s \bar s$ member of the $J^{PC}=1^{++}$ nonet~\cite{Aston:1987ak}.
The relative phases of the  $J^{PC}=1^{-+}$ resonances 
are found to be different in the two final states as a result of the inversion of the population of neutral and charged $K^*$ in the \kskpi Dalitz plot for \bkskkpip and \bkskkpim decays.

A strong $J^{PC}=0^{-+}$ contribution is found, composed of $\eta(1405)$, $\eta(1475)$ and $\eta(1760)$ resonances.  As these states are also observed in $J/\psi$ radiative decays, this suggests
that their production mechanisms are possibly similar to those acting in \Bp decays.

The understanding of the $I=0$, $J^{PC}=0^{-+}$ nonet is still incomplete, with each of the three $\eta$ resonances being proposed as candidates for the pseudoscalar glueball~\cite{Li:2021gsx} (see the review on the spectroscopy of light-meson resonances in Ref.~\cite{PDG2024}).

In this paper, the first measurements of branching fractions for exclusive $\Bu \to R^0 \Kp$ decays are reported, where $R^0$ is an $I=0$ resonance having $u \bar u$, $s \bar s$ or possibly $gg$ content. 
 These results provide new information on light-meson spectroscopy in the low-mass region, offering insights into gluonium physics, a fundamental aspect of QCD.
 New experimental inputs may arise in the near future from current experiments studying charmonium decays or central production, or from future proton-antiproton colliders.

\clearpage

\section*{Acknowledgements}
%
%
\noindent We express our gratitude to our colleagues in the CERN
accelerator departments for the excellent performance of the LHC. We
thank the technical and administrative staff at the LHCb
institutes.
We acknowledge support from CERN and from the national agencies:
CAPES, CNPq, FAPERJ and FINEP (Brazil); 
MOST and NSFC (China); 
CNRS/IN2P3 (France); 
BMBF, DFG and MPG (Germany); 
INFN (Italy); 
NWO (Netherlands); 
MNiSW and NCN (Poland); 
MCID/IFA (Romania); 
MICIU and AEI (Spain);
SNSF and SER (Switzerland); 
NASU (Ukraine); 
STFC (United Kingdom); 
DOE NP and NSF (USA).
We acknowledge the computing resources that are provided by CERN, IN2P3
(France), KIT and DESY (Germany), INFN (Italy), SURF (Netherlands),
PIC (Spain), GridPP (United Kingdom), 
CSCS (Switzerland), IFIN-HH (Romania), CBPF (Brazil),
and Polish WLCG (Poland).
We are indebted to the communities behind the multiple open-source
software packages on which we depend.
Individual groups or members have received support from
ARC and ARDC (Australia);
Key Research Program of Frontier Sciences of CAS, CAS PIFI, CAS CCEPP, 
Fundamental Research Funds for the Central Universities, 
and Sci. \& Tech. Program of Guangzhou (China);
Minciencias (Colombia);
EPLANET, Marie Sk\l{}odowska-Curie Actions, ERC and NextGenerationEU (European Union);
A*MIDEX, ANR, IPhU and Labex P2IO, and R\'{e}gion Auvergne-Rh\^{o}ne-Alpes (France);
AvH Foundation (Germany);
ICSC (Italy); 
Severo Ochoa and Mar\'ia de Maeztu Units of Excellence, GVA, XuntaGal, GENCAT, InTalent-Inditex and Prog. ~Atracci\'on Talento CM (Spain);
SRC (Sweden);
the Leverhulme Trust, the Royal Society
 and UKRI (United Kingdom).

\section*{Data availability statement}

Data associated to the plots in this publication as well as in supplementary materials are made available on the CERN document server in~Ref.\cite{data}.

\clearpage

\appendix

\section{Appendix A}
\label{sec:app1}

An amplitude analysis is performed on candidates in the \Bu signal region (see Sec.~\ref{sec:evsel}) and $m(\kskpi)<1.85$ \gev to describe the two \Bu decays and to obtain information on the resonances contributing to the \kskpi final state. Amplitudes are modeled by the nonrelativistic Zemach-tensor formalism~\cite{Zemach:1963bc,Dionisi:1980hi,Filippini:1995yc}.
The particles involved in the decay are labeled as
\begin{equation}
\nonumber
  B \to \ \pi_1 K_2 K^0_3 K_4,
\end{equation}
and the decay is assumed to proceed as
\begin{equation}
\nonumber
  B \to R^0 K_4,
\end{equation}
where $R^0$ indicates an intermediate resonance which decays as
\begin{equation}
  R^0 \to \pi_1 K_2 K^0_3,
  \label{eq:part}
\end{equation}
and $K_4$ is assumed to be a spectator.
An isobar model is assumed for the decay of the resonance as $R^0 \to X c$, where $X$ indicates a two-body $X \to a b$ resonance. 
While the $B \to R^0 K_4$ process is a weak decay and conserves only angular momentum, the $R^0 \to \pi_1 K_2 K^0_3$ transition is a strong decay and conserves angular momentum and parity.
In the decay $R^0 \to \pi_1 K_2 K^0_3$, $p_i$ ($i=1,2,3$) indicate the momenta of the three particles in the rest frame of the $\pi_1 K_2 K^0_3$ system.
The amplitudes are defined as follows.
\begin{itemize}
\item{}
Symmetric and traceless tensors of rank $L$ constructed with $p_i$ are used to describe orbital angular momenta $L$ between the resonance $X$ and $c$.
\item{}
Symmetric and traceless tensors of rank $S$ constructed with  $t_i$ are used to describe the spin of the intermediate resonance $X$.
For a resonance $X$, decaying to the $a$ and $b$ final states, having 3-momenta $p_j$, $p_k$ with masses $m_j$ and $m_k$, the $t_i$ are defined as
\begin{equation}
t_i = p_j - p_k - (p_j + p_k)\frac{m_j^2-m_k^2}{m_{jk}^2},
\end{equation}
with cyclic $i,j,k$ indices.
\item{}
  The tensors are then combined into a tensor $T_J$ of rank $J$ to obtain the spin $J$ of the $\pi_1 K_2 K^0_3$ system.
\item{}
  To describe the decay $B \to R^0 K_4$, $q_4$ indicates the momentum of the kaon $K_4$ in the $B$ rest frame.
  A symmetric and traceless tensor $Q_J$ of rank $J$ made with $q_4$ is used to describe the orbital angular momentum between $R^0$ and $K_4$.
\item{}
  Finally, the two $T_J$ and $Q_J$ tensors are contracted to a scalar to obtain the spin of the $B$ meson.
\end{itemize}

Resonances that can contribute to the decay of the \kskpi system for \mbox{$m(\kskpi)<1.85 \gev$}  are: $K^*(892)^+$, $K^*(892)^0$, and $a_0(980)$.
  The $K^*(892)$ resonance is described by a complex spin-1 relativistic BW function with standard Blatt--Weisskopf~\cite{Blatt:1952ije} form factors and a radius $r$ fixed to $2.5\gev^{-1}$~\cite{LHCb-PAPER-2022-051}. The symbol $K^*(ij)$ in Table~\ref{tab:Tab15} indicates the $K^*(892)$ BW function formed with the particles combination $ij$ listed in Eq.~\ref{eq:part}.
  The $a_0(980)$ resonance is described by a complex coupled-channel Breit--Wigner function
  \begin{equation}
    a_0(m) = \frac{g_2}{m_0^2 - m^2 - i(\rho_1g_1^2 + \rho_2g_2^2)},
    \label{eq:a0}
  \end{equation}
  where $\rho_1$ and $\rho_2$, and $g_1$ and $g_2$ are phase-space factors and couplings, respectively, to the $\eta \pi$~(1) and $K \Kb$~(2) final states.
  The $a_0(980)$ parameters are fixed to the values \mbox{$m_0=999 \pm 2 \mev$}, $g_1=324 \pm 15\mev$, and $g^2_2/g^2_1=1.03 \pm 0.14$~\cite{Abele:1998qd} and the corresponding BW function is indicated with $a_0(23)$ in Table~\ref{tab:Tab15}.
  
 For isospin $I=0$ resonances, the interference between neutral $K^*(12)$ and charged $K^*(13)$ is fixed by $G$-parity, being constructive for $G=+1$ and destructive for $G=-1$. 
 Note, however, that as $G$-parity is related to $C$-parity by $G = C(-1)^I$, an additional minus sign is added to the above definitions of the three-momenta to ensure correct transformation under $C$-parity instead, such that constructive interference is 
 obtained with $G=-1$ and vice versa.
 
\begin{table}      
\caption{\small\label{tab:Tab15} Amplitudes used in the Dalitz plot analysis of $B \to (R^0 \to \pi_1 K_2 K^0_3) K_4$ decays, where $J^P$ indicate the $R^0$ spin-parity, and $L$ the angular momentum between the subresonance ($K^*$ or $a_0$) and the other final-state particle. Bold symbols represent three-vectors.}
  \centering
\resizebox{1.0\columnwidth}{!}{
\begin{tabular}{l|c|l}
  \hline\\ [-2.3ex]
   $J^P$  & L & Amplitude ($W$)\cr
      \hline\\ [-2.3ex]
      $0^-[S]$ & 0 & $a_0(23)$ \cr
      \hline\\ [-2.3ex]
        $0^-[P]$  & 1 & $K^*(12)({\bf t_3 \cdot p_3}) + GK^*(13)({\bf t_2 \cdot p_2})$ \cr
         \hline\\ [-2.3ex]
           $1^+[S]$ & 0 & $[K^*(12){\bf t_3} + GK^*(13){\bf t_2}] \cdot {\bf q_4}$ \cr
              \hline\\ [-2.3ex]
      $1^+[P]$ & 1 & $a_0(23)({\bf p_1 \cdot q_4})$ \cr
      \hline\\ [-2.3ex]
            $1^+[D]$ & 2 & $[K^*(12)\{{\bf p_3(t_3 \cdot p_3) - \frac{1}{3}(p_3 \cdot p_3)t_3}\} + G K^*(13)\{{\bf p_2(t_2 \cdot p_2) - \frac{1}{3}(p_2 \cdot p_2)t_2\}]\cdot q_4}$ \cr
           \hline\\ [-2.3ex]
       $1^-$ & 1 & $[K^*(12){\bf (t_3 \times p_3)} + G K^*(13){\bf (t_2 \times p_2)]\cdot q_4}$ \cr
        \hline\\ [-2.3ex]
$2^-[P]$ & 1 &  $[K^*(12)\{\frac{1}{2}(t_3^i\cdot p_3^j + t_3^j\cdot p_3^i)-\frac{1}{3}({\bf t_3\cdot p_3}) \delta^{ij}\}$ \cr
       & &   $+ G K^*(13)\{\frac{1}{2}(t_2^i\cdot p_2^j + t_2^j\cdot p_2^i)-\frac{1}{3}({\bf t_2\cdot p_2}) \delta^{ij}\}] \cdot [q^i_4q^j_4 - \frac{1}{3}|{\bf q_4}|^2\delta^{ij}]$\cr
             \hline\\ [-2.3ex]
        $2^-[D]$ & 2 & $a_0(23)(p_1^ip_1^j - \frac{1}{3}(|{\bf p_1}|^2)\cdot(q^i_4q^j_4 - \frac{1}{3}|{\bf q_4}|^2\delta^{ij})$ \cr
            \hline  \\ [-2.3ex]         
$2^-[F]$ & 3 &  [$K^*(12)\{({\bf t_3} \cdot {\bf p_3}) (p^i_3p^j_3 
            - \frac{1}{3}|{\bf p_3}|^2\delta^{ij})\}$ \cr 
       &   &   $+ GK^*(13)\{({\bf t_2} \cdot {\bf p_2}) (p^i_2p^j_2 - \frac{1}{3}|{\bf p_2}|^2\delta^{ij})\}]\cdot[q^i_4q^j_4 - \frac{1}{3}|{\bf q_4}|^2\delta^{ij}]$\cr
          \hline\\ [-2.3ex]
$2^+$ & 2 &  $[K^*(12)\{\frac{1}{2}({\bf t_3}\times {\bf p_3})^ip^j_3+
            p^i_3({\bf t_3}\times {\bf p_3})^j\} - \frac{1}{3}\{{\bf
               (t_3} \times {\bf p_3}) \cdot {\bf p_3}\}\delta^{ij} +$ \cr
      &   &   $G K^*(13)\{\frac{1}{2}[{\bf t_2}\times {\bf p_2})^ip^j_2+
            p^i_2({\bf t_2}\times {\bf p_2})^j\} - \frac{1}{3}\{{\bf
               (t_2} \times {\bf p_2}) \cdot {\bf p_2}\}\delta^{ij}]\cdot[q^i_4q^j_4 - \frac{1}{3}|{\bf q_4}|^2\delta^{ij}]$\cr
               \hline
\end{tabular}}
\end{table}

Table~\ref{tab:Tab15} lists the amplitudes used in this analysis.
Asymmetries between the charged and neutral $K^*$ can be generated by interference between $I=0$ amplitudes having opposite $C$-parities or by interfering $I=0$ and $I=1$ amplitudes. In this analysis, no evidence is found of contributing $I=1$ amplitudes.

 In the $B \to R^0 K_4$ decay, $R^0$ resonances are described with the BW function given in Eq.~\ref{eq:bw} 
  multiplied by the appropriate spin-parity terms $W(\vec x)$ described in Table~\ref{tab:Tab15}.
\section{Appendix B}
\label{sec:app2}

Systematic uncertainties on the measured fit fractions and phases are listed in Tables~\ref{tab:Tab16} and \ref{tab:Tab17}.

\begin{table} [ht]
  \centering
  \caption{\small\label{tab:Tab16} Absolute systematic uncertainties on fractions and phases for the \bkskkpip decay.}
  \resizebox{1.0\columnwidth}{!}{
\begin{tabular}{lc|rrrrrrr|r}
  \hline\\ [-2.3ex]
    & & \multicolumn{8}{c}{Fractions [\%] } \cr
  Contribution & Decay mode & $r$ & BW & $a_0$ & eff & bkg & int & trig & Total \cr
\hline \\[-2.3ex]   
$\eta(1475)$& $K^*\Kb$ & 0.07 & 1.04 & 0.05 & 0.07 & 0.02 & 0.28 & 0.02 & 1.08 \cr 
& $a_0 \pi$ & 0.10 & 0.30 & 0.08 & 0.04 & 0.04 & 0.02 & 0.17 & 0.37 \cr
& $PS$ & 0.26 & 1.99 & 0.30 & 0.57 & 0.08 & 0.16 & 0.22 & 2.13 \cr
\hline\\[-2.3ex]
$\eta(1760)$ & $K^*\Kb$ & 0.02 & 0.31 & 0.01 & 0.01 & 0.02 & 0.06 & 0.01 & 0.32 \cr
 & $a_0 \pi$& 0.08 & 0.19 & 0.05 & 0.04 & 0.03 & 0.01 & 0.21 & 0.30 \cr 
 & $PS$ & 0.42 & 2.60 & 0.09 & 0.05 & 0.11 & 0.26 & 0.05 & 2.65 \cr
\hline\\[-2.3ex]
$\eta(1405)$ & $K^*\Kb$ & 0.03 & 1.86 & 0.04 & 0.14 & 0.02 & 0.08 & 0.03 & 1.87 \cr
 & $PS$ & 0.03 & 0.80 & $<0.01$ & 0.09 & 0.02 & 0.02 & 0.06 & 0.81 \cr
\hline\\[-2.3ex]
$f_1(1285)$ & $a_0 \pi$ & $<0.01$ & 0.14 & 0.01 & 0.04 & $<0.01$ & 0.04 & 0.03 & 0.15 \cr
$f_1(1420)$ & $K^*\Kb$& 0.06 & 2.10 & 0.03 & 0.23 & 0.02 & 0.13 & 0.32 & 2.14 \cr 
\hline\\[-2.3ex]
$h_1(1415)$ & $K^*\Kb [S]$& 0.06 & 1.86 & 0.01 & 0.16 & 0.02 & 0.19 & 0.56 & 1.96 \cr
& $K^*\Kb [D]$& 0.04 & 0.11 & $<0.01$ & 0.11 & $<0.01$ & 0.09 & 0.08 & 0.20 \cr
\hline\\[-2.3ex]
$f_1(1510)$ & $K^*\Kb$ & $<0.01$ & 1.18 & 0.01 & 0.10 & 0.01 & 0.06 & 0.03 & 1.19 \cr
$h_1(1595)$ & $K^*\Kb [S]$ & 0.09 & 0.82 & 0.02 & 0.19 & 0.23 & 0.19 & 1.02 & 1.36 \cr
$\eta_2(1645)$ &$K^*\Kb$ & 0.03 & 0.80 & $<0.01$ & 0.08 & $<0.01$& 0.02 & 0.07 & 0.81 \cr
\hline\\[-2.3ex]
$PS$ & & 0.82 & 2.71 & 0.04 & 0.17 & 0.10 & 0.28 & 0.42 & 2.88 \cr
$K^*_{\rm ne}$ & & 0.11 & 0.19 & 0.01 & 0.03 & 0.02 & 0.07 & 0.11 & 0.26 \cr 
$K^*_{\rm ch}$ & & 0.16 & 0.80 & 0.02 & 0.19 & 0.04 & 0.25 & 0.26 & 0.91 \cr
\hline\\[-2.3ex]
  & & \multicolumn{8}{c}{Phases [rad]} \cr
Contribution & Decay mode & $r$ & BW & $a_0$ & eff & bkg & int & trig & Total \cr 
\hline \\[-2.3ex]   
$\eta(1475)$& $a_0 \pi$ & 0.01 & 0.09 & 0.04 & 0.01 & 0.01 & 0.04 & 0.11 & 0.16 \cr
& $PS$ & 0.01 & 0.10 & 0.01 & 0.03 & 0.01 & $<0.01$ & 0.06 & 0.12 \cr
\hline\\[-2.3ex]
$\eta(1760)$ & $K^*\Kb$ & $<0.01$ & 0.20 & 0.01 & 0.16 & 0.09 &0.02 & 0.06 & 0.28 \cr  
 & $a_0 \pi$& 0.01 & 0.16 & 0.04 & 0.10 & 0.01 & 0.01 & 0.03 & 0.21 \cr
& $PS$ & 0.01 & 0.19 & 0.01 & 0.02 & $<0.01$ & 0.01 & 0.01 & 0.20 \cr
\hline\\[-2.3ex]
$\eta(1405)$ & $K^*\Kb$ & 0.02 & 0.13 & $<0.01$ & 0.14 & $<0.01$ & 0.02 & 0.03 & 0.20 \cr
 & $PS$ & $<0.01$ & 0.19 & $<0.01$ & 0.19 & $<0.01$ & 0.01 & 0.08 & 0.31 \cr
\hline\\[-2.3ex]
$f_1(1285)$ & $a_0 \pi$ & 0.01 & 0.24 & 0.04 & 0.05 & 0.01 & $<0.01$ & 0.08 & 0.26 \cr
$f_1(1420)$ & $K^*\Kb$& $<0.01$ & 0.22 & $<0.01$ & 0.05 & $<0.01$ & 0.02 & 0.05 & 0.23 \cr
\hline\\[-2.3ex]
$h_1(1415)$ & $K^*\Kb [S]$& 0.01 & 0.18 & $<0.01$ & 0.05 & $<0.01$ & 0.02 & 0.06 & 0.20 \cr
& $K^*\Kb [D]$& $<0.01$ & 0.07 & $<0.01$ & 0.14 & $<0.01$ & 0.01 & 0.05 & 0.16 \cr
\hline\\[-2.3ex]
$f_1(1510)$ & $K^*\Kb$ & $<0.01$ & 0.20 & $<0.01$ & 0.42 & $<0.01$ & 0.02 & 0.09  & 0.48 \cr
$h_1(1595)$ & $K^*\Kb [S]$ & $<0.01$ & 0.16 & $<0.01$ & 0.10 & $<0.01$ & 0.01 & $<0.01$ & 0.19 \cr
$\eta_2(1645)$ &$K^*\Kb$ & 0.01 & 0.07 & $<0.01$ & $<0.01$ & $<0.01$ & $<0.01$ &0.05 &0.09\cr 
\hline\\[-2.3ex]
 $PS$ & & 0.01 & 0.14 & $<0.01$ & 0.03 & $<0.01$ & 0.01 & 0.04 & 0.15 \cr
 $K^*_{\rm ne}$ & & 0.01 & 0.08 & $<0.01$ & 0.19 & $<0.01$ & 0.02 & 0.06 & 0.22 \cr
 $K^*_{\rm ch}$ & & $<0.01$ & 0.06 & $<0.01$ & 0.18 & $<0.01$ & 0.01 & 0.01 & 0.19 \cr
\hline
\end{tabular}
}
\end{table}
    
 \begin{table} [ht]
  \centering
  \caption{\small\label{tab:Tab17} Absolute systematic uncertainties on fractions and phases for the \bkskkpim decay.}
          \resizebox{1.0\columnwidth}{!}{
 \begin{tabular}{lc|rrrrrrr|r}
  \hline\\ [-2.3ex]
    & & \multicolumn{8}{c}{ Fractions[\%]} \cr           
  Contribution & Decay mode & $r$ & BW & $a_0$ & eff& bkg & int & trig & Total \cr
\hline  \\[-2.3ex]  
$\eta(1475)$& $K^*\Kb$ & 0.13 &1.28 & 0.11 & 0.09 & 0.03 & 0.07 & 0.59 & 1.43 \cr
& $a_0 \pi$ & 0.14 & 0.30 & 0.11 & 0.04 & 0.02 & 0.08 & 0.01 & 0.62 \cr
& $PS$ & 0.12 & 2.23 & 1.17 & 0.31 & 0.07 & 0.08 & 1.03 & 2.74 \cr
\hline\\[-2.3ex]
$\eta(1760)$ & $K^*\Kb$ & 0.09 & 0.40 & 0.19 & 0.02 & 0.01 & 0.02 & 0.07 & 0.53 \cr
& $a_0 \pi$& 0.07 & 0.23 & 0.11 & 0.03 & 0.01 & 0.03 & 0.08 & 0.29 \cr
& $PS$ & 0.48 & 4.57 & 2.04 & 0.09 & 0.08 & 0.55 & 0.77 & 6.15 \cr
\hline\\[-2.3ex]
$\eta(1405)$ & $K^*\Kb$ & 0.08 & 0.57 & 0.22 & 0.08 & 0.01 & 0.03 & 0.23 & 0.73 \cr
& $PS$ & 0.17 & 0.77 & 0.19 & 0.10 & 0.01 & 0.05 & 0.41 & 3.54 \cr
\hline\\[-2.3ex]
$f_1(1285)$ & $a_0 \pi$ & 0.01 & 0.19 & 0.02 & 0.05 & $<0.01$ & 0.03 & 0.14 & 0.54 \cr
$f_1(1420)$ & $K^*\Kb$& 0.01 & 1.49 & 0.80 & 0.09 & 0.01 & 0.12 & 0.01 & 1.71 \cr
\hline\\[-2.3ex]
$h_1(1415)$ & $K^*\Kb [S]$& 0.14 & 3.19 & 1.47 & 0.25 & 0.02 & 0.27 & 0.16 & 3.54 \cr
& $K^*\Kb [D]$& 0.01 & 0.11 & 0.16 & 0.09 & 0.01 & 0.03 & 0.09 & 0.24 \cr
\hline\\[-2.3ex]
$f_1(1510)$ & $K^*\Kb$ & 0.08 & 2.72 & 0.06 & 0.10 & $<0.01$ & 0.07 & 0.26 & 2.74 \cr
$h_1(1595)$ & $K^*\Kb [S]$ & 0.08 & 2.17 & 1.64 & 0.44 & 0.03 & 0.16 & 0.09 & 2.77 \cr
$\eta_2(1645)$ &$K^*\Kb$ & 0.02 & 0.05 & 0.13 & 0.05 & $<0.01$ & 0.02 & 0.10 & 0.18 \cr
\hline\\[-2.3ex]
$PS$ & & 0.81 & 4.79 & 1.19 & 0.07 & 0.07 & 0.42 & 0.26 & 7.33 \cr
$K^*_{\rm ne}$ & & 0.06 & 0.25 & 0.19 & 0.03 & 0.03 & 0.01 & 0.05 & 0.37 \cr
$K^*_{\rm ch}$ & & 0.09 & 0.21 & 0.94 & 0.08 & 0.04 & 0.07 & 0.08 & 0.97 \cr
\hline\\[-2.3ex]
  & & \multicolumn{8}{c}{ Phases [rad]} \cr
Contribution & Decay mode & $r$ & BW & $a_0$ & eff & bkg & int & trig & Total\cr
\hline \\[-2.3ex]   
$\eta(1475)$& $a_0 \pi$ & 0.02 & 0.11 & 0.07 & $<0.01$ & 0.01 & 0.02  & 0.01 & 0.13 \cr
& $PS$ & 0.01 & 0.12 & 0.05 & $<0.01$ & $<0.01$ & 0.02 & $<0.01$ & 0.14 \cr
\hline\\[-2.3ex]
$\eta(1760)$ & $K^*\Kb$ & 0.01 & 0.19 & 0.07 & 0.01 & $<0.01$ & 0.02 & 0.01 & 0.21 \cr
& $a_0 \pi$& 0.01 & 0.13 & 0.19 & 0.01 & 0.01 & 0.02  & 0.07 & 0.25 \cr
& $PS$ & 0.01 & 0.20 & 0.06 & $<0.01$ & $<0.01$ & 0.01 & 0.06 & 0.22 \cr
\hline\\[-2.3ex]
$\eta(1405)$ & $K^*\Kb$ & 0.02 & 0.18 & 0.02 & $<0.01$ & $<0.01$ & 0.02 & 0.01 & 0.18 \cr
& $PS$ & 0.02 & 0.19 & 0.03 & $<0.01$ & 0.01 & 0.01 & 0.06 & 0.20 \cr
\hline\\[-2.3ex]
$f_1(1285)$ & $a_0 \pi$ & 0.01 & 0.09 & 0.04 & 0.01 & 0.01 & 0.01 & 0.07 & 0.14 \cr
$f_1(1420)$ & $K^*\Kb$& 0.02 & 0.26 & 0.07 & $<0.01$ & $<0.01$ & 0.02 & 0.02 & 0.27 \cr
\hline\\[-2.3ex]
$h_1(1415)$ & $K^*\Kb [S]$& 0.01 & 0.19 & 0.51 & $<0.01$ & $<0.01$ & 0.01 & 0.03 & 0.55 \cr
& $K^*\Kb [D]$& 0.01& 0.08 & 0.41 & $<0.01$ & $<0.01$ & 0.03 & 0.06 & 0.42 \cr
\hline\\[-2.3ex]
$f_1(1510)$ & $K^*\Kb$ & $<0.01$ & 0.19 & 0.06 & 0.02 & $<0.01$ & 0.02 & 0.12 & 0.24 \cr
$h_1(1595)$ & $K^*\Kb [S]$ & 0.01 & 0.15 & 0.46 & 0.01 & $<0.01$ & 0.01 & 0.01 & 0.48 \cr
$\eta_2(1645)$ &$K^*\Kb$ & 0.01 & 0.07 & 0.02 & 0.03 & $<0.01$ & 0.03 & 0.01 & 0.09 \cr
\hline\\[-2.3ex]
$PS$ & & 0.11 & 0.11 & 0.05 & 0.01 & $<0.01$ & 0.01 & 0.01 & 0.17 \cr
 $K^*_{\rm ne}$& & 0.07 & 0.07 & 0.08 & $<0.01$ & $<0.01$ & 0.01 & 0.10 & 0.16 \cr
 $K^*_{\rm ch}$ & & 0.07 & 0.07 & 0.01 & 0.01 & $<0.01$ & 0.02 & 0.02 & 0.10 \cr
\hline
\end{tabular}
}
 \end{table}
 
\section{Appendix C}
\label{sec:app3}

\begin{figure}[htb]
\centering
\small
\includegraphics[width=0.95\textwidth]{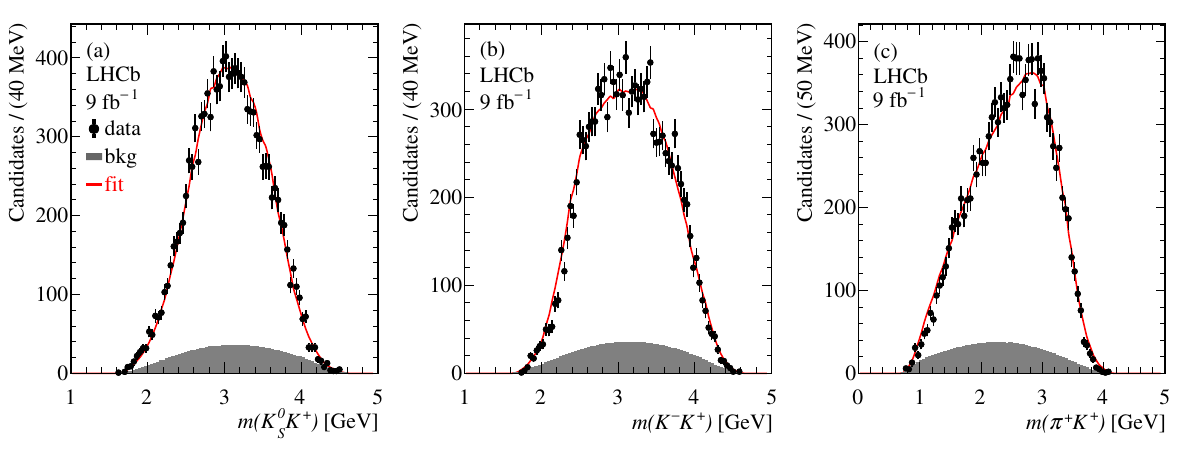}
\includegraphics[width=0.95\textwidth]{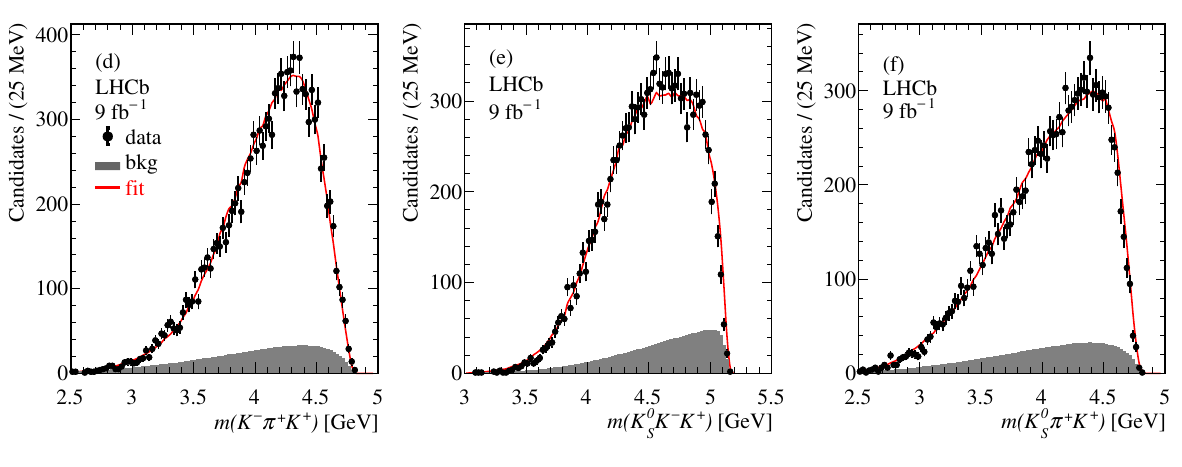}
\caption{\small\label{fig:Fig19} Invariant-mass distributions and fit projections of (a)--(c) two- and (d)--(f) three-body  final-state particles involving the spectator kaon for \bkskkpip data.
}
\end{figure}

\begin{figure}[htb]
\centering
\small
\includegraphics[width=0.95\textwidth]{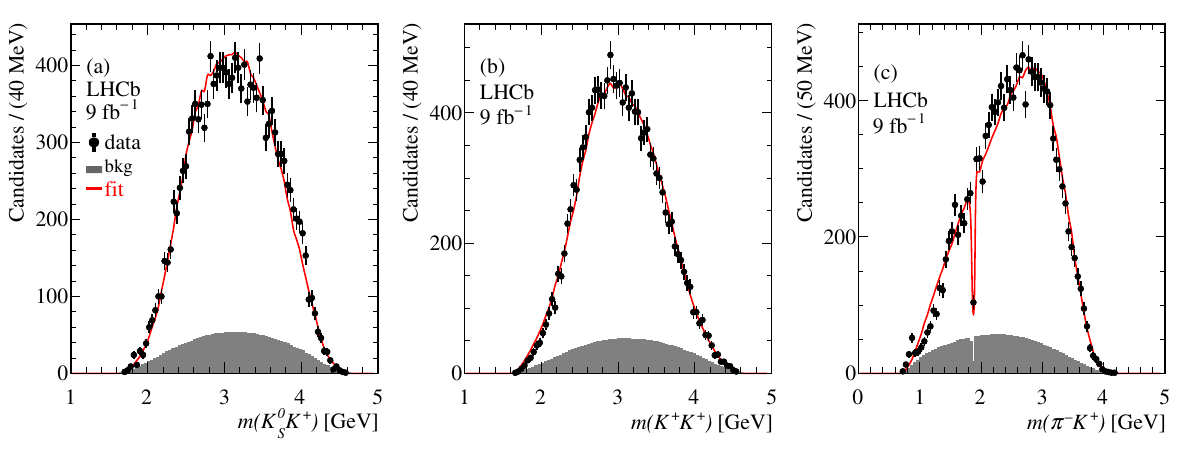}
\includegraphics[width=0.95\textwidth]{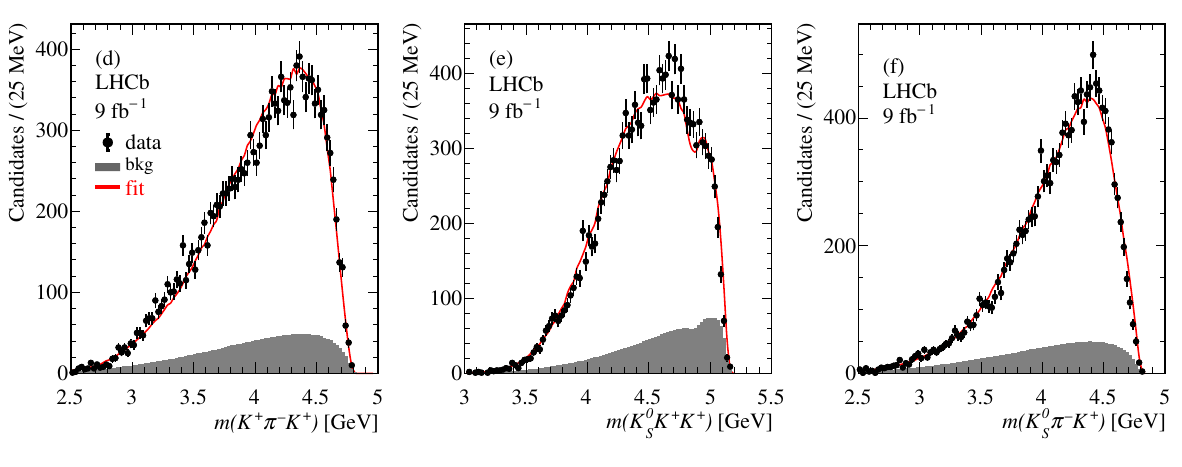}
\caption{\small\label{fig:Fig20} Invariant-mass distribution of (a)--(c) two- and (d)--(f) three-body  final state particles involving the spectator kaon for \bkskkpim data. The depletion in (c) is due to the removal of the open-charm contribution due to $\Dzb \to \Kp \pim$ decays.}
\end{figure}

Figures~\ref{fig:Fig19} and ~\ref{fig:Fig20} show the mass distributions of 
two- and three-body final-state particles involving the spectator kaon for \bkskkpip and \mbox{\bkskkpim} data. The agreement between the fit and data is reasonable with all distributions and fit projections being consistent with phase space, as expected.

\clearpage


\addcontentsline{toc}{section}{References}
\bibliographystyle{LHCb}
\bibliography{main,standard,LHCb-PAPER,LHCb-CONF,LHCb-DP,LHCb-TDR}

\newpage
\centerline
{\large\bf LHCb collaboration}
\begin
{flushleft}
\small
R.~Aaij$^{38}$\lhcborcid{0000-0003-0533-1952},
A.S.W.~Abdelmotteleb$^{57}$\lhcborcid{0000-0001-7905-0542},
C.~Abellan~Beteta$^{51}$,
F.~Abudin{\'e}n$^{57}$\lhcborcid{0000-0002-6737-3528},
T.~Ackernley$^{61}$\lhcborcid{0000-0002-5951-3498},
A. A. ~Adefisoye$^{69}$\lhcborcid{0000-0003-2448-1550},
B.~Adeva$^{47}$\lhcborcid{0000-0001-9756-3712},
M.~Adinolfi$^{55}$\lhcborcid{0000-0002-1326-1264},
P.~Adlarson$^{82}$\lhcborcid{0000-0001-6280-3851},
C.~Agapopoulou$^{14}$\lhcborcid{0000-0002-2368-0147}
% ...

{\footnotesize \it

% ...
$^{14}$Universit{\'e} Paris-Saclay, CNRS/IN2P3, IJCLab, Orsay, France\\
% ...
$^{38}$Nikhef National Institute for Subatomic Physics, Amsterdam, Netherlands\\
% ...
$^{47}$Instituto Galego de F{\'\i}sica de Altas Enerx{\'\i}as (IGFAE), Universidade de Santiago de Compostela, Santiago de Compostela, Spain\\
% ...
$^{51}$Physik-Institut, Universit{\"a}t Z{\"u}rich, Z{\"u}rich, Switzerland\\
% ...
$^{55}$H.H. Wills Physics Laboratory, University of Bristol, Bristol, United Kingdom\\
% ...
$^{57}$Department of Physics, University of Warwick, Coventry, United Kingdom\\
% ...
$^{60}$School of Physics and Astronomy, University of Glasgow, Glasgow, United Kingdom\\
$^{61}$Oliver Lodge Laboratory, University of Liverpool, Liverpool, United Kingdom\\
% ...
$^{69}$Syracuse University, Syracuse, NY, United States\\
% ...
$^{82}$Department of Physics and Astronomy, Uppsala University, Uppsala, Sweden, associated to $^{60}$\\
% ...
}
\end{flushleft}

\end{document}